\documentclass[fleqn,usenatbib]{mnras}
\usepackage{booktabs}
\usepackage{comment}
\usepackage{tabularx}

\usepackage{newtxtext,newtxmath}

\usepackage{xcolor}
\usepackage{hyperref}
\usepackage[T1]{fontenc}

\DeclareRobustCommand{\VAN}[3]{#2}
\let\VANthebibliography\thebibliography
\def\thebibliography{\DeclareRobustCommand{\VAN}[3]{##3}\VANthebibliography}

\usepackage{graphicx}	% Including figure files
\usepackage{amsmath}	% Advanced maths commands
\usepackage{tikz}
\usetikzlibrary{shapes.geometric, arrows.meta, positioning, calc}

\def \HI{{\sc Hi}}

\def \myr {\rm Myrs}
\def \fsiii {f_{\star,\mathrm{III}}}
\def \fxiii {f_{X}}

\title[Impact of XRB Stochasticity on 21-cm Signal]{Impact of Stochastic Pop~III X-ray Binaries on the Cosmological 21-cm Signal}

\author[Dasgupta et al.]{
Saswata Dasgupta,\textsuperscript{\href{https://orcid.org/0000-0001-6461-769X}{\includegraphics[width=2.5mm]{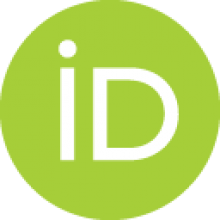}}}$^{1,2}$\thanks{E-mail: saswata.iiti@gmail.com}
Boyuan Liu,\textsuperscript{\href{https://orcid.org/0000-0002-4966-7450}{\includegraphics[width=2.5mm]{orcid.png}}}$^{1,3}$\thanks{E-mail: boyuan.liu@uni-heidelberg.de}
Anastasia Fialkov,\textsuperscript{\href{https://orcid.org/0000-0002-1369-633}{\includegraphics[width=2.5mm]{orcid.png}}}$^{1,2}$
Furen Deng,$^{4,5}$
Jiten Dhandha,$^{1,2}$
\newauthor
~Rennan Barkana$^{6}$   
\\
$^{1}$Institute of Astronomy, University of Cambridge, Madingley Road, Cambridge, CB3 0HA, UK\\
$^{2}$Kavli Institute for Cosmology, Madingley Road, Cambridge, CB3 0HA, UK\\
$^{3}$Institut für Theoretische Astrophysik, Zentrum für Astronomie, Universität Heidelberg, Albert Ueberle Straße 2, D-69120 Heidelberg, Germany\\
$^{4}$National Astronomical Observatories, Chinese Academy of Sciences, 20A Datun Road, Chaoyang District, Beijing 100101, China\\
$^{5}$School of Astronomy and Space Science, University of Chinese Academy of Sciences, No.1 Yanqihu East Rd, Huairou District, Beijing 101408, China\\
$^{6}$School of Physics and Astronomy, Tel Aviv University, Tel Aviv 69978, Israel
}

\date{Accepted XXX. Received YYY; in original form ZZZ}

\pubyear{\the\year{}}

\begin{document}
\label{firstpage}
\pagerange{\pageref{firstpage}--\pageref{lastpage}}
\maketitle

% Abstract of the paper
\begin{abstract}
High-mass X-ray binaries are one of the primary drivers of the 21-cm signal from Cosmic Dawn and Reionization, playing a leading role in the thermal history of the intergalactic medium. In traditional semi-numerical simulations, a deterministic scaling relation between the total X-ray luminosity of high-mass X-ray binaries, $L_{\rm X}$, and star formation rate (SFR) is usually adopted. However, this assumption is inaccurate for high-redshift low-SFR regions hosting few sources. The spatial variation in the number and luminosity of these sources is expected to enhance fluctuations in the Cosmic Dawn 21-cm signal. Here we quantify this effect by introducing a stochastic $L_{\rm X}$ model sampled from a power-law X-ray luminosity function. Implementing this in \texttt{21cmSPACE}, a large-scale simulation framework of Cosmic Dawn and Reionization, and adopting a soft X-ray spectral energy distribution ($\beta=1.5$, $E_{\rm min}=0.1\,\mathrm{keV}$) that leads to a relatively strong stochastic signature, we find that the stochasticity leads to enhanced fluctuations in X-ray heating rate fields, and affects the 21-cm power spectrum on small scales ($k>0.3~ \mathrm{cMpc^{-1}}$). The impact of stochasticity on the global 21-cm signal and on the large-scale power spectrum is found to be negligible. Our results suggest these effects will remain undetected by the upcoming Square Kilometer Array. However, large-scale lunar-based experiments may be sensitive to the signatures of stochastic X-ray heating at $z\sim 25$. Quantifying these corrections is a vital step toward robust 21-cm modeling and ensuring that future precision data interpretation is free from astrophysical biases.
\end{abstract}


%%%%%%%%%%%%%%%%%%%%%%%%%%%%%%%%%%%%%%%%%%%%%%%%%%

%%%%%%%%%%%%%%%%% BODY OF PAPER %%%%%%%%%%%%%%%%%%

\section{Introduction}

The 21-cm line of neutral hydrogen is expected to be one of the most powerful probes of the early Universe, offering a unique window onto the earliest epochs of cosmic evolution, including the Dark Ages (DA), Cosmic Dawn (CD), and the Epoch of Reionization (EoR) \citep[e.g.][]{madau97,furlanetto06}. Radiation from early sources such as the first generations of stars, i.e. Population III (Pop~III) stars, and the subsequent population of X-ray binaries (XRBs) affects the hydrogen signal \citep[e.g.][and references therein]{wouthuysen52, field58, fialkov14a, gesseyjones25}, thereby imprinting their signatures onto the high-redshift observable. This dependence makes the 21-cm signal one of the best probes of the high-redshift heating and ionization of the intergalactic medium (IGM).

The immense scientific promise of the 21-cm signal in unveiling the early Universe has motivated extensive observational efforts with low-frequency radio telescopes. Currently, observations concentrate on two summary statistics: the sky-averaged global signal targeted by instruments such as EDGES \citep{bowman18}, SARAS \citep{singh22}, REACH \citep{eloy22}, MIST \citep{monslave24}, RHINO \citep{bull24}, GINAN \citep{McKay2025} and PRI$^{\rm{Z}}$M \citep{Philip2019} among others, and the power spectrum of spatial fluctuations observed by interferometric arrays such as MWA \citep{tingay13}, LOFAR \citep{van15}, NenuFAR \citep{mertens21,munshi25} and HERA \citep{deboer15}. Future experiments might also attempt to construct three-dimensional tomographic maps of the 21-cm signal \citep[e.g. the under-construction Square Kilometer Array, SKA][]{koopmans15}, and extend the measurements into the cosmic DA using moon-based telescopes \citep[e.g.][and references therein]{bale23, burns20, chen21, brinkerink25}.

Despite the dedicated observational effort, the detection of the cosmological 21-cm signal remains extremely challenging due to its faint nature, particularly in comparison to the bright Galactic and extragalactic foregrounds \citep{shaver99,bharadwaj05,jelic08,jelic10,datta10}. Additional hurdles include the telescope beam modeling \citep{kim22,dasgupta23}, instrumental noise \citep{nasirudin20, pal25}, the effect of the ionosphere \citep{datta16,shen21}, calibration error \citep{sims20}, horizon effect \citep{pattinson24}, among others \citep{nasirudin20, ohara24,carucci20,cunnington21,rath25,ohara25}.

A key advance in the field came from the EDGES Low-Band experiment, which detected a tentative cosmological signal—a deep absorption feature centered at a redshift of $z\sim 17$ \citep{bowman18}. The following SARAS 3 attempt to verify this signal resulted in the best-fit EDGES profile being rejected with 95.5\% confidence \citep{singh22}.
In addition to the observational discrepancy, the best-fit EDGES Low-Band absorption profile is too deep to be explained by conventional astrophysical modelling, requiring exotic astrophysical scenarios beyond the $\Lambda$CDM paradigm \citep[e.g.][]{barkana18,berlin18,feng18,fialkov18b,hills18,munoz18,brandenberger19,Liu2019,mittal22}.
However, it is not yet clear whether the EDGES Low-Band signal is truly cosmological, with studies such as \citet{hills18b, singh19, bradley19, sims20} questioning its origin and suggesting systematic uncertainties as a potential source of the features in the data. The quest for experimental verification continues with other global experiments, including SARAS 3, REACH, MIST and GINAN; while interferometers \citep[e.g.][]{Gehlot2020} are searching for the corresponding enhancement in the power spectrum \citep[e.g.][]{fialkov19, reis20}.
While a robust detection remains elusive, the 21-cm signal parameter space is being systematically narrowed by increasingly stringent upper limits from current experiments \citep{heraupper22, Ghara2025, Trott2025}, alongside growing synergies with JWST and other high-redshift probes \citep{Dhandha2025b, Sims2025}. Looking forward, the SKA, currently under construction, is designed to detect the 21-cm signal and map its evolution across a vast range of cosmic redshifts and spatial scales. 

The committed observational effort requires a matched theoretical development of accurate signal models. The feedback exerted by the first sources on the IGM during CD significantly affects the 21-cm signal. This enables constraints on different properties of Pop~III stars like their formation history, initial mass function, X-ray and UV emission efficiencies \citep[e.g.][and references therein]{fialkov13, madau18, Schauer2019, gesseyjones22}. For instance, recent 21-cm observations from HERA and SARAS 3  constrain the average star formation efficiency of Pop~III stars to be below 5.5\% at 68\% confidence \citep{pochinda24}. It has been shown by \citet{gesseyjones25} that future observations achieving a $25~\rm mK$ sensitivity measurement of the global 21-cm signal can provide $3\sigma$ constraints on the initial mass function of Pop~III stars owing to its strong correlation with the X-ray efficiency \citep{sartorio23} and Lyman-band luminosity \citep{gesseyjones22}. Besides, \cite{dhandha25} found that the average X-ray efficiency of high-redshift ($z>6$) galaxies (including Pop~III galaxies) can be constrained with $\sim0.7~\rm dex$ errors at 68\% confidence levels by exploiting synergies between JWST and 21-cm observations, with the total X-ray luminosity, $L_X$, per star formation rate (SFR), $f_X = L_X/\rm{SFR}$, constrained to $f_X = 0.8^{+9.7}_{-0.4}$. 

The existing analysis and parameter inference rely on the deterministic linear scaling relations between SFR and $L_X$ based on local observations \citep[e.g.][]{Grimm2003,Antoniou2016} and binary population synthesis (BPS) studies \citep[e.g.][]{fragos13a, sartorio23, liu23} of XRBs. These studies include high-mass X-ray binaries (HMXBs) in low-metallicity regions, which act as proxies for high-redshift XRB populations \citep{Brorby2016} and inform 21-cm modeling \citep[e.g.][]{pacucci14, fialkov14a, ross17, kaur22}. However, this deterministic approach is accurate only when applied to large samples of XRBs, while the total luminosities of a small cluster of XRBs can show significant scatter around the mean luminosity predicted by the $L_X-$SFR scaling relation.  

In reality, HMXBs are discrete, short-lived objects, and their luminosity and number density depend on the environment  \citep[e.g.][]{mineo12, Gilfanov2022}. The observed luminosity function of HMXBs in the local Universe can be approximated as a power law, i.e. $dN/dL \propto L^{-\alpha}$, where $\alpha\sim 1.6$ \citep{Gilfanov2022}, meaning that most X-rays come from a small number of luminous objects\footnote{The power-law slope is smaller than 2, which means that the total luminosity is dominated by the most luminous objects, as the integrated luminosity of XRBs per logarithm of luminosity $L^{2}(dN/dL)\propto L^{2-\alpha}$ increases with $L$ given $2-\alpha>0$.}. The bright HMXBs become more dominant in more metal-poor regions \citep[see their fig.~3]{lehmer21}. 
Thus, we expect a stochastic relation described by an X-ray Luminosity Function (XLF) to be a better representation of the CD XRB population than the deterministic $L_X-$SFR relation. 

To provide a more realistic description of the thermal history of the IGM, we incorporate a stochastic XLF into the large-scale cosmological semi-numerical framework 
\href{https://www.cosmicdawnlab.com/21cmSPACE}{\texttt{21cmSPACE}} \citep[e.g.][]{visbal12, fialkov13, gesseyjones25}, which models CD and the EoR observable signals, including the 21-cm signal of neutral hydrogen, JWST UVLFs, as well as X-ray and radio backgrounds. While previous work has examined stochasticity in star formation efficiency, photon escape, and star formation scatter—primarily during the EoR \citep[e.g.][]{reis22, nikolic24, murmu24} our study is the first to quantify the impact of the XLF itself on the 21-cm signal. We show a joint treatment combining XLF stochasticity with one of these channels in Appendix~\ref{apdx:poisson_survival}.

In this work, we develop a novel stochastic model to generate $L_{\rm X}$ from the XLF and SFR across the simulation box, effectively capturing the discrete and localized nature of the first sources of heat: Population~III XRBs\footnote{In this work, we focus on short-lived Pop~III HMXBs, which are expected to produce significantly stronger X-ray emission than low-mass XRBs \citep{sartorio23}, although the Pop~III initial mass function remains a subject of considerable debate \citep{klessen23}. As our approach is applicable to all X-ray binaries, we use the term `XRB' henceforth.}. By introducing these stochastic sources into our simulations, we account for significant spatial fluctuations in the IGM temperature—particularly in low-SFR regions—that are not captured by traditional deterministic scaling relations. We discuss how variations in the shape and normalization of the Pop III XLF affect 21-cm observables, contributing to a robust theoretical framework for interpreting next-generation observations with the SKA and upcoming lunar-based missions.

This paper is organized as follows: In \textbf{Section~\ref{sec:method}}, we explain the approach to stochastic modeling of Pop~III XRBs. It entails sampling XRB luminosities from an XLF and accounting for the small-number statistics that are characteristic of regions with low star formation rates. The results are compared with the results of BPS simulations from \citet[]{liu23}. In \textbf{Section~\ref{sec:results}} we present our results of the effect of stochastic heating on the 21-cm signal maps and the 21-cm power spectrum, comparing simulations with stochastic and deterministic modeling of XRBs.
In \textbf{Section~\ref{sec:obs}}, we discuss the implications of our findings for future 21-cm observations, particularly with instruments such as the SKA and future lunar experiments. 

Finally, we summarize our findings and conclusions in \textbf{Section~\ref{sec:conclusion}}.

The simulation used in this study assumes a flat $\Lambda$CDM cosmology with parameters consistent with \textit{Planck} 2013 observations: $h = 0.6704$, $\Omega_b = 0.0490$, $\Omega_c = 0.2678$, $\Omega_m = 0.3169$, $\Omega_\Lambda = 0.6831$, and $n_s = 0.9619$ \citep{planck14}.

\section{Methodology}
\label{sec:method}

The intensity and fluctuations of the 21-cm signal are influenced by both cosmological and astrophysical phenomena \citep[e.g.][]{furlanetto06}. One critical astrophysical process is the injection of energy into the IGM by X-rays. XRBs, particularly black-hole binaries, are believed to be the dominant sources of X-ray photons during CD and the EoR, at $z \approx 6-20$ \citep[e.g.][]{fragos13b}, and are widely recognized as key players in shaping the thermal evolution of the Universe throughout these epochs \citep{fragos13b, fialkov14a, eide18, Gilfanov2022, sartorio23}. In this section, we develop the methodology, first discussing the stochastic modeling of XRB luminosities in subsection \ref{modeling}. We then test the importance of XRB stochasticity and its implications for CD-EoR observables by introducing this prescription into \texttt{21cmSPACE} in subsection \ref{sec:sims}.

\subsection{Stochastic modeling of XRB luminosities}
\label{modeling}
Previous theoretical works have suggested that the nature of XRBs in high-redshift environments, such as those containing Pop~III stars \citep{sartorio23}, differs from more metal-rich stellar populations  \citep[e.g.][]{fragos13a,fragos13b,liu23}. These Pop~III stars are thought to have higher X-ray luminosity per star formation rate than their metal-rich counterparts, which can lead to significant X-ray feedback on the IGM, influencing both heating and ionization \citep{fialkov2017}. During the onset of CD, characterized by the small-number statistics of XRBs, it might become important to model the stochastic Pop~III X-ray luminosity, rather than assuming a deterministic $L_X-$SFR scaling relation. We provide a summary of all the symbols used for stochastic XLB modeling in Table \ref{table:symbols}.
\begin{table}
    \centering
    \caption{Summary of symbols used in subsection \ref{modeling} for stochastic XRB luminosity modeling.}
    \label{table:symbols}
    \resizebox{\columnwidth}{!}{
    \begin{tabular}{l c l}
        \toprule
        Symbol & Meaning \\
        \midrule
        $L_X$ & Total X-ray luminosity of an XRB population\\
        $\hat{L}_X$ & Expected total X-ray luminosity of an XRB population\\
        $f_X$ & X-ray efficiency parameter \\
        $N$ & Number of XRBs in a given region \\
        $\hat{N} = \hat{L}_X/\langle \mathcal{L} \rangle$ & Expected number of XRBs in a given region\\
        $\psi_X = dN/dL$ & X-ray luminosity function (XLF) \\
        $\alpha$ & Power-law index of the XLF \\
        $l = L_X / \hat{L}_X$ & Normalized X-ray luminosity of an XRB population \\
        $\bar{l}$ & Expected value of $l$ from the numerical PDF of $l$\\
        $p_{l>0}$ & Probability of finding non-zero $l$ \\
        $\mathcal{L}_i$ & Individual XRB luminosity sampled from the XLF \\
        $\langle \mathcal{L} \rangle$ & Average luminosity per XRB \\
        $t_X$ & Average lifetime of an XRB\\
        $N_s$ & Number of Monte Carlo realizations of XRB populations \\
        \bottomrule
    \end{tabular}
    }
\end{table}

As mentioned above, the traditional deterministic approaches use the linear scaling relation between the SFR and the total X-ray luminosity of a star-forming region \citep{grimm03, mineo12, fragos13a, sartorio23}
\begin{equation}
    L_X/SFR= 3 \times 10^{40} \times f_X ~\rm{erg~s}^{-1}\rm{M}_{\odot}^{-1}\rm{yr},
    \label{eq:lx}
\end{equation}
where $f_X$ is an efficiency parameter with a standard value of $f_X=1$ \citep[although large uncertainties at high redshifts are considered][]{pochinda24,dhandha25}. Eq. \ref{eq:lx} is motivated by the observed linear relationship between X-ray luminosity and SFR in galaxies with relatively constant star formation rate sustained for at least tens to hundreds of Myr—a duration sufficient for the HMXB population to reach statistical equilibrium over multiple formation–death cycles
\citep{Gilfanov2022}. 

In reality, the total X-ray luminosity in a given region is a composite effect of an ensemble of individual sources, whose luminosities follow an XLF, and we should replace $L_X$ in Eq.~\ref{eq:lx} by its expectation value $\hat{L}_X$. For small XRB populations where the XLF is sampled sparsely, such as expected at CD, the local X-ray luminosity can exhibit considerable stochasticity, leading to substantial deviations from the globally-averaged expectation value, $\hat{L}_X$.

Moreover, CD populations are expected to be metal-poor and produce XRBs with higher $f_X$, as shown by both observations \citep[e.g.][]{Brorby2016} and simulations \citep[e.g.][]{fragos13b, sartorio23, liu23}. The XLF is also expected to evolve with redshift: compared to the local Universe, where the XLF follows a power-law slope of $\alpha \sim 1.6$ \citep{Gilfanov2022}, the higher-redshift XLF of metal-poor XRBs is dominated by a small number of very bright XRBs \citep{lehmer21}, potentialling calling for a lower value of $\alpha$.  

A simple order-of-magnitude calculation is insightful to illustrate the picture at CD, especially in the context of our \texttt{21cmSPACE} simulations  (as described in subsection~\ref{sec:sims}) with a typical simulation cell volume $V = 27$ cMpc$^3$. The SFR within this cell can be calculated as $\rm SFR = SFRD \times V$, where SFRD is the SFR density. The expected number of XRBs within this cell can be expressed as $\hat{N} = \hat{L}_X/\langle \mathcal{L} \rangle$ where $\langle \mathcal{L} \rangle$ is the average luminosity per XRB derived as follows:
\begin{equation}
    \langle \mathcal{L} \rangle = \frac{\int \psi_X L dL}{\int \psi_X dL},
    \label{eq:avg_l}
\end{equation}
with the XLF $\psi_X = dN/dL$ describing the distribution of XRB luminosities. Assuming a typical SFRD of $\sim 10^{-4}~\rm M_{\odot}$ yr$^{-1}$ cMpc$^{-3}$ at $z = 20$ \citep{klessen23}, the SFR in a simulation cell is $27 \times 10^{-4}\ M_{\odot}$ yr$^{-1}$. For $f_X = 1$, and using a typical XLF parameterized as a power law $\psi_X \propto L^{-1.6}$ over the interval $L \in \left[10^{35}, 10^{40.3}\right]~\rm{erg~s}^{-1}$ corresponding to the luminosity range of XRBs in local observations \citep[see their fig.~5]{Gilfanov2022}, the expected number of XRBs in the cell is $ \hat{N}  \approx 4$ with $\langle \mathcal{L}\rangle\simeq 2\times 10^{37}\ \rm erg\ s^{-1}$. This relatively small expected number of XRBs, which is representative of what we might find at high redshifts, indicates the need for stochastic modeling. The small-number statistics imply that the actual number (and total X-ray luminosity) of XRBs in each local star-forming region (and, thus, a simulation cell) can vary significantly from region to region, potentially impacting the overall X-ray heating and ionization processes. In reality, we expect an even stronger variation due to large-scale fluctuations in density and velocity fields, which bias local star formation \citep{tseliakhovich11, fialkov12}. Hence, incorporating this stochastic nature  might become important when simulating the impact of XRBs on the IGM at early cosmic times.

We now move on to develop a prescription for stochastic XRB populations at CD that can be used as a subgrid model in cosmological simulations. Consider an ensemble of $N_s$ star-forming regions with the same SFR and $f_X$, each containing a random number $N_j$, $j\in[1,..,N_s]$, of XRBs with individual luminosities $\mathcal{L}_i$, $i\in[1,..,N_j]$, distributed according to an XLF, $\psi_X$. As in the example above, we assume the XLF is parameterized by a power law $\psi_X \propto L^{-\alpha}$ with $\alpha$ in the range from $0$ (high stochasticity) to $2$ (low stochasticity) and $L \in [10^{35}, 10^{40.3}] ~\mathrm{erg~s^{-1}}$. In this scenario, the total luminosity of each region $L_{X,j} = \sum_{i=1}^{N_j} \mathcal{L}_i$ is a random variable, and is a function of both $\psi_X$ and $\hat{N}$. Assuming that the number of XRBs follows a Poisson distribution with the expectation value $\hat{N}$, we can now calculate the PDF of $L_X$ using Monte Carlo sampling of $N_j$ and $\mathcal{L}_i$. It is important to note that a large sample size $N_s$ must be used to calculate the PDF accurately.
In practice, we use $N_{s}=\max[(\alpha+2)\times 10^6/\hat{N},10^4]$. For the ease of interpretability of the results, we use the normalized total luminosity $l_j \equiv L_{X,j} / \hat{L}_X$ for each realization, with the mean value $\bar{l}$ calculated over the $N_s$ realizations. For a large enough $N_s$, we expect $\bar{l}\rightarrow 1$ by definition. Here, we have effectively treated XRB formation as independent stochastic events for simplicity\footnote{This approach is supported by the results of STARFORGE simulations showing that random sampling of individual stars from the IMF (as if they form as independent events) well describe the statistics of star clusters formed in $\sim 2000\ \rm M_\odot$ clouds \citep{Grudic2023}. Since a few $1000\ \rm M_\odot$ of stars is needed to produce one XRB, our model is mainly relevant for even larger star-forming systems, where Poisson distribution + random sampling of XLF will be more suitable. We must point out that whether the sampling of IMF is completely random in star clusters is still in debate \citep[see, e.g.][]{Smith2023,Yan2023}. We defer a detailed investigation to this issue to future work.}.

\begin{figure}
    \includegraphics[width=\columnwidth]{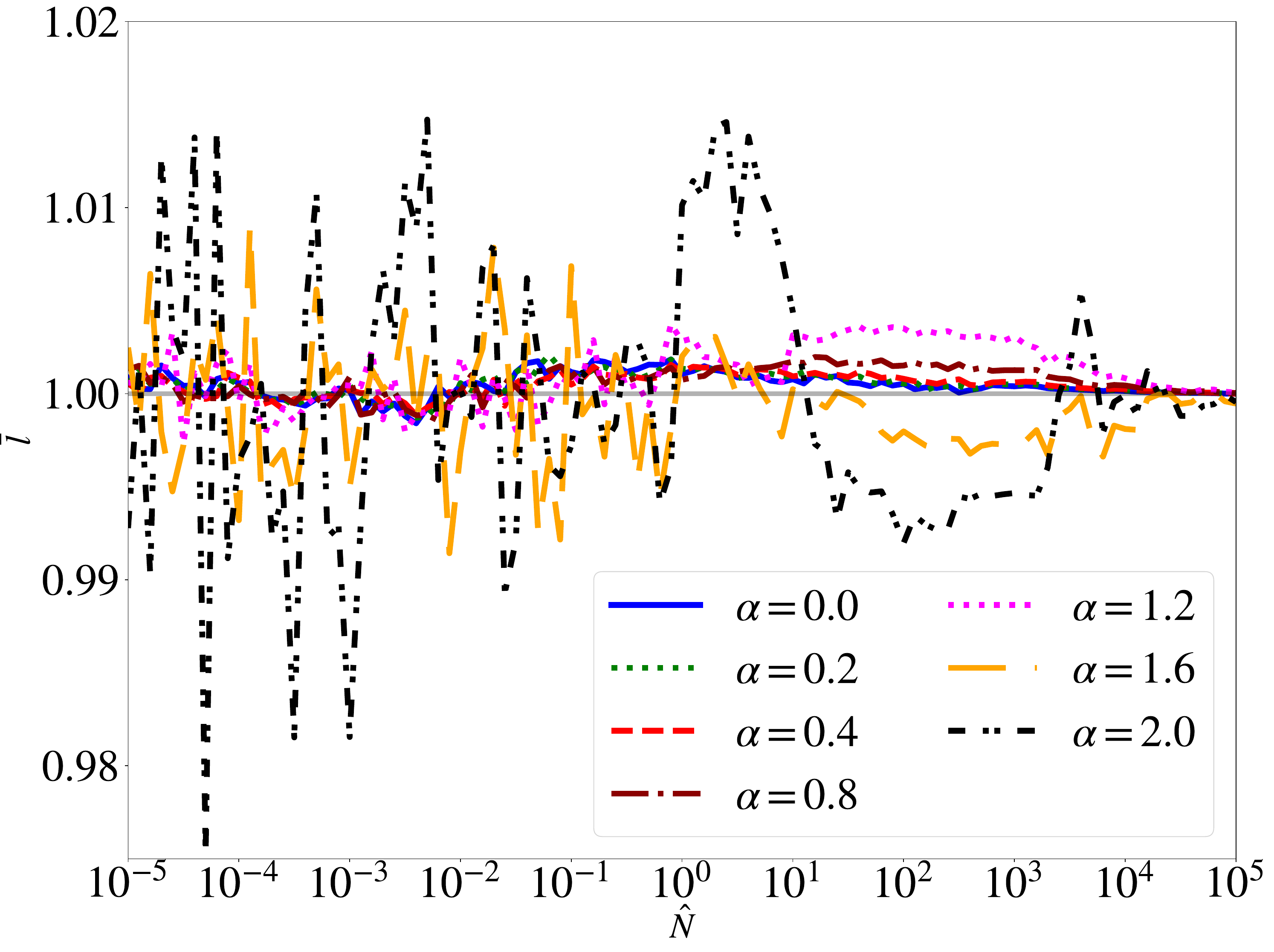}
    \caption{Mean normalized luminosity $\bar{l} = \langle L_X / \hat{L}_X \rangle$ as a function of the expected number of X-ray binaries $\hat{N}$ for various values of the XLF power-law slope $\alpha$. We compare the numerically calculated $\bar{l}$ with $N_{s}=\max[(\alpha+2)\times 10^6/\hat{N},10^4]$ to the theoretical expectation of $\bar{l} = 1$  ($N_s\rightarrow \infty$, solid grey). Each curve corresponds to a different value of $\alpha$, ranging from $0$  (high stochasticity) to $2$ (low stochasticity).}
    \label{fig:CLT}
\end{figure}

Figure~\ref{fig:CLT} provides an essential validation of our stochastic sampling framework, justifying the use of numerically generated PDFs for $l$ across different regimes of $\hat{N}$ and $\alpha$. The figure shows that the mean normalized luminosity $\bar l$  correctly recovers its theoretical value of $\bar{l}=1$ with errors $\lesssim 0.02$ for all tested values of $\alpha\in[0,2]$ and $\hat{N}\in [10^{-5},10^5]$.  The lower bound of $\hat{N}=10^{-5}$ is set to correspond to the lowest-SFR cells in the simulation (an example is given above for a typical value of SFRD $\sim 10^{-4}~\rm M_{\odot}$ yr$^{-1}$ cMpc$^{-3}$  at $z = 20$), while the upper limit $\hat{N}=10^{5}$ covers the most actively star-forming cells at lower redshifts. The strongest agreement between the simulated and the expected values of $\bar l$ is achieved for systems with a large number of binaries, $\hat{N}\sim 10^4-10^5$.

\begin{figure}
    \includegraphics[width=\columnwidth]{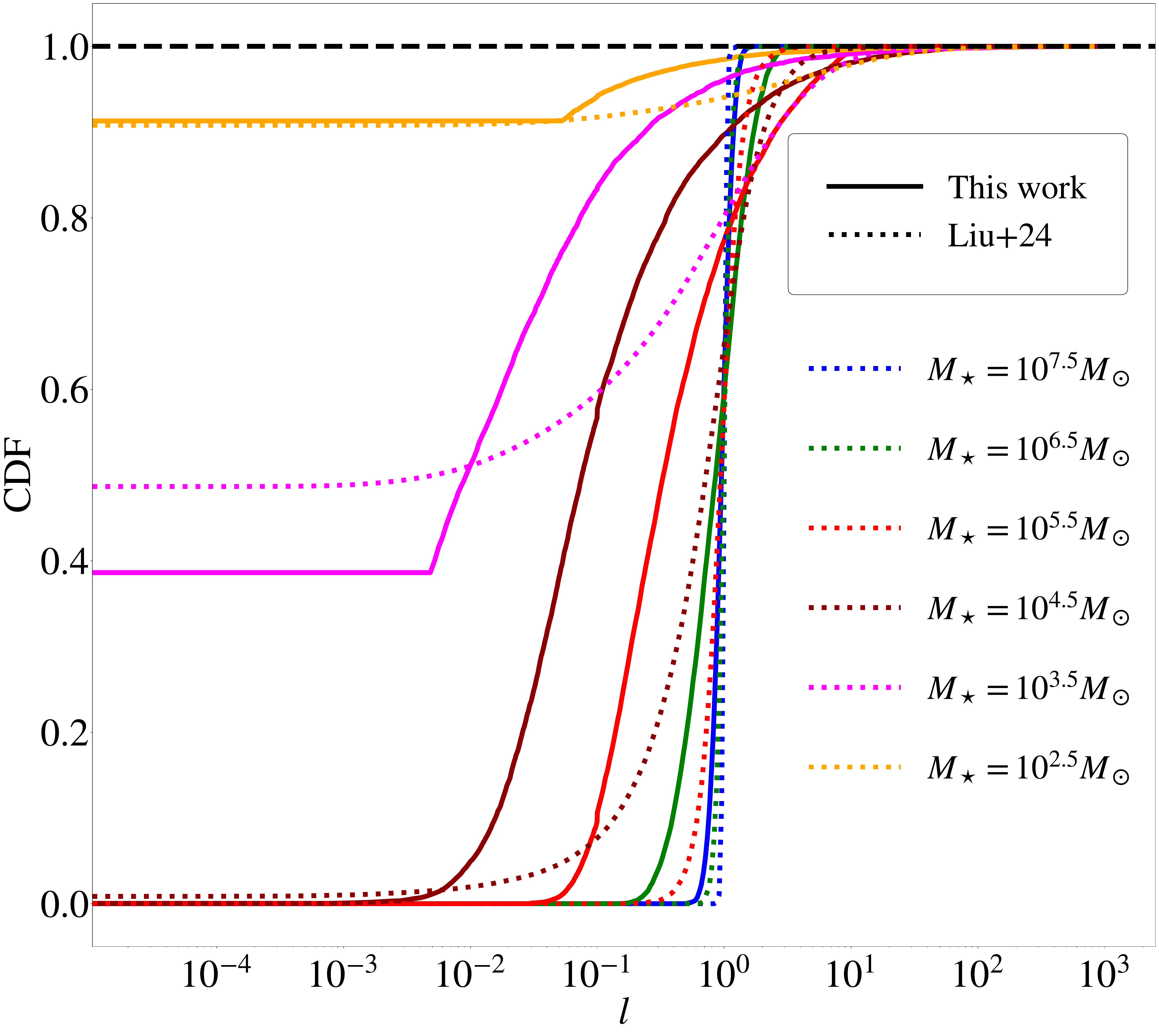}
     \caption{ Comparison of cumulative distribution functions (CDFs) of the normalized X-ray luminosity $l = L_X/\hat{L}_X$ from our stochastic model (solid lines) and BPS simulations from \citet{liu23} (dotted lines). The stochastic model results are produced for a fixed XLF slope $\alpha = 1.6$ and the expected number of XRBs $\hat N = 10^{-1}$ (orange), $\hat N = 1$ (magenta), $\hat N = 10$ (brown), $\hat N = 10^2$ (red), $\hat N = 10^3$ (green), $\hat N = 10^4$ (blue). The BPS results are defined originally by the total stellar mass $M_\star$ underlying the simulated XRB population. To select the BPS case corresponding to a specific $\hat N$ value, we use the relation $\hat N \simeq M_\star / 3000\,M_\odot$ predicted by the BPS run, i.e. $\hat N = 10^{x}$ matches $M_\star=10^{3.5+x}\,M_\odot$. 
        The comparison illustrates that our phenomenological XLF sampling reproduces the general stochastic behavior seen in BPS simulations across a wide dynamic range of total stellar mass, i.e. the distribution of $l$ becomes wider as $\hat{N}$ decreases. Exact agreement is not expected, as the BPS simulations include detailed binary evolution physics, while our model assumes a simple power-law XLF with Poisson sampling.}
    \label{fig:BPS}
\end{figure}

In Figure~\ref{fig:BPS} we show the cumulative distribution functions (CDFs) of the normalized X-ray luminosity $l = L_X / \hat{L}_X$ for different values of the expected number of XRBs $\hat{N}$ and for a fixed XLF slope $\alpha = 1.6$ (solid lines). As $\hat{N}$ increases, the CDF transitions from a broad shape to a steep step-like function centered near $l=1$, indicating a reduced stochasticity in the total X-ray luminosity. For low $\hat{N}$ values ($\hat{N} \lesssim 1$), the CDF is broader and is shifted toward lower $l$, demonstrating the small-number statistics and the possibility of having no luminous XRBs in a given region. The probability of having no XRBs at all is given by the  Poisson distribution and corresponds to the initial value of the CDF (at $l\rightarrow0$).  
These Monte Carlo results are compared with the results of BPS simulations (dotted lines) from \citet[]{liu23}, produced using the publicly available code \texttt{binary\_c} \citep[][]{izzard04, izzard06, izzard23} considering their prescriptions for rotationally-driven mass transfer and critical spin-up (SR\_CS) at  $Z=10^{-4}$. In the BPS case, the XRB sample size is not directly defined by $\hat{N}$ but by the total mass $M_\star$ of the underlying stellar population. 

We can estimate the corresponding value of  $\hat{N}$ from $M_\star$ using the familiar relation $\hat{N}=\hat{L}_{\rm X}/\langle \mathcal{L} \rangle$, with $\hat{L}_X = 3 \times 10^{40} \times f_X \times \rm SFR$ and SFR$=M_{\star}/t_X$. The values of $f_X$, $\langle \mathcal{L}\rangle$, and $t_X$ are predicted by the BPS run, where $t_X\sim 10~\rm Myr$ is the average lifetime of XRBs. In the end, we obtain the relation $\hat{N}\simeq M_{\star}/(3000\ \rm M_\odot)$, which is used to find the matching $\hat{N}$ value for our phenomenological stochastic model based on XLFs.  The BPS results are shown for stellar masses between $10^{2.5}$ and $10^{7.5}\rm \, M_\odot$. % 
These BPS-derived CDFs show general consistency with our simulated CDFs for the matched $\hat{N}$ values, validating our numerical approach across a wide range of stellar populations. % and a large range of values of $\hat{N}$. 
We do not expect a perfect match, as the Monte Carlo approach is based on the simple power-law XLF with Poisson
sampling, while the BPS simulations include detailed binary evolution physics.
The agreement in the overall stochastic behavior is therefore sufficient to justify our approach for the purposes of large-scale 21-cm simulations.

To explore cosmological implications of stochastic XLF using \texttt{21cmSPACE} (Section \ref{sec:sims}), we use a well-known inverse transform sampling method \citep[e.g.][]{gentle2003} to sample $l$ on the fly with parameters that vary across the simulation box and as a function of redshift. 

To save computational time, we pre-compute two functions: (1) the probability $p_{l>0}$ of obtaining a non-zero $l$ from the Poisson distribution and (2) the Inverse Cumulative Distribution Function (ICDF) for $l>0$. Both quantities are evaluated as functions of $\hat{N}'$ and $\alpha'$ on a grid 
$\log\hat{N}'\in [-5,5]$ and $\alpha'\in [0,2]$ of a size $101\times 11$.   
These precomputed grids will be utilized in \texttt{21cmSPACE} as follows: for a simulation run, we assume an input value of $\alpha$, while the expected number of XRBs $\hat{N}_{\rm cell}$ per cell varies from cell to cell, and as a function of redshift, according to the local SFR (analogous to the BPS example above, see Sec.~\ref{sec:sims} below for details). For the given $\alpha$ and $\hat N_{\rm cell}$ in each simulation cell, we determine the value of $p_{l>0}$ by interpolating the first precomputed grid. We then draw a random number $r_1$ from a uniform distribution  $U[0,1]$ and compare it to $p_{l>0}$. If $r_1 > p_{l>0}$, we set $l=0$ in this simulation cell. 
Otherwise, we employ the inverse transform sampling method to evaluate the local value of $l$ by using the interpolated ICDF.

\begin{figure*}
    \includegraphics[width=\textwidth]{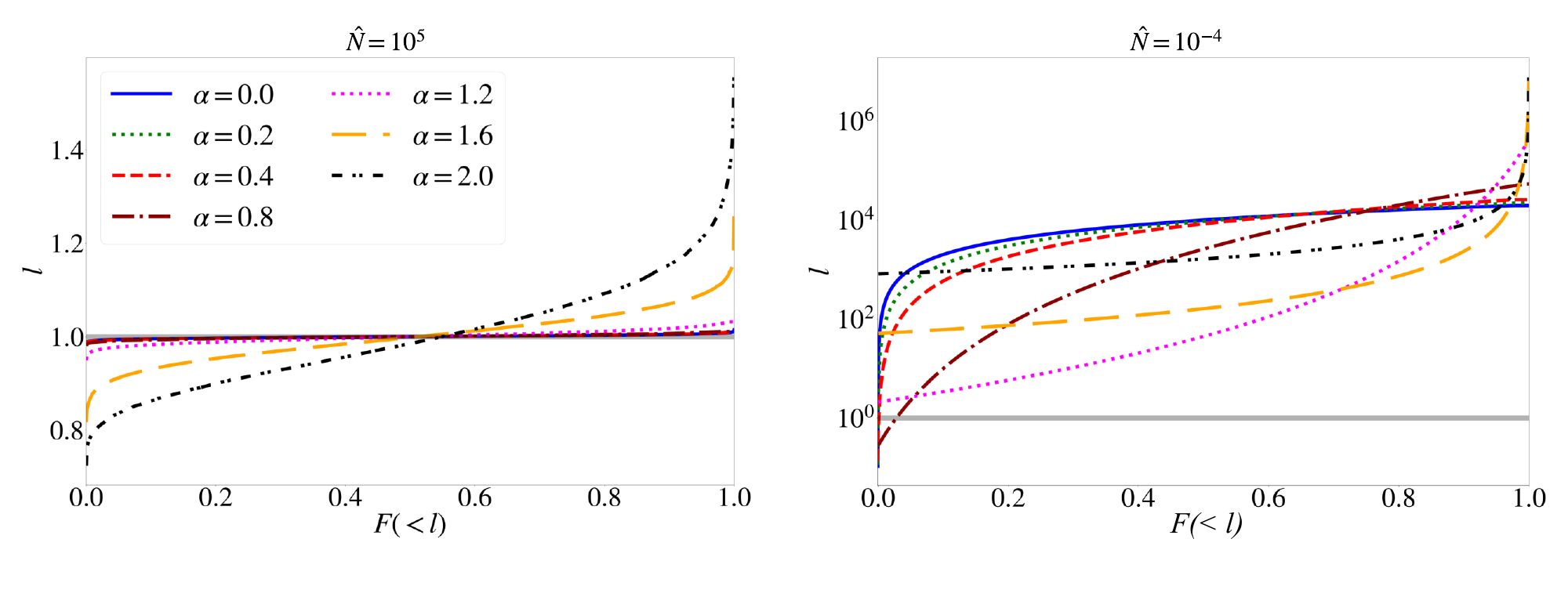}
    \caption{Inverse cumulative distribution functions (ICDFs) of \textit{non-zero} normalized total luminosity $l = L_X / \hat{L}_X$ for large ($\hat{N} = 10^5$, left) and small ($\hat{N} = 10^{-4}$, right) $\hat{N}$ across varying XLF slopes $\alpha$.
\textbf{Left:} In the large-$\hat{N}$ case, the distributions remain sharply peaked and nearly symmetric around $l=1$  For flatter XLFs (e.g.\ $\alpha \leq 1.2$), closely approximating the deterministic behavior assumed in standard 21-cm simulations. As the XLF steepens (e.g.\ $\alpha = 1.6$--$2.0$), the ICDFs broaden slightly, developing asymmetric tails due to enhanced contributions from rare but extremely luminous XRBs. \textbf{Right:} Unlike the large-$\hat{N}$ case, the distributions for small $\hat{N}$ are highly skewed and heavy–tailed, reflecting the fact that rare, high-luminosity XRBs dominate the total luminosity. The sensitivity to $\alpha$ is much stronger in this regime: steeper XLFs yield lower average luminosities (per XRB) and stronger outlier contributions, underscoring the importance of accurately modelling these tails in stochastic 21-cm simulations.}
\label{fig:ICDF_combined}
\end{figure*}

In the left panel of Figure~\ref{fig:ICDF_combined} we show the ICDF of \textit{non-zero} normalized X-ray luminosity $l = L_X / \hat{L}_X$ for a fixed expected number of X-ray binaries $\hat{N} = 10^5$ and a range of $\alpha$ from 0 to 2. This plot demonstrates the effect of the XLF shape on the sampling distribution in the large $\hat N$ limit. The Central Limit Theorem implies that the distribution of $l$ should approach a Gaussian distribution that is more narrowly peaked around $l = 1$ as $\hat N$ increases. In fact, given $\hat{N} = 10^5$, the curves for shallow slopes ($\alpha \leq 1.2$) are very close to the horizontal line $l = 1$ showing only a slight deviation. However, as the slope steepens ($\alpha = 1.6$ and $2$), the deviation from $l = 1$ increases with a visible asymmetry due to the increasingly more extreme luminous outliers from the power-law tail of the XLF. 

This behavior showcases the sensitivity of the distribution of total luminosity to the shape of the XLF, especially for steep slopes. Capturing these subtleties in the ICDF ensures high-fidelity stochastic modeling in our simulations.

The right panel of Figure~\ref{fig:ICDF_combined} illustrates the ICDF of \textit{non-zero} normalized total luminosity $l = L_X / \hat{L}_X$ for a very small expected number of X-ray binaries, $\hat{N} = 10^{-4}$.
This scenario corresponds to the extreme low SFR regimes of early, low-mass halos during CD, where only a tiny number of XRBs may exist in a given volume.

At such low $\hat{N}$, the majority of realizations yield $l = 0$, and only a small fraction of realizations with $l>0$ contribute significantly to heating. The ICDF of non-zero $l$ is highly sensitive to the slope of the XLF as steeper slopes suppress the likelihood of drawing high-luminosity sources and meanwhile reduces $\langle \mathcal{L}\rangle$ (and $\hat L_{\rm X}$ for a fixed $\hat N$). 

In general, the evolution of the ICDF with the XLF slope $\alpha$ is non-trivial, which can leave imprints in the 21-cm signal, as discussed in Sec.~\ref{sec:results}.

\subsection{Implementation in \texttt{21cmSPACE}}
\label{sec:sims}

A key ingredient in the 21-cm signal modelling is an accurate treatment of the thermal and ionization state of the intergalactic gas strongly affected by heating sources such as XRBs \citep{pritchard07, mesinger11, fialkov14a}. To model the impact of stochastic XRBs on the 21-cm signal, we implement the stochastic XLF approach discussed in  Sec.~\ref{modeling} into \texttt{21cmSPACE} \citep{fialkov12, visbal12}\footnote{For details, please check \href{https://www.cosmicdawnlab.com/21cmSPACE/}{https://www.cosmicdawnlab.com/21cmSPACE/}.}, which models a universe across the cosmic DA, CD and EoR. 

The simulations are performed in a box of $(384~\text{cMpc})^3$ resolved into $128^3$ cells, with a cell size of $3\ \rm cMpc$. The simulations are initialized at redshift $z = 50$ with initial conditions for density and velocity fields generated using power spectra from \texttt{CAMB} \citep{lewis11}. 
The evolution of baryons and dark matter density fields on scales larger than the simulation cell is followed using linear theory. On smaller (sub-grid) scales, halo formation and the subsequent gas collapse are addressed using a hybrid halo mass function  \citep{barkanaloeb05b,tseliakhovich11}, and accounting for suppression due to streaming velocities \citep{tseliakhovich10, fialkov12}. A fraction $f_{\star}$ of the collapsed gas in halos above the threshold circular velocity $V_c$ is assumed to form stars.

\texttt{21cmSPACE} adopts a star formation model of the primordial Pop III and the metal-enriched Pop II stars with Pop III stellar spectra calculated using an in-house model \citep{gesseyjones22}, assuming a log-flat Initial Mass Function (IMF) described by a power-law with a slop $\alpha_{\text{III}}$ for stars of masses $\in[M_{\text{min,III}}, M_{\text{max,III}}]$   and accounting for the impact of Lyman-Werner radiation and relative velocity between dark matter and gas on star formation efficiency \citep{fialkov13}. The Pop~II spectra are calculated using \texttt{Starburst99} \citep{leitherer99} as described in \cite{barkana05}. The transition between Pop III and Pop II star formation occurs with a delay $t_{\text{recov}}$, which mimics the impact of supernovae on the recollapse of the star-forming gas  \citet{Magg2022}. Pop III and Pop II XRBs trace corresponding SFRs and have X-ray efficiencies of $f_{X,\text{III}}$ and  $f_{X,\text{II}}$, respectively. In this work, XRB stochasticity is implemented exclusively for Pop~III stars where low-$\hat{N}$ effects are most dominant (see further discussion in Appendix~\ref{apdx:pop2}). For simplicity, we fix $f_{X,\text{II}}$ at its fiducial value of $1$ and hereafter denote the Pop III X-ray efficiency $f_{X,\text{III}}$ simply as $f_X$. Furthermore, the resulting populations of XRBs are assumed to have a power-law spectral energy distributions (SED) with a spectral index $\beta = 1.5$ and a minimum energy cutoff of $E_{\mathrm{min}} = 0.1\,\mathrm{keV}$, for both Pop~II and Pop~III XRBs\footnote{We tested several SEDs for both Pop~II and Pop~III. In this work, our main results are shown for the case with the strongest XRB stochasticity.}. Star formation at late times is susceptible to the photoheating feedback \citep{madau97,barkana04,sobacchi13,cohen16}. Although we focus on the high-redshift regime of CD, for completeness, we include reionization implemented using the excursion-set formalism \citep{furlanetto04} with the ionizing efficiency $\zeta$ and assuming the mean free path of ionizing photons $R_{\text{mfp}}$. In this work, we assume the radio background is entirely attributable to the Cosmic Microwave Background (CMB), with no contribution from galaxies. We refer the reader to our recent publications for further details of  \texttt{21cmSPACE} \citep[e.g.][]{gesseyjones25, dhandha25}.  The astrophysical input parameters of \texttt{21cmSPACE} are listed in Table~\ref{table:params}. 

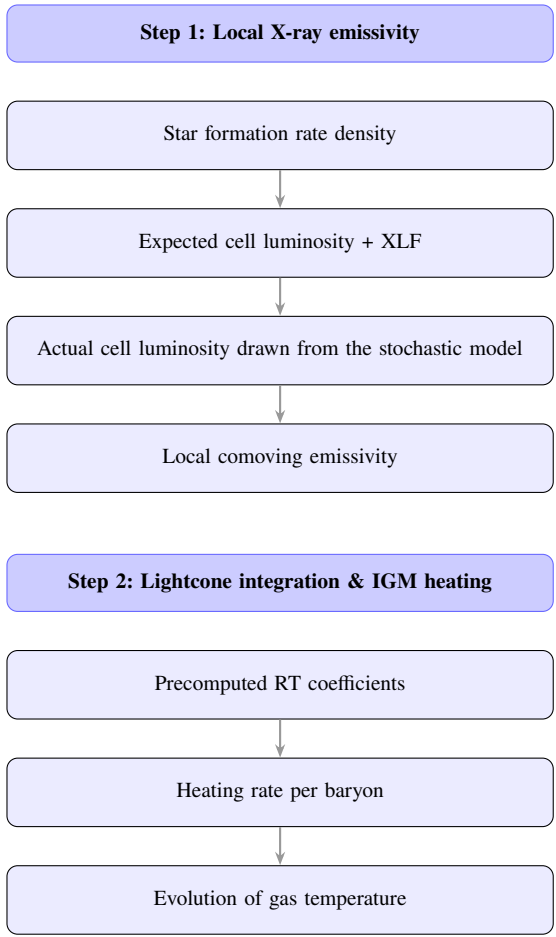
\begin{figure}
    \centering
    \begin{tikzpicture}[
        node distance = 0.5cm and 0.1cm,
        box/.style     = {rectangle, rounded corners=4pt, draw=black,
                          fill=blue!8, text width=7.0cm, align=center,
                          minimum height=0.9cm, font=\small},
        arr/.style     = {-{Stealth[length=5pt]}, thick, gray!80},
        hdr/.style     = {rectangle, draw=blue!60, fill=blue!20,
                          text width=7.0cm, align=center,
                          minimum height=0.75cm, font=\small\bfseries,
                          rounded corners=4pt},
    ]
    
    \node[hdr] (h1) {Step 1: Local X-ray emissivity};

    \node[box, below=of h1] (sfrd)
        {Star formation rate density}; 

    \node[box, below=of sfrd] (lhat)
        {Expected cell luminosity + XLF};

    \node[box, below=of lhat] (stoch)
        {Actual cell luminosity drawn from the stochastic model};
        
    \node[box, below=of stoch] (emiss)
        {Local comoving emissivity};
        
    % --- Step 2 header ---
    \node[hdr, below=0.8cm of emiss] (h2)
        {Step 2: Lightcone integration \& IGM heating};

    \node[box, below=of h2] (coef)
        {Precomputed RT coefficients};

    \node[box, below=of coef] (heat)
        {Heating rate per baryon};
        
    \node[box, below=of heat] (tk)
        {Evolution of gas temperature};
        
    % --- Arrows ---
    % \draw[arr] (h1)    -- (sfrd);
    \draw[arr] (sfrd)  -- (lhat);
    \draw[arr] (lhat)  -- (stoch);
    \draw[arr] (stoch) -- (emiss);
    % \draw[arr] (emiss) -- (h2);
    % \draw[arr] (h2)    -- (coef);
    \draw[arr] (coef)  -- (heat);
    \draw[arr] (heat)  -- (tk);
    \end{tikzpicture}
    \caption{Schematic describing the propagation of the stochastic Pop~III XRB model to the 21-cm signal in \texttt{21cmSPACE}. 
    Step~1 (top) computes the local X-ray emissivity per cell from the local SFRD and the stochastic XRB model.
    Step~2 (bottom) derives the heating rate deposited in each cell via a
    lightcone summation over precomputed radiative-transfer coefficient grids,
    and evolves the gas temperature accordingly (see Eq.~\ref{eq:thermal_new}).}
    \label{fig:pipeline}
\end{figure}

\begin{table*}
    \centering
    \caption{List of astrophysical parameters and their values used in our simulations with \texttt{21cmSPACE}. The X-ray spectral index $\beta$ and the minimum energy cutoff $E_{\mathrm{min}}$ are assumed to be the same for Pop~II and Pop~III XRBs.}
    \label{table:params}
    \begin{tabularx}{\textwidth}{l c X c}
        \toprule
        Parameter & Value & Description & Relevant citation\\
        \midrule
        $V_c$ & 4.2 $\rm km~s^{-1} $& Minimum halo virial circular velocity for star formation & \cite{fialkov12, schauer21, gesseyjones23}\\
        $f_{\star,\text{II}}$ & 0.03 & Star formation efficiency of Pop~II stars & \cite{mirocha17,munoz22,pochinda24}\\
        $f_{\star,\text{III}}$ & 0.003 & Star formation efficiency of Pop~III stars & \cite{gesseyjones22,munoz22,gurian24} \\
        $t_{\text{recov}}$ & 30 \myr & Recovery time after Pop~III feedback & \cite{chiaki18, Magg2022} \\
        $\zeta$ & 15 & UV ionizing efficiency per baryon & \cite{furlanetto06,planck18} \\
        $R_{\text{mfp}}$ & 50 cMpc & Mean free path of ionizing photons & \cite{wyithe04}\\
        $f_{X,\text{II}}$ & 1 & X-ray heating efficiency for Pop~II XRBs & \cite{fragos13b,fialkov14a}\\
        $f_{X,\text{III}}\equiv f_{X}$ & [1, 100] & X-ray heating efficiency for Pop~III XRBs & \cite{sartorio23} \\
        $\beta,\, E_{\mathrm{min}}$ & 1.5,\ $0.1\,\mathrm{keV}$ & X-ray SED spectral index and min. energy cutoff & [chosen to maximise the stochastic 21-cm signal; see Sec.~\ref{sec:sims}] \\
        $f_{\text{rad}}$ & 0 & Radio production  efficiency by galaxies & \cite{visbal12,cohen16}\\
        $\alpha_{\text{III}}$ & Log-flat & Pop~III IMF slope & \cite{bromm02,sartorio23} \\
        $M_{\text{min,III}}$ & 2 $M_{\odot}$ & Minimum Pop~III stellar mass & \cite{sartorio23} \\
        $M_{\text{max,III}}$ & 180 $M_{\odot}$ & Maximum Pop~III stellar mass & \cite{hirano14}\\
        \bottomrule
    \end{tabularx}
\end{table*}
The algorithm for connecting stochastic Pop~III XRBs to the IGM heating in \texttt{21cmSPACE} is outlined in  Fig.~\ref{fig:pipeline} and discussed below. The pipeline consists of two major steps: (i) compute X-ray emissivity of each simulation cell using the SFRD and the stochastic XRB model, and (ii) use the existing \texttt{21cmSPACE} infrastructure (identical to the one employed for deterministic heating) to calculate X-ray heating and ionization by the population of sources. The simulation outputs three-dimensional cubes of the 21-cm signal, gas temperature (discussed in {\it Step 2} below), ionization fraction, and other derived properties.  

\subsubsection*{Step 1: Local X-ray emissivity of Pop III XRBs}

At every redshift step, \texttt{21cmSPACE} calculates the comoving Pop III\footnote{The treatment of Pop II XRB heating closely follows the deterministic Pop III heating case with differences in star formation feedback and the computation of Pop II SFR \citep[e.g.][]{gesseyjones22}.} SFRD $\dot{\rho}_*(z,\vec{x})$ %$M_\odot\,\mathrm{yr}^{-1}\,\mathrm{cMpc}^{-3}
in every cell of volume $V =27\,\mathrm{cMpc}^3$, and the total Pop III SFR within a cell is $\dot{M}_*=\dot{\rho}_*\,V$. This quantity varies from cell to cell, affected by location-dependent factors such as the large-scale density and streaming velocity, background Lyman-Werner radiation and photoheating feedback. 

Similarly to how we treated the BPS example in Sec.~\ref{modeling}, we can convert the total Pop III SFR into the expected respective X-ray luminosities of the cell  
\begin{equation}
    \hat{L}_X = 3\times10^{40}\,f_X\times\dot{M}_*(z,\vec{x})
    \quad\mathrm{erg}{\ \rm s^{-1}},
    \label{eq:Lhat}
\end{equation}
where $f_X$ is a dimensionless X-ray heating efficiency parameter (Table~\ref{table:params}). The stochasticity is introduced through the cell-specific normalized X-ray luminosity $l$ as defined in Sec.~\ref{modeling}. The actual total X-ray luminosity of the cell is given by $L_X = l \times\hat{L}_X$, and the local comoving X-ray emissivity is
\begin{equation}
    \varepsilon_{\rm XRB}(z,\vec{x})
    = \frac{L_X}{V}
    = l\times3\times10^{40}\,f_X\,\dot{\rho}_*(z,\vec{x})\, \quad\mathrm{erg\,s}^{-1}\,\mathrm{cMpc}^{-3}.
    \label{eq:epsxrb_new}
\end{equation}
For simplicity, we do not model the stochastic variations in the spectral shape (i.e. $l$ is achromatic). The factor $l$ is sampled via its ICDF as described in Sec.~\ref{modeling}, using the effective number $\hat N_{\rm cell}$ of Pop III XRBs contributing during a simulation time-step $\Delta t$, and for the input value of the XLF slope $\alpha$. The typical lifetime $t_X$ of an individual Pop~III XRB determines how many statistically independent XRB generations contribute within $\Delta t$:
\begin{equation}
% \label{eq:tx}
    \hat{N}_{\rm cell} = \max\!\left(\frac{\Delta t}{t_X},\,1\right)\hat{N},
    \label{eq:Nhat_tx_new}
\end{equation}
where $\hat{N}=\hat{L}_X/\langle\mathcal{L}\rangle$ is the expected number of XRBs for one independent population and $\langle\mathcal{L}\rangle$ is the mean XRB luminosity averaged over the XLF. The time-step $\Delta t$ is redshift-dependent and corresponds to the fixed redshift step of $\Delta z=1$  (e.g.  $\Delta t = 8$ Myrs at $z = 25$, $\Delta t = 13$ Myrs at $z = 20$ and $\Delta t = 25$ Myrs at $z = 15$). At high redshifts, when $t_X > \Delta t$, a single draw is made and the random seed is fixed until a full lifetime has elapsed to ensure that the surviving XRBs from previous time-steps are consistently represented. At lower redshifts $t_X < \Delta t$ and multiple independent XRB populations can form within one time-step, reducing the effective stochastic variance through temporal averaging.  We adopt a fiducial value $t_X=10\,\mathrm{Myr}$, within an order of magnitude of Pop~III XRB lifetimes derived by \citet{sartorio23}; the impact of varying $t_X$ is explored in Appendix~\ref{apdx:t_X}.

The deterministic Pop III and the Pop II cases assume $L_X = \hat L_X$ in Eq. \ref{eq:epsxrb_new} to compute corresponding X-ray emissivities.

\subsubsection*{Step 2: Lightcone integration and IGM heating}

X-ray photons emitted by each cell will travel through the IGM, get redshifted and partially absorbed before depositing their energy as heat, ionization and Ly-$\alpha$ radiation \citep[e.g.][]{pritchard07}.

The X-ray heating rate per baryon at each location $\vec{x}$ in the simulation box and at redshift $z$ is a result of a lightcone integration over all contributing emitters in the past and can be written as:
\begin{equation}
    \epsilon_X(z,\vec{x})
    = \sum_{z'>z, x'}
      \varepsilon_{\rm XRB}(z',\vec{x}')\;
      \mathcal{C}(z,z',x_{\rm HI},x_e)
    \quad \mathrm{erg \,s}^{-1}\,
    \label{eq:epsx_new}
\end{equation}
where $\mathcal{C}(z,z',x_{\rm HI},x_e)$ are the precomputed radiative-transfer coefficient grids. Here, $x_{\rm HI}(z)$ is the neutral hydrogen fraction and $x_e$ is the electron fraction. Each of the coefficients encapsulates, for a unit emissivity at $z'$, the energy deposited per baryon at $z$. It is calculated by integrating the product of the normalised X-ray SED $\phi(E)\propto E^{-\beta}$ over photon energies  (in our case, $\beta=1.5$, $E_{\rm min}=0.1\,\mathrm{keV}$, $E_{\rm max}=95.65\,\mathrm{keV}$, $\int\phi(E)\,dE=1$), the IGM transmission $e^{-\tau(E,z,z', x_{\rm HI}, x_e)}$, the photoionization cross-sections of \HI~ and He\,\textsc{i}; and the fraction of photon energy converted to heat \citep{shull85, Furlanetto2010}. Since $l$ enters multiplicatively in $\varepsilon_{\rm XRB}$ (Eq.~\ref{eq:epsxrb_new}), it propagates linearly through Eq.~\ref{eq:epsx_new}, modifying the heating rate in each cell according to the XRB population in its past lightcone.

The heating rate $\epsilon_X$ drives the thermal evolution of the gas in the IGM outside the ionized bubbles via the following differential equation for the kinetic temperature.%. The kinetic temperature evolves as
\begin{align}
     \frac{dT_K}{dz}
    &= \frac{2}{3k_B(1+x_e)}\frac{dt}{dz}\,\Sigma_p \epsilon_p \notag\\ 
    & +\frac{2}{3}\frac{T_K}{n_b}\frac{dn_b}{dz}-\frac{T_K}{1+x_e}\frac{dx_e}{dz}
      + \frac{2T_K}{1+z},
    \label{eq:thermal_new}
\end{align}
where $k_B$ is the Boltzmann constant, $n_b$ the number density of baryons, and $dt/dz=-[(1+z)H(z)]^{-1}$. The first term in this equation contains the heating/cooling rate per baryon, $\epsilon_p$, due to various astrophysical processes (including X-ray heating, $\epsilon_X$). In addition to X-ray heating we include Compton heating \citep{madau97}, Ly-$\alpha$ heating/cooling \citep{reis2021}, and CMB heating \citep{venumadhav18,fialkov19, reis2021}. The second term is heating due to structure formation, the third term represents cooling due to the charged particle creation during the EoR, and the fourth term is the adiabatic cooling of the gas due to the cosmic expansion. This equation is solved together with the coupled equation for $x_e$ \citep[e.g.][]{mesinger11, gesseyjones23}.

The observed 21-cm brightness temperature is calculated in post-processing using the well-known relation:
\begin{equation}
T_{b}(z) = \left(1 - e^{-\tau_{21}}\right) \frac{T_S - T_r}{1 + z}, \label{eq:T21}
\end{equation}
where $T_S$ is the hydrogen spin temperature, $T_r$ is the radio background temperature (typically CMB), and $\tau_{21}$ is the 21-cm optical depth given by:
\begin{equation}
\tau_{21}(z) \approx \frac{3h_{\rm P}c^3 A_{10}}{32 \pi k_B \nu_{21}^2} \frac{x_{\rm HI}(z) n_H(z)}{T_S (1+z)(dv_\parallel/dr_\parallel)}. \label{eq:tau21}
\end{equation}
Here $h_{\rm P}$ is the Planck constant, $c$ is the speed of light, $A_{10} = 2.869\times10^{-15}~\mathrm{s}^{-1}$ is Einstein coefficient of the hyperfine transition, $\nu_{21} = 1420.4,\mathrm{MHz}$ is the rest-frame frequency of the 21-cm line, $n_H(z)$ is the total (physical) hydrogen number density, and $dv_\parallel/dr_\parallel$ is the line-of-sight velocity gradient (Hubble flow plus peculiar velocities).
The spin temperature is regulated by coupling to $T_r$, the kinetic temperature $T_K$, and the color temperature $T_c \approx T_K$ and is expressed as:
\begin{equation}
T_S^{-1} = \frac{x_r T_r^{-1} + x_\alpha T_c^{-1} + x_c T_K^{-1}}{x_r + x_\alpha + x_c},
\label{eq:Tspin}
\end{equation}
where $x_r$, $x_\alpha$, and $x_c$ are the radiative, Lyman-$\alpha$, and collisional coupling coefficients \citep{field58, madau97, furlanetto06, barkana16} , respectively, that quantify the efficiency of each process in driving $T_S$ toward $T_r$, $T_c$, and $T_K$.

\section{Results}
\label{sec:results} 
In this section, we focus on the impact of stochastic Pop III XRBs on the fluctuations in the 21-cm signal and compare these results to the ones obtained via the traditional deterministic approach. 

As discussed in Section~\ref{sec:method}, in regions of low metallicity at high redshifts the XLF of the XRB populations can be different from the observed XLF in the local Universe, with the total luminosity driven by a small number of extremely luminous sources. This, as a result, can lead to strong stochastic effects in the 21-cm field at small scales, especially during CD when the SFRD is relatively low.

To demonstrate the impact of stochasticity on the 21-cm signal, we compare the deterministic case with three stochastic models assuming an extremely stochastic XLF with $\alpha = 0.2$, a moderately stochastic scenario with $\alpha = 1.5$, and a nearly deterministic (weakly stochastic) scenario with $\alpha = 2$. These values are within the range of our precomputed grid spanning $0 \leq \alpha \leq 2$, with $ \alpha =2$ and $\alpha =0.2$ on the grid. While lower values of $\alpha$ than the selected $0.2$ would further enhance the stochasticity, they are also computationally expensive. Our choice of $\alpha = 0.2$ provides a reasonable trade-off between numerical feasibility and the strength of the stochasticity.  The intermediate case of $\alpha = 1.5$ is not on the precomputed grid, allowing us to demonstrate the robustness of our interpolation scheme (see the end of Sec.~\ref{modeling}). This case also confirms that our method can be generalized to arbitrary XLF slopes within the grid boundaries, which will be crucial for future work aiming to constrain $\alpha$ observationally. 

The \texttt{21cmSPACE} runs are performed with the input astrophysical parameters listed in Table \ref{table:params}. 
The specific shape of Pop III  X-ray SED was selected by experimentation to highlight the effect of XRB stochasticity. We demonstrate the results for two cases $f_X=1$ and $f_X=100$, justified by the values obtained by \citet{gesseyjones25} for populations of Pop III XRB with different IMFs (Salpeter-like and log-flat, respectively).

\subsection{Redshift evolution of X-ray  heating rate}
\label{sec:heating_field_z}

\begin{figure*}
    \includegraphics[width=\textwidth,height=\textheight,keepaspectratio]{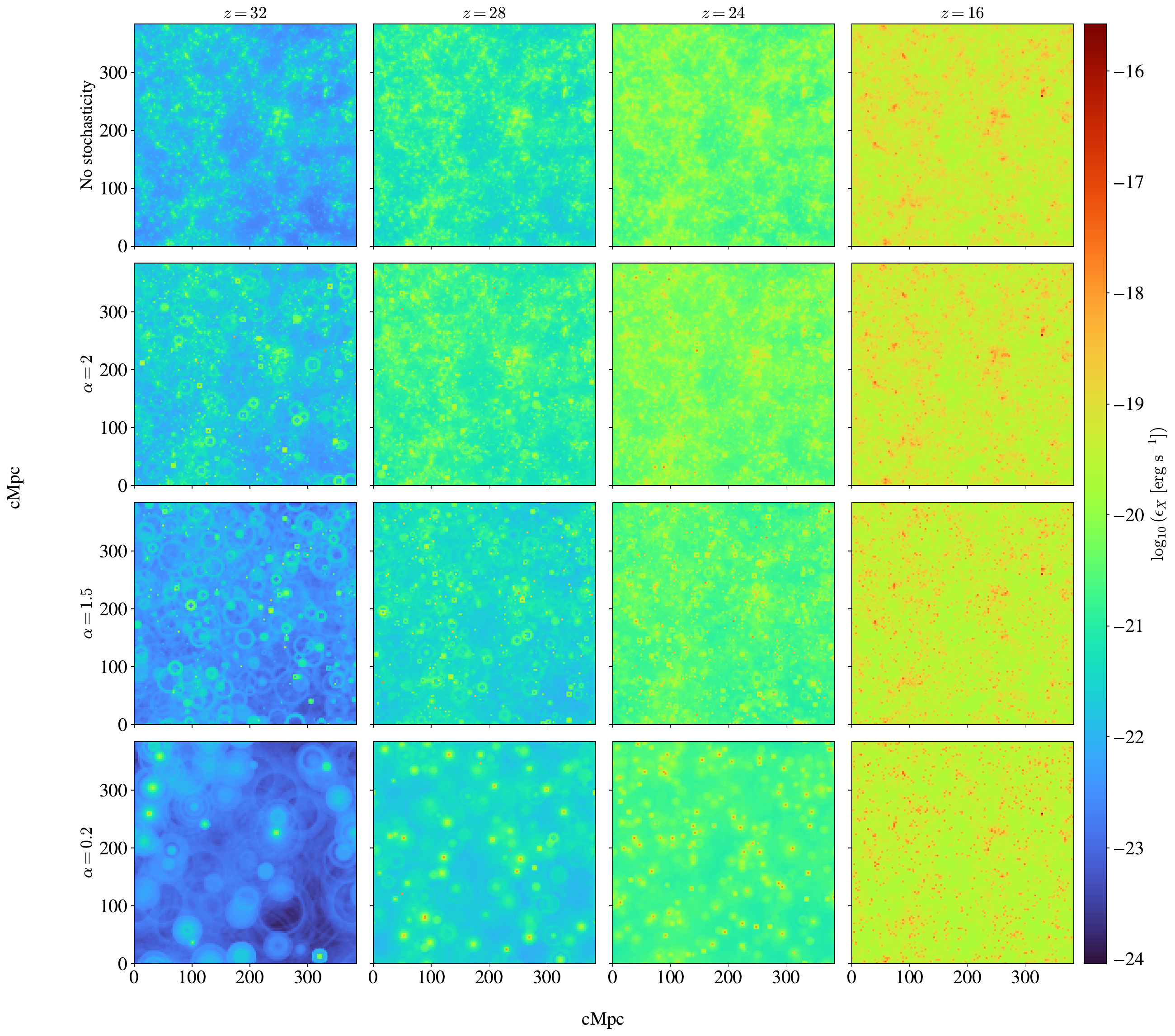}
    \caption{Evolution of the X-ray heating rate per baryon, $\log_{10}(\epsilon_X)$, shown at redshifts $z = 32, 28, 24,$ and $16$ (columns) for one slice of depth $3\,\rm cMpc$ of the heating rate fields. The rows correspond to different cases for the stochastic XRB models: a deterministic case with no stochasticity (top row), followed by stochastic models with XLF slopes $\alpha = 2, 1.5,$ and $0.2$, where smaller $\alpha$ corresponds to stronger stochasticity. At high redshifts, stochastic Pop~III XRBs produce pronounced shell-like heating structures tracing the finite lifetimes of individual XRB populations and the propagation of X-ray photons through the IGM. These features become progressively weaker at lower redshifts as Pop~II star formation begins to dominate the global X-ray emissivity, yielding a heating field largely insensitive to Pop~III stochasticity. Here we assumed $\fxiii=1$, with the rest of the parameters set to their fiducial values (Table \ref{table:params}).}
    \label{fig:eps_evolution}
\end{figure*}

As we alluded to in  Sec.~\ref{sec:sims}, X-ray emissivity (Eq.~\ref{eq:epsxrb_new}) and the resulting X-ray heating rate (Eq.~\ref{eq:epsx_new}) are the key quantities that encode the XRB stochasticity. Figure \ref{fig:eps_evolution} shows the evolution of the X-ray heating rate per baryon with redshift (columns) and for different values of the stochasticity index $\alpha$ (rows). All of the simulations discussed in this subsection were run with the same initial conditions and assuming astrophysical parameters from Table \ref{table:params} with  $f_{X} = 1$.

Due to the finite lifetime of Pop~III X-ray binaries and the inherent stochasticity in their formation, the X-ray luminosity of each simulation cell exhibits temporal fluctuations. %, even under a fixed star formation rate. 
These fluctuations have a characteristic correlation timescale set by the typical lifetime of Pop~III X-ray binaries. As the X-ray photons propagate outward from their production sites, temporal variations in the local luminosity are manifested as spatial modulations in the heating rate per baryon, which are visible in Figure \ref{fig:eps_evolution}. At high redshifts ($z \gtrsim 25$), the impact of stochasticity is more pronounced as very few luminous XRBs appear, and there are ``dark" gaps in the X-ray emissivity fields between the death of one source and the appearance of another. The corresponding heating rate fields inherit this behavior which is manifested as concentric shell-like structures around star-forming regions. The width of the shells reflects the typical lifetime of the X-ray binaries ($10~\myr$ in our case). The intensity of the ring-like features are sensitive to the stochasticity parameter $\alpha$ of the Pop~III XRBs. Models with smaller $\alpha$ exhibit stronger stochasticity, which leads to fewer but more luminous XRBs and correspondingly more pronounced shell-like heating structures, whereas larger values of $\alpha$ are more similar to the deterministic case (top row in Figure \ref{fig:eps_evolution}).

The heating rate is also affected by our choice of the X-ray SEDs. We have selected a relatively soft X-ray SED with spectral index $\beta = 1.5$ and minimum energy cutoff of $E_{\rm min} =0.1$ keV which gives strong stochasticity. For this SED, 50\% of energy is emitted in photons with  $E\lesssim 0.375$ keV and  mean free paths $\lesssim 3.2$ Mpc \citep{pritchard07}. In other words, approximately half of the X-ray energy is injected on sub-grid scales, contributing to the temperature of the host cell. This explains the hot red pixels observed on some of the maps in the centres of the concentric rings and heated bubbles.

As redshift decreases, the star formation rate increases, leading to a more abundant population of X-ray sources with no significant gaps in X-ray emissivity, which explains the gradual disappearance of the shell-like features in the heating rate field.  
This transition is facilitated by the shift in the stellar population from Pop~III to Pop~II at around $z\sim 23$ for the selected astrophysical parameters\footnote{This redshift is model-dependent and strongly depends on the choices of $f_{\star,\text{III}}$, $f_{\star,\text{II}}$, $V_c$ and $t_{\rm recov}$ as well as the Pop III IMF parameters.}, beyond which Pop II XRBs dominate the X-ray emissivity and stochastic effects are expected to be small. We refer the readers to Appendix \ref{apdx:pop2} for more details about the roles of Pop~II XRBs. 

At low redshifts, we expect the stochastic and deterministic cases to converge fully. This trend is visible at the low-redshift end (last column) of Figure~\ref{fig:eps_evolution} where we show the results for $z=16$.

\subsection{Impact of X-ray stochasticity on the 21-cm brightness temperature}

\begin{figure*}
\includegraphics[width=\textwidth,height=\textheight,keepaspectratio]{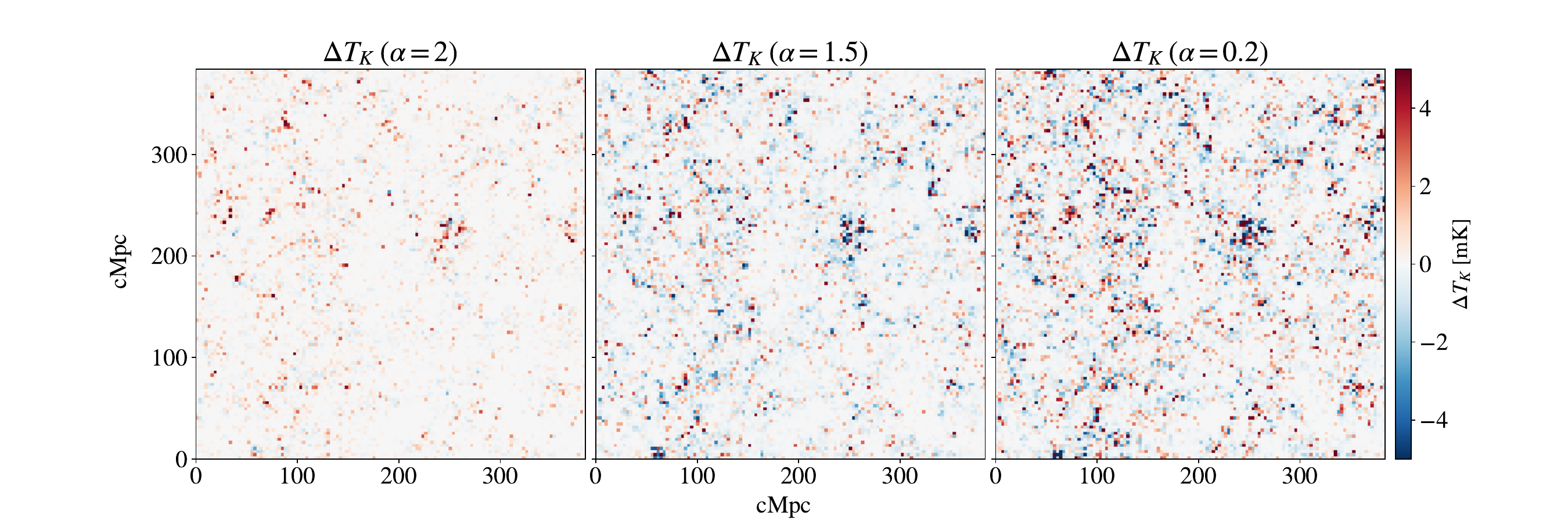}
\caption{Temperature difference maps (with a slice depth of 3~cMpc), $\Delta T_K \equiv T_K^{\mathrm{stoch}} - T_K^{\mathrm{det}}$, between stochastic cases with $\alpha = 2, 1.5,$ and $0.2$ (left to right) and the corresponding deterministic case at $z = 24$. All cases assume the same astrophysical model with $f_{X} = 100$, see Table \ref{table:params} for the values of other model parameters.  The colour scale is symmetric about zero, highlighting local temperature enhancements and differences relative to the deterministic case. Smaller values of $\alpha$, corresponding to stronger stochasticity, produce larger-amplitude but spatially localized temperature fluctuations, while the large-scale temperature features remain unchanged.}
\label{fig:tk_diff}
\end{figure*}

The stochasticity in the X-ray heating rate propagates into the 21-cm signal fluctuations via its effect on the gas temperature (Eq.~\ref{eq:thermal_new}). This link is first demonstrated in Figure \ref{fig:tk_diff}, where we show the contrast in gas kinetic temperature between the stochastic and deterministic cases, $\Delta T_K \equiv T_K^{\mathrm{stoch}} - T_K^{\mathrm{det}}$, for different XLF slopes at $z=24$. Unlike in the previous subsection, here we intentionally use a higher value of Pop~III heating efficiency $\fxiii = 100$ to demonstrate the strongest expected impact of X-ray stochasticity for the selected star formation history (Table \ref{table:params}).

In all the cases shown in Figure \ref{fig:tk_diff}, the temperature fluctuations induced by stochasticity are imprinted close to the source, consistent with the short mean free path of the soft X-ray SED discussed in Sec.~\ref{sec:heating_field_z}, with positive and negative deviations with respect to the deterministic case tracing regions that experience transient over- or under-heating due to the random occurrence and finite lifetimes of individual Pop~III XRBs. The amplitude of these fluctuations increases systematically as the stochasticity parameter $\alpha$ decreases, reflecting that the rare but luminous sources dominate the heating budget. Importantly, the $\Delta T_K$ maps do not exhibit large variations in the large-scale structures with changing stochasticity, indicating that stochastic XRB heating redistributes thermal energy locally without significantly altering the global or large-scale temperature fields.

\begin{figure}
    \centering
    \includegraphics[width=\columnwidth]{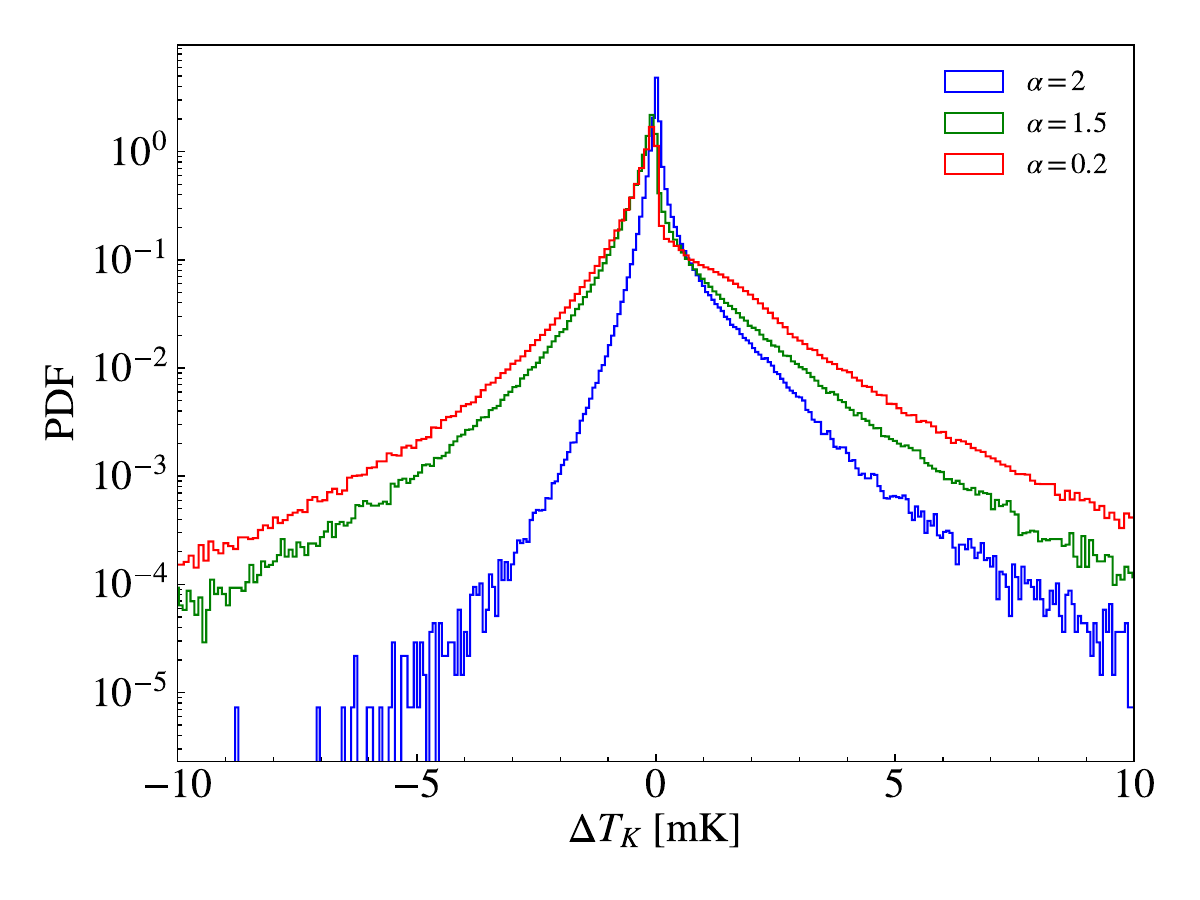}
    \caption{Probability distribution functions (PDFs) of the kinetic temperature differences, $\Delta T_K \equiv T_K^{\mathrm{stoch}} - T_K^{\mathrm{det}}$, at $z = 24$ for Pop III X-ray heating with $f_{X} = 100$. The results are shown for stochastic XRB luminosity distributions with power-law indices $\alpha = 2, 1.5,$ and $0.2$. All PDFs peak close to zero, indicating very little bias in the global temperature, while decreasing $\alpha$ produces broader distributions with pronounced non-Gaussian tails, reflecting stronger local temperature fluctuations induced by rare, luminous X-ray sources.}
    \label{fig:tk_pdf}
\end{figure}

This effect is further solidified by Figure \ref{fig:tk_pdf}, which quantifies the impact of stochastic Pop~III XRB heating on the thermal state of the IGM through the 1D probability distribution functions (PDFs) of temperature differences $\Delta T_K$ relative to the deterministic case. The $\Delta T_K$ distributions are centered around zero, as expected, showing that stochasticity does not introduce a large systematic difference in the global gas kinetic temperature. This validation confirms our design requirement that the average X-ray luminosity be kept constant across all cases. On the other hand, the width of the probability distribution strongly depends on the stochasticity parameter $\alpha$. As $\alpha$ decreases (i.e. for cases with stronger stochasticity), we observe a broader distribution with extended tails, consistent with the boosted small-scale structure seen in the $\Delta T_K$ maps. This is because in such cases, heating is dominated by rare, luminous XRBs. 

\begin{figure*}
    \includegraphics[width=\textwidth,height=\textheight,keepaspectratio]{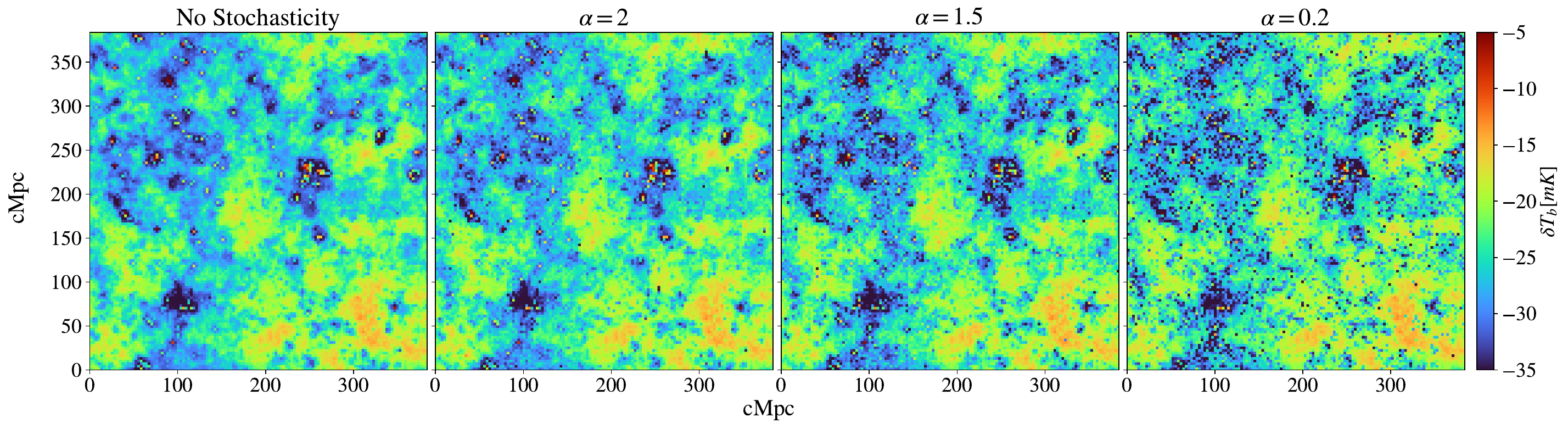}
    \caption{The 21-cm differential brightness temperature maps (with a slice depth of 3~cMpc) at $z = 24$ shown for the deterministic case (left) and for stochastic Pop~III XRB models with $\alpha = 2, 1.5,$ and $0.2$ (left to right) for $\fxiii = 100$. All maps are displayed on the same colour scale for consistency. While the large-scale morphology of $\delta T_b$ remains similar across models, decreasing $\alpha$ enhances the small-scale fluctuations, producing visibly more fine-grained `speckles' in the 21-cm field.}
    \label{fig:temperature_maps}
\end{figure*}

Figure \ref{fig:temperature_maps} shows the resulting 21-cm brightness temperature maps at $z = 24$ for the same astrophysical parameters (with $f_X = 100$) and stochastic models discussed above. The overall large-scale morphology is very similar across stochastic and deterministic models, showing that stochastic Pop~III XRB heating does not introduce large-scale fluctuations in the 21-cm brightness temperature. However, the impact of stochasticity is visible in the small-scale feature of the 21-cm maps: as $\alpha$ decreases, the signal becomes increasingly fine-grained, with enhanced small-scale fluctuations. 

This trend is a direct effect of the stochastic parameter $\alpha$ as well as of our choice of the X-ray SEDs, as discussed in the previous subsection. 
This behavior anticipates the power-spectrum result: stochasticity primarily increases power at high $k$ (small scales), with little effect on the large scales (low $k$). We discuss this in the next subsection.

\subsection{Impact on the 21-cm power spectrum}
\label{sec:stochastic_ps}

\begin{figure}
    \includegraphics[width=\columnwidth,keepaspectratio]{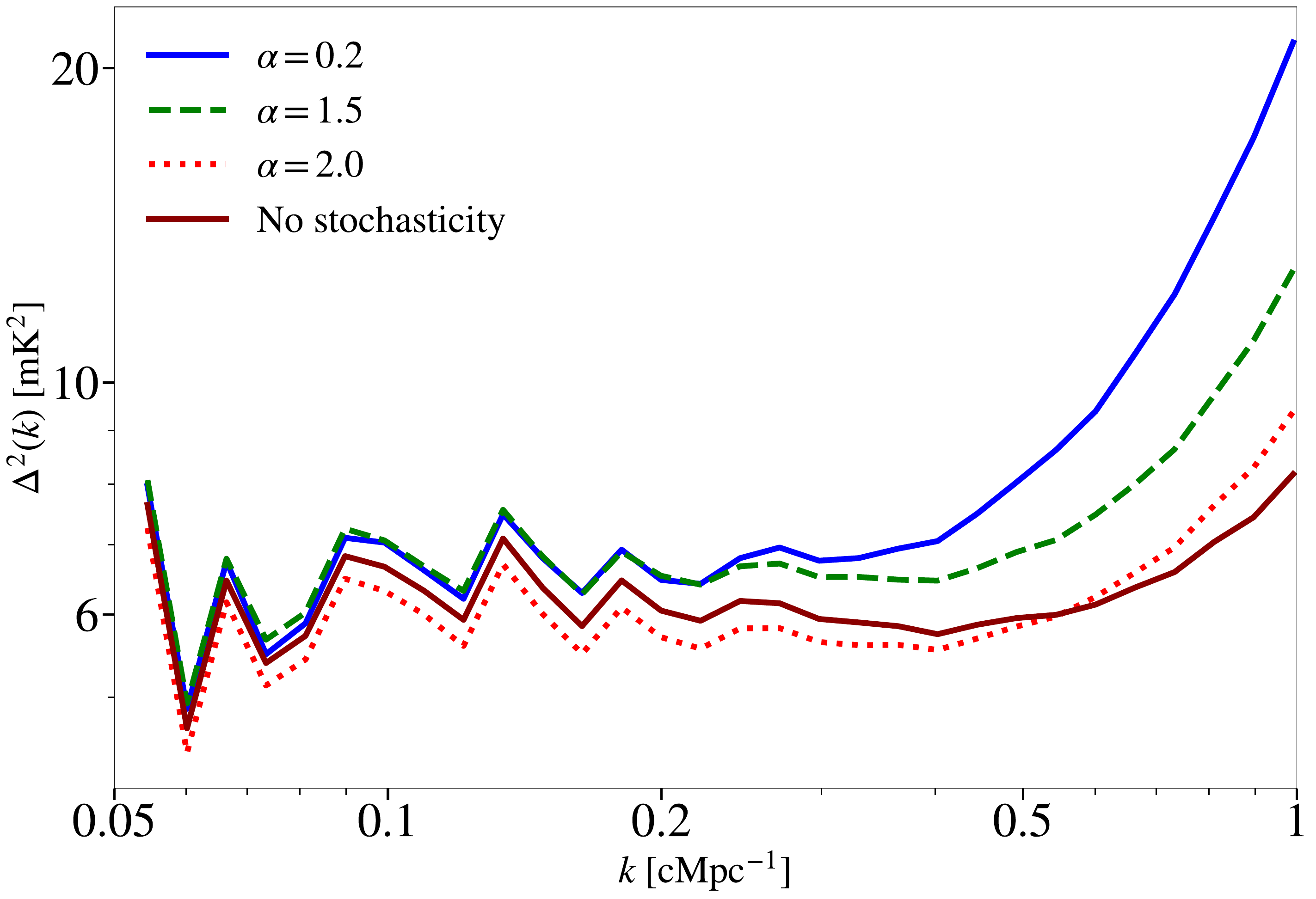
%Images/ps_z24_fx100_fs0003_linear_k.pdf
}
    \caption{21-cm power spectra, $\Delta^2(k)$, at $z = 24$ generated with \texttt{21cmSPACE} assuming the value $\fxiii = 100$ with the rest of the parameters adopting their standard values from Table \ref{table:params}. The power spectra are shown for stochastic Pop~III XRB models with $\alpha = 0.2, 1.5,$ and $2.0$, as well as the case with no stochasticity (as specified in the legend). Decreasing $\alpha$ enhances power on small scales ($k \gtrsim 0.3,\mathrm{cMpc}^{-1}$), while the large-scale power remains largely unchanged. This demonstrates that stochastic Pop~III XRB heating primarily boosts the small-scale 21-cm fluctuations without affecting the large-scale $k$ values.}
    \label{fig:ps_k}
\end{figure}

In Figure \ref{fig:ps_k}, we examine the scale dependence of the 21-cm power spectrum $\Delta^2(k)$ at a fixed redshift ($z=24$) to understand how different $k$-modes respond to X-ray heating fluctuations induced by stochastic XLF of Pop~III X-ray binaries. The results are shown for the same astrophysical parameters as used to produce the 21-cm maps (Fig.~\ref{fig:temperature_maps}), i.e. $\fxiii = 100$ and the values in Table \ref{table:params}. 

While all models exhibit similar power at large scales ($k \le 0.3\, \rm cMpc^{-1}$) and converge  at $k \le 0.1\, \rm cMpc^{-1}$, a clear enhancement of fluctuation in the 21-cm field emerges at high wavenumbers ($k \gtrsim 0.3\, \rm cMpc^{-1}$) as the stochasticity parameter $\alpha$ decreases. The most stochastic model ($\alpha = 0.2$) shows a sharp rise in $\Delta^2(k)$ towards the small scales, whereas models with larger $\alpha$ approach the deterministic limit and have weaker fluctuations at these scales. This behavior is in agreement with the observations made directly from the 21-cm maps, namely that the stochastic nature of Pop III binaries (aided by the selected soft SED) results in relatively localized heating signatures. 

The monotonic increase in small-scale power with decreasing $\alpha$ further highlights the role of rare, luminous XRBs in shaping the fine-grained structure of the 21-cm signal, while reinforcing the conclusion that Pop III XRB stochasticity constitutes a higher-order correction to large-scale 21-cm observables.

\begin{figure*}
    \includegraphics[width=\textwidth,height=\textheight,keepaspectratio]{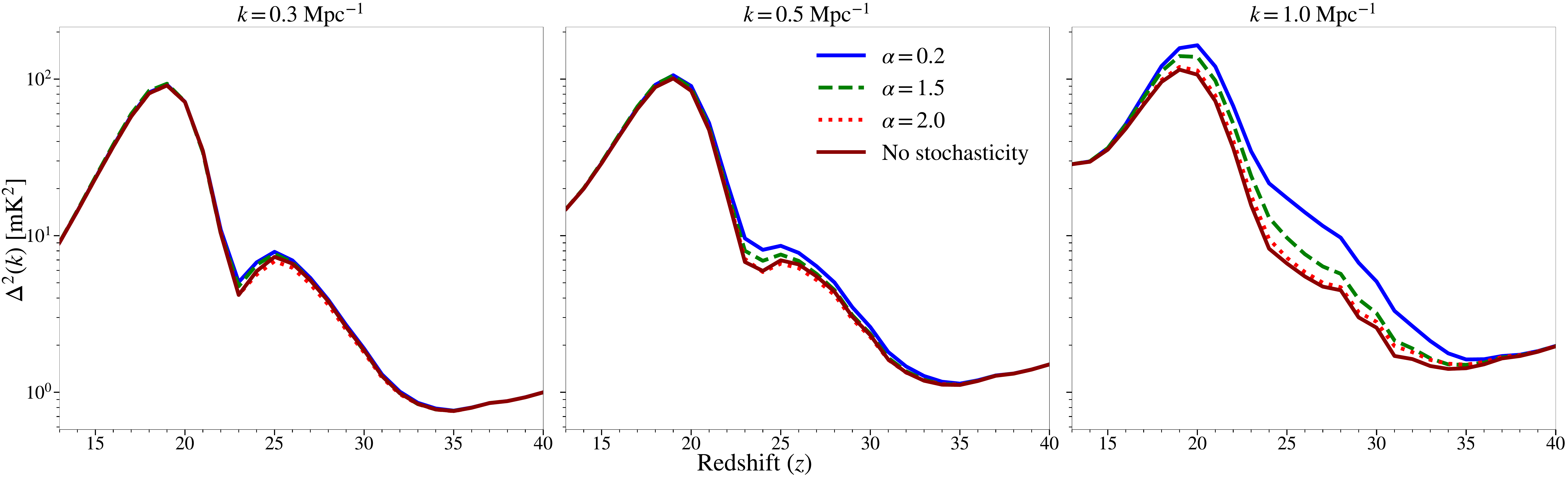}
    \caption{Evolution of the 21-cm power spectrum, $\Delta^2(k)$, as a function of redshift for different comoving wavenumbers $k = 0.3$, $0.5$, and $1.0\ \mathrm{cMpc}^{-1}$ (left to right). The results are shown for stochastic Pop~III XRB models with XLF slope $\alpha = 0.2, 1.5,$ and $2.0$, as well as for the deterministic case with no stochasticity, assuming  $\fxiii = 100$ and the standard values of the astrophysical parameters from Table \ref{table:params}. At all $k$, the curves converge at high redshifts ($z\gtrsim 35$), during Dark Ages in which X-ray heating plays no role and the 21-cm signal is dominated by density fluctuations. At lower redshifts, decreasing $\alpha$ leads to a progressive enhancement of power on small scales ($k \ge 0.3\ \mathrm{cMpc}^{-1}$). The curves converge again at low redshifts ($z\lesssim 16$), where abundant X-ray sources smooth out the stochasticity in X-ray heating (see Fig.~\ref{fig:eps_evolution}). }
    \label{fig:ps_z}
\end{figure*}

Figure \ref{fig:ps_z} shows the redshift evolution of the 21-cm power spectrum at three representative spatial scales, highlighting both the scale-dependence of stochastic Pop~III XRB heating and its redshift evolution. At high redshifts ($z \gtrsim 35$), all stochastic and deterministic models converge at each $k$, in agreement with our expectations for the cosmic DA before the emergence of astrophysical sources. 
In this regime, the thermal state of the IGM is determined by adiabatic cooling and is naturally insensitive to the details of the XRB population, which appears later. As redshift decreases, differences between models begin to emerge in a strongly scale-dependent manner. On large scales (represented by $k = 0.3\ \mathrm{cMpc}^{-1}$), the power spectra remain nearly identical even at $z\sim 23$ when a significant Pop III population is present, indicating that stochastic Pop~III XRB heating does not impact the large-scale features in the 21-cm fields for our choice of X-ray SED. At $k = 0.3\ \mathrm{cMpc}^{-1}$ we find the maximal enhancement of 19\% at $z = 23$. However, at the intermediate ($k = 0.5\ \mathrm{cMpc}^{-1}$) and small ($k = 1.0\ \mathrm{cMpc}^{-1}$) scales, the enhancement of the power spectra by stochastic X-ray population becomes apparent, with decreasing XLF slope $\alpha$ producing the strongest enhancement across the entirety of Cosmic Dawn ($z\sim 15-35$). We find the maximal enhancement of 34\% at z = 23 and 89\% at $z = 25$, $k = 0.5\ \mathrm{cMpc}^{-1}$ and $k = 1.0\ \mathrm{cMpc}^{-1}$ respectively. 
The right-most panel of Figure \ref{fig:ps_z} clearly shows that stochastic X-ray sources have a non-negligible impact on the 21-cm signal even at redshifts traditionally associated with Ly-$\alpha$ physics and well before the build-up of a significant X-ray population that drives cosmic heating. Owing to the small-scale nature of this impact, the signatures do not show up in global quantities such as the evolution of the box-average IGM temperature or the global 21-cm signal. As expected, the stochastic and deterministic cases converge fully again at low redshifts ($z\lesssim 16$), where abundant X-ray sources lead to the disappearance of the stochastic effects in X-ray heating (see Fig.~\ref{fig:eps_evolution}).

\section{Observability of the stochastic X-ray fluctuations}
\label{sec:obs}
\subsection{Prospects with the SKA-Low}
\label{sub:ska}
To quantify the observational relevance of stochastic Pop~III XRB heating, we compute the redshift evolution of the difference signal-to-noise ratio (SNR),
\begin{equation}
\delta\mathrm{SNR}(z,k) \equiv 
\frac{\Delta^2_{\rm stoch}(z,k) - \Delta^2_{\rm det}(z,k)}{\Delta^2_{\rm noise}(z,k)},
\end{equation}
which measures the discrepancy between stochastic and deterministic Pop~III XRB models at a fixed spatial scale $k$, normalized by the SKA-Low thermal noise. We compute the noise power spectrum $\Delta^2_{\rm noise}(k)$ for $\alpha\sim 0.2-2$ in the same ranges of redshifts and scales covered by Fig.~\ref{fig:ps_z}, i.e. $z\sim 13-40$ and $k\sim 0.3-1\ \rm cMpc^{-1}$, following \citet{koopmans15}, assuming 5000 hours of integration time, a 10 MHz bandwidth{\footnote{This bandwidth corresponds to a comoving depth of $\sim220$--$330$~Mpc across $z=15$--35, consistent with the $\sim250$~Mpc redshift-bin depth adopted in SKA-Low Cosmic Dawn/EoR power-spectrum forecasts \citep{greig20}, and well within the 30\% fractional-bandwidth limit for keeping non-linear evolution sub-dominant within a bin identified by \citet{bull24b}.}, and SKA-Low system specifications with the AA4 layout. We stress that $\delta\mathrm{SNR}$ is not a detection SNR for the 21-cm power spectrum itself, but rather a relative metric that quantifies the magnitude of stochastic signatures compared to instrumental noise.
The sign of $\delta\mathrm{SNR}$ indicates whether stochasticity boosts or suppresses power relative to the deterministic case, while its absolute value reflects the magnitude of this difference relative to the thermal noise in dimensionless units. We adopt the power-spectral resolution $\Delta k$ corresponding to the spacing of the spherical $k$-bins used in the simulation output to do this analysis. 

We find that the amplitude of \textbf{$\delta\mathrm{SNR}$} remains below 0.0003, with the maximum of \textbf{$\delta\mathrm{SNR}$} achieved by $\alpha=0.2$ on large scales ($k\sim 0.3\ \rm cMpc^{-1}$) at $z\sim 17$, where SKA-Low exhibits excellent sensitivity. The reason is that the stochastic XRB effects are intrinsically weak on large scales at this relatively low redshift, as shown in Fig.~\ref{fig:ps_z}. On the other hand, while the stochastic effects are intrinsically strongest on small scales ($k\sim 1\ \rm cMpc^{-1}$) at $z\sim 22-30$, the thermal noise of SKA-Low is significantly larger in this regime. Therefore, the amplitude of \textbf{$\delta\mathrm{SNR}$} remains negligible ($\lesssim 4\times 10^{5}$), indicating that these nearly 100\% differences are not robustly detectable with SKA-Low.

Our results demonstrate that while stochastic XRB heating leaves clear imprints on the small-scale structure of the 21-cm signal in noiseless simulations, its detectability with SKA-Low is severely limited by the mismatch between the scales (and epochs) of maximal intrinsic stochasticity and instrumental sensitivity. Furthermore, in our analysis, we have not considered other hindering factors for detectability, e.g. the effect of the array synthesized point spread function, foregrounds, ionosphere etc. These effects will further reduce the $\delta \rm SNR$ and limit the ability to discern the two cases. This motivates us for the exploration of even more advanced instrumental configurations with enhanced sensitivity to small-scale fluctuations and higher-redshift signals, such as low-frequency radio arrays deployed on the lunar far side, which we investigate in the following subsection.

\subsection{Prospects for Lunar-Based Observatories}
\label{sec:lunar}
Given the limitations of SKA-Low in accessing the high wavenumber regime ($k > 0.3~\mathrm{cMpc}^{-1}$ at $z\sim 25$), lunar-based observatories may offer a promising path forward enabling CD and DA science \citep{Mondal2023, fialkov24}. The radio-quiet far side of the Moon provides a pristine environment for low frequency radio astronomy, free from terrestrial radio frequency interference and ionospheric distortions. Despite the premature termination of the NASA's ROLSES-1 mission due to a hard
landing, the experiment returned valuable power spectral density measurements,
demonstrating the feasibility of radio observations from the lunar surface
\citep{burns21}. Future missions like ROLSES-2  \citep{hibbard25} and LuSEE-Night \citep{bale23} are expected to provide first-of-their-kind measurements of the low-frequency sky, including constraints on the global 21-cm signal \citep{hibbard25}. Ambitious proposals, such as \textit{FarView} \citep{polidan24}, \textit{DSL/HONGMENG} \citep{chen21}, \textit{PRATUSH} \citep{rao23} and \textit{DEX/ALO} \citep{brinkerink25} among others, will offer a promising way towards probing the small-scale structure in the CD 21-cm signal that are crucial for detecting the imprint of the stochastic XRBs.

Figure~\ref{fig:snr_diff_lunar} shows the redshift evolution of the difference signal-to-noise ratio, $\delta\mathrm{SNR} = (\Delta^2_{\rm stoch}-\Delta^2_{\rm det})/\Delta^2_{\rm noise, lunar}$, for three $k$-modes and for three stages of a hypothetical lunar far side interferometer as described in \citet{bull24b}. 
The denominator $\Delta^2_{\rm noise, lunar}$ is constructed to match exactly the error model
used in the preceding SKA analysis in subsection \ref{sub:ska}. For each $(z,k)$ and array stage, we calculate the total power $\Delta^2_{\rm tot} = \Delta^2_{\rm signal} + \sigma^2_{\rm noise}$ and define the $1-\sigma$ uncertainty as
\begin{equation}
\Delta^2_{\rm noise, lunar}(z,k) =
\Delta^2_{\rm tot}(z,k)\,
\frac{2\pi}{\sqrt{V_{\rm survey}(z)\,k^2\,\Delta k}} ,
\end{equation}
where $V_{\rm survey}(z)$ is the comoving survey volume corresponding to a bandwidth of $\Delta\nu=5$~MHz\footnote{Corresponding to a comoving depth of $\sim110$--$170$~Mpc across $z=15$--35 -- roughly half the SKA bin depth (Sec.~\ref{sub:ska}), chosen to better resolve the more rapidly-evolving Dark Ages regime targeted by the lunar array.} and an effective sky coverage of $20{,}000~{\rm deg}^2$, scaled from the full-sky comoving volume.  The spectral resolution $\Delta k$ is set to the spacing of the $k$ bins used in the simulation output of \texttt{21cmSPACE}. This definition of $\Delta^2_{\rm noise, lunar}$ includes both thermal noise and sample variance and is identical to the denominator used in our SKA $\delta\mathrm{SNR}$ calculation, allowing us to have a direct comparison between ground-based and lunar-based experiments.

\begin{figure*}
    \includegraphics[width=\textwidth,height=\textheight,keepaspectratio]{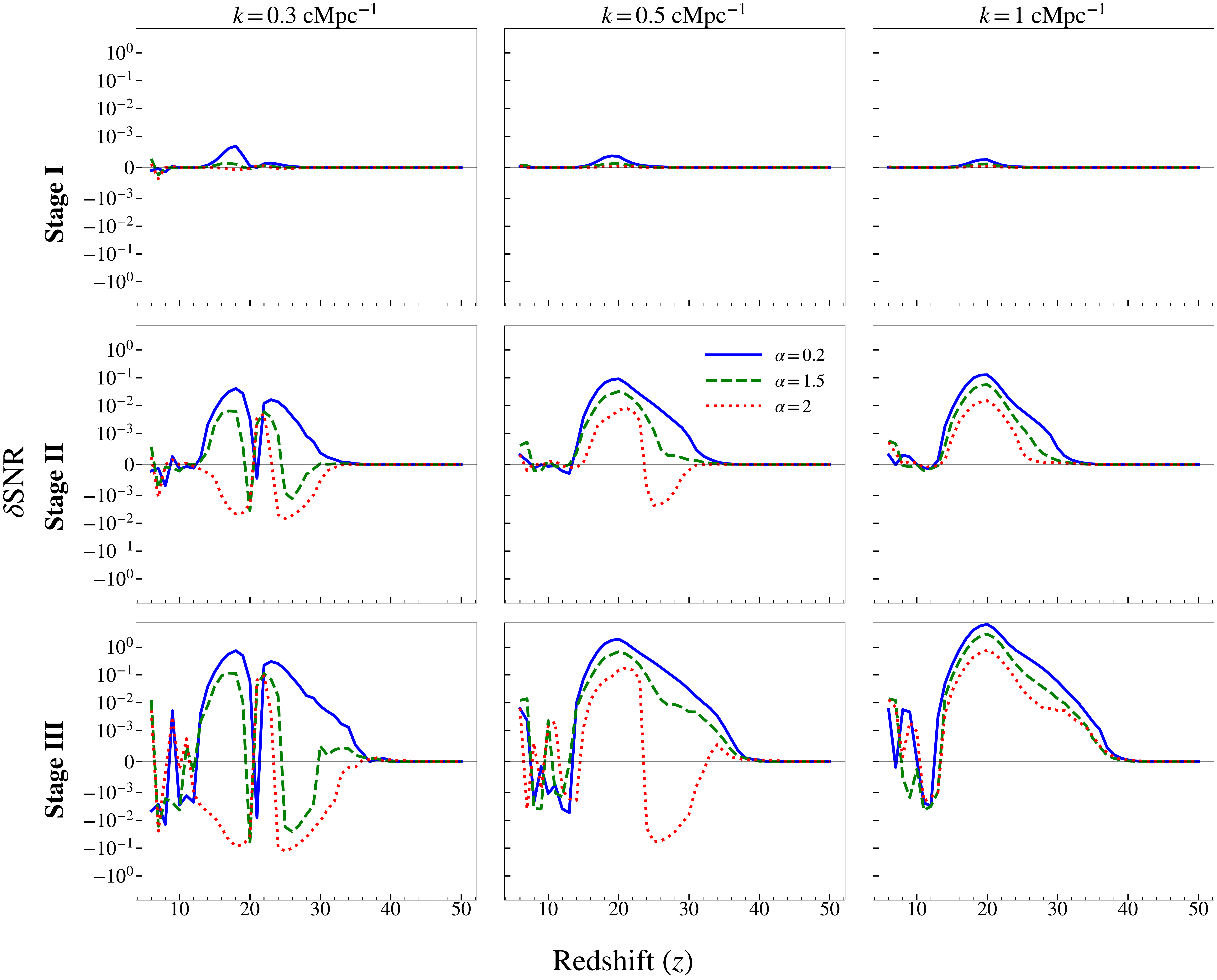}
    \caption{Redshift evolution of the difference signal-to-noise ratio, $\delta\mathrm{SNR} \equiv (\Delta^2_{\rm stoch}-\Delta^2_{\rm det})/\Delta^2_{\rm noise, lunar}$, for three representative length-scales, $k = 0.3$, $0.5$, and $1.0~\mathrm{cMpc}^{-1}$ (columns), and for three successive stages of a proposed lunar far side interferometer (rows: Stage~I, II, and III). Each panel shows the stochastic models with XLF slopes $\alpha = 0.2$, $1.5$, and $2.0$ (solid, dashed, and dotted curves, respectively) for $f_X = 100$. The denominator $\Delta^2_{\rm noise, lunar}$ is the full $1-\sigma$ uncertainty on the power spectrum, constructed from the total power $\Delta^2_{\rm tot} = \Delta^2_{\rm signal}+\sigma^2_{\rm noise}$ and the survey volume, following the same formalism as in the SKA analysis.}
    \label{fig:snr_diff_lunar}
\end{figure*}

The three lunar stages shown in Figure \ref{fig:snr_diff_lunar} correspond to the baseline distributions and antenna numbers proposed in \citet{bull24b}, and are implemented as follows: For each stage, the baseline distribution is characterized by the characteristic radius $r_0$ and width $w$ in the $uv$-plane. Stage~I consists of $N_{\rm ant}=512$ antennas with a compact baseline having $(r_0,w)=(-180~{\rm m},310~{\rm m})$. Stage~II increases the array to $N_{\rm ant}=10^4$ antennas with $(r_0,w)=(-1000~{\rm m},3100~{\rm m})$, and finally Stage~III represents a full-scale array with $N_{\rm ant}=10^5$ antennas and $(r_0,w)=(-2100~{\rm m},6200~{\rm m})$. For all the stages, baselines are truncated between $r_{\rm min}=20~{\rm m}$ and $r_{\rm max}=100~{\rm km}$, and the baseline density is spherically averaged before calculating the noise power spectrum. Here, the negative $r_0$ values indicate a centrally concentrated configuration with maximum density near zero baseline. These configurations progressively improve sensitivity at higher wavenumbers by increasing the density of long baselines.

The thermal noise power spectrum is computed using the same formalism as in
our previous SKA sensitivity calculations shown in subsection \ref{sub:ska}. At each redshift, the system temperature ($T_{\rm sys}$) is modeled as $T_{\rm sys} = 5000 (\nu/50\,{\rm MHz})^{-2.5}$~K. The effective collecting area is assumed as $A_{\rm eff} = \lambda^2/(4\pi)$ per antenna (including a gain factor of 2), and the noise power spectrum $\Delta^2_{\rm noise}(z,k)$ is then computed from the baseline density in the $uv$-plane. A total observing time of $22{,}000$~hours is assumed, distributed over multiple pointings such that the effective integration time per field is $t_{\rm obs} \simeq 1.13\times10^4$~hours. All noise calculations are performed at a fixed redshift, without assuming
foreground avoidance or a wedge cut.

Owing to the intrinsic faintness of the stochastic signatures and the large noise, 
we find $\delta\mathrm{SNR} \lesssim 0.1$ for both Stage~I and Stage~II across all redshifts and scales, despite non-negligible intrinsic differences in the power spectrum.
The stochastic signature becomes marginally detectable with $\delta\rm SNR\sim 1$ in Stage III at large scales $k\lesssim 0.5~\mathrm{cMpc}^{-1}$. %all the proposed experimental stages (I, II and III). 
At the smallest spatial scale explored here, $k=1.0~\mathrm{cMpc}^{-1}$, the intrinsic imprint of stochastic XRB heating is the strongest, but it is also the regime most strongly limited by thermal noise. At these scales, Stage~III achieves $\delta\mathrm{SNR} \gtrsim 1$ near $z \sim 18-22$ for $\alpha \lesssim 1.5$ (up to $\delta\rm SNR\sim10$ at $z\sim 20$ for $\alpha=0.2$), indicating that a sufficiently large lunar array could unambiguously distinguish stochastic and deterministic heating scenarios on these small scales.

As an additional example, the case-D setup of a lunar array in \cite{Mondal2023} is a similarly sized configuration as our Stage~III, albeit with a lower integration time. Thus, we expect lower signal-to-noise in the case D setup compared to the best-case scenario presented here. Another example of lunar arrays considered in literature is the DEX/ALO \cite{brinkerink25},  which, however, is significantly smaller than Stage~III \citep{bull24b}.

Using the above proof-of-concept formalism and bearing in mind that the observational noise can be further scaled down by increased integration time and collecting area, we demonstrated that a sufficiently sensitive lunar far side interferometer might enable identification of the stochastic XRB signal, particularly at $k \gtrsim 0.5~\mathrm{cMpc}^{-1}$.  This finding further motivates lunar-based experiments as a natural extension beyond the SKA-class facilities for studies of faint but science-rich CD signatures. In reality, the interpretation of the stochastic signal will have an inference bias due, which we partially address in Appendix~\ref{apdx:inference_bias}.

\section{Conclusions}
\label{sec:conclusion}

In this work, we have explored the impact of stochastic X-ray heating from X-ray binaries (XRBs), particularly those arising from Population~III (Pop~III) stars on the redshifted 21-cm signal from Cosmic Dawn. Traditional semi-numerical simulations of Cosmic Dawn and the Epoch of Reionization often assume a fixed, deterministic scaling between star formation rate (SFR) and X-ray luminosity. However, this simplification fails to capture the discrete and short-lived nature of X-ray sources in the early Universe, especially in low-mass, low-SFR environments where small-number statistics dominate. In such regimes, the presence or absence of even a single bright source can significantly influence the local thermal and ionization history of the intergalactic medium (IGM), calling for a stochastic treatment.

To address this, we model the  X-ray luminosity of a source as a stochastic variable, sampling from a power-law X-ray luminosity function (XLF) with slope $\alpha$ and treating the number of XRBs per simulation cell as a Poisson-distributed quantity with mean $\hat{N}$ that is determined by SFR, X-ray efficiency $f_{X}$ and average luminosity per XRB $\langle \mathcal{L} \rangle$ (determined by the XLF, see Eq.~\ref{eq:avg_l}). The total luminosity in each cell is then computed via Monte Carlo realizations of individual XRB luminosities, normalized by the expected mean luminosity $\hat{L}_X$, yielding a dimensionless variable $l = L_X / \hat{L}_X$. 
This sampling procedure reproduces the expected statistical behavior of XRB populations across different stellar environments, and is validated against binary population synthesis (BPS) results \citep{liu23}. The resulting distribution of $l$ is used during the simulation runtime to draw stochastic luminosities on a cell-by-cell basis within the \texttt{21cmSPACE} simulation framework.

Our simulations reveal that while the stochasticity has a negligible impact on the thermal history of the universe, the global 21-cm signal and the large-scale power spectrum at $k < 0.3~\mathrm{cMpc}^{-1}$, it enhances power at smaller scales over a broad redshift range $z\sim 15-35$. Owing to the scarcity of sources at Cosmic Dawn, the effects are most salient at $z \gtrsim 25$, where we find an enhancement of up to 100\% at $k=1$ cMpc$^{-1}$ with $\alpha=0.2$ and a Pop III XRB efficiency of $f_X=100$, motivated by recent BPS simulations with a log-flat Pop III stellar initial mass function \citep{sartorio23}.
The stochastic signatures gradually fade away at lower redshifts as source abundance increases. The impact of stochasticity is strongest when the XLF slope ($\alpha$) is shallow. In contrast, high values of $\alpha$ result in X-ray heating fields more similar to the deterministic case. 

Despite these theoretical discrepancies between stochastic and deterministic models, and despite adopting X-ray spectral parameters (Sec.~\ref{sec:sims}) chosen to highlight the stochastic signature, we find that existing and upcoming ground-based interferometers such as SKA-Low may not be able to discriminate between stochastic and deterministic X-ray heating models due to their limited sensitivity and high thermal noise at small scales, where the impact of the stochasticity is mostly pronounced. 
However, some of the proposed lunar-based radio observatories may offer a viable path to accessing the high-resolution observations at low radio frequencies required to detect the subtle features imprinted by stochastic heating. Our results underscore the importance of modeling the stochasticity of the first sources and its impact on the 21-cm signal in the range of redshifts ($z\gtrsim 20$) and scales ($k>0.5\ \rm cMpc^{-1}$) accessible from the Moon. Incorporating stochastic behavior into simulations will be essential for an unbiased interpretation of forthcoming lunar and space observations.

\section*{Acknowledgements}

SD acknowledges the Cambridge Trust and Isaac Newton Studentship for funding his Ph.D. BL acknowledges the funding of the Royal Society University Research Fellowship and the Deutsche Forschungsgemeinschaft (DFG, German Research Foundation) under Germany's Excellence Strategy EXC 2181/1 - 390900948 (the Heidelberg STRUCTURES Excellence Cluster). JD acknowledges support from the Boustany Foundation and Cambridge Commonwealth Trust in the form of an Isaac Newton Studentship. RB acknowledges the support of the Israel Science Foundation (grant no. 1078/24).
%%%%%%%%%%%%%%%%%%%%%%%%%%%%%%%%%%%%%%%%%%%%%%%%%%
\section*{Data Availability}

Snapshot data and post-processed results are made available upon reasonable request.

%%%%%%%%%%%%%%%%%%%% REFERENCES %%%%%%%%%%%%%%%%%%

\bibliographystyle{mnras}
\bibliography{example} % if your bibtex file is called example.bib

@book{gentle2003,
  title={Random Number Generation and Monte Carlo Methods},
  author={Gentle, James E.},
  series={Statistics and Computing},
  edition={2nd},
  year={2003},
  publisher={Springer},
  address={New York},
  doi={10.1007/b97336},
  isbn={978-0-387-00178-4}
}

@ARTICLE{Mondal2023,
       author = {{Mondal}, Rajesh and {Barkana}, Rennan},
        title = "{Prospects for precision cosmology with the 21 cm signal from the dark ages}",
      journal = {Nature Astronomy},
         year = 2023,
        month = sep,
       volume = {7},
        pages = {1025-1030},
          doi = {10.1038/s41550-023-02057-y},
archivePrefix = {arXiv},
       eprint = {2305.08593},
 primaryClass = {astro-ph.CO},
       adsurl = {https://ui.adsabs.harvard.edu/abs/2023NatAs...7.1025M}
}

@ARTICLE{reis2021,
       author = {{Reis}, Itamar and {Fialkov}, Anastasia and {Barkana}, Rennan},
        title = "{The subtlety of Ly {\ensuremath{\alpha}} photons: changing the expected range of the 21-cm signal}",
      journal = {\mnras},
         year = 2021,
        month = oct,
       volume = {506},
       number = {4},
        pages = {5479-5493},
          doi = {10.1093/mnras/stab2089},
archivePrefix = {arXiv},
       eprint = {2101.01777},
 primaryClass = {astro-ph.CO},
       adsurl = {https://ui.adsabs.harvard.edu/abs/2021MNRAS.506.5479R}
}

@ARTICLE{Furlanetto2010,
       author = {{Furlanetto}, Steven R. and {Stoever}, Samuel Johnson},
        title = "{Secondary ionization and heating by fast electrons}",
      journal = {\mnras},
         year = 2010,
        month = jun,
       volume = {404},
       number = {4},
        pages = {1869-1878},
          doi = {10.1111/j.1365-2966.2010.16401.x},
archivePrefix = {arXiv},
       eprint = {0910.4410},
 primaryClass = {astro-ph.CO},
       adsurl = {https://ui.adsabs.harvard.edu/abs/2010MNRAS.404.1869F}
}

@ARTICLE{Trott2025,
       author = {{Trott}, Cathryn M. and {Nunhokee}, C.~D. and {Null}, D. and {Barry}, N. and {Qin}, Y. and {Wayth}, R.~B. and {Line}, J.~L.~B. and {Jordan}, C.~H. and {Pindor}, B. and {Cook}, J.~H. and {Bowman}, J. and {Chokshi}, A. and {Ducharme}, J. and {Elder}, K. and {Guo}, Q. and {Hazelton}, B.~J. and {Hidayat}, W. and {Ito}, T. and {Jacobs}, D. and {Jong}, E. and {Kolopanis}, M. and {Kunicki}, T. and {Lilleskov}, E. and {Morales}, M.~F. and {Pober}, J. and {Selvaraj}, A. and {Shi}, R. and {Takahashi}, K. and {Tingay}, S.~J. and {Webster}, R.~L. and {Yoshiura}, S. and {Zheng}, Q.},
        title = "{Improved Limits on the 21 cm Signal at z = 6.5{\textendash}7.0 with the Murchison Widefield Array Using Gaussian Information}",
      journal = {\apj},
         year = 2025,
        month = oct,
       volume = {991},
       number = {2},
          eid = {211},
        pages = {211},
          doi = {10.3847/1538-4357/adff80},
       adsurl = {https://ui.adsabs.harvard.edu/abs/2025ApJ...991..211T}
}

@ARTICLE{Ghara2025,
       author = {{Ghara}, R. and {Zaroubi}, S. and {Ciardi}, B. and {Mellema}, G. and {Giri}, S.~K. and {Mertens}, F.~G. and {Mevius}, M. and {Koopmans}, L.~V.~E. and {Iliev}, I.~T. and {Acharya}, A. and {Brackenhoff}, S.~A. and {Ceccotti}, E. and {Chege}, K. and {Georgiev}, I. and {Ghosh}, S. and {Hothi}, I. and {H{\"o}fer}, C. and {Ma}, Q. and {Munshi}, S. and {Offringa}, A.~R. and {Shaw}, A.~K. and {Pandey}, V.~N. and {Yatawatta}, S. and {Choudhury}, M.},
        title = "{Constraints on the state of the intergalactic medium at z{\ensuremath{\sim}}8 ‑ 10 using redshifted 21 cm observations with LOFAR}",
      journal = {\aap},
         year = 2025,
        month = jul,
       volume = {699},
          eid = {A109},
        pages = {A109},
          doi = {10.1051/0004-6361/202554163},
archivePrefix = {arXiv},
       eprint = {2505.00373},
 primaryClass = {astro-ph.CO},
       adsurl = {https://ui.adsabs.harvard.edu/abs/2025A&A...699A.109G}
}

@ARTICLE{Sims2025,
       author = {{Sims}, Peter H. and {Bevins}, Harry T.~J. and {Fialkov}, Anastasia and {Anstey}, Dominic and {Handley}, Will J. and {Heimersheim}, Stefan and {de Lera Acedo}, Eloy and {Mondal}, Rajesh and {Barkana}, Rennan},
        title = "{Rapid and late cosmic reionization driven by massive galaxies: a joint analysis of constraints from 21-cm, Lyman line, and CMB data sets}",
      journal = {\mnras},
         year = 2025,
        month = dec,
       volume = {544},
       number = {4},
        pages = {3856-3882},
          doi = {10.1093/mnras/staf1864},
archivePrefix = {arXiv},
       eprint = {2504.09725},
 primaryClass = {astro-ph.CO},
       adsurl = {https://ui.adsabs.harvard.edu/abs/2025MNRAS.544.3856S}
}

@ARTICLE{McKay2025,
       author = {{McKay}, Luke and {Subrahmanyan}, Ravi and {Chippendale}, Aaron and {Bolli}, Pietro and {Kyriakou}, Georgios and {Dunning}, Alex and {Ekers}, Ronald},
        title = "{Precise Measurement of the Absolute Sky Brightness at 60 to 350 MHz}",
      journal = {arXiv e-prints},
         year = 2025,
        month = sep,
          eid = {arXiv:2509.11846},
        pages = {arXiv:2509.11846},
          doi = {10.48550/arXiv.2509.11846},
archivePrefix = {arXiv},
       eprint = {2509.11846},
 primaryClass = {astro-ph.CO},
       adsurl = {https://ui.adsabs.harvard.edu/abs/2025arXiv250911846M}
}

@ARTICLE{Magg2022,
       author = {{Magg}, Mattis and {Reis}, Itamar and {Fialkov}, Anastasia and {Barkana}, Rennan and {Klessen}, Ralf S. and {Glover}, Simon C.~O. and {Chen}, Li-Hsin and {Hartwig}, Tilman and {Schauer}, Anna T.~P.},
        title = "{Effect of the cosmological transition to metal-enriched star formation on the hydrogen 21-cm signal}",
      journal = {\mnras},
         year = 2022,
        month = aug,
       volume = {514},
       number = {3},
        pages = {4433-4449},
          doi = {10.1093/mnras/stac1664},
archivePrefix = {arXiv},
       eprint = {2110.15948},
 primaryClass = {astro-ph.CO},
       adsurl = {https://ui.adsabs.harvard.edu/abs/2022MNRAS.514.4433M}
}

@ARTICLE{Gehlot2020,
       author = {{Gehlot}, B.~K. and {Mertens}, F.~G. and {Koopmans}, L.~V.~E. and {Offringa}, A.~R. and {Shulevski}, A. and {Mevius}, M. and {Brentjens}, M.~A. and {Kuiack}, M. and {Pandey}, V.~N. and {Rowlinson}, A. and {Sardarabadi}, A.~M. and {Vedantham}, H.~K. and {Wijers}, R.~A.~M.~J. and {Yatawatta}, S. and {Zaroubi}, S.},
        title = "{The AARTFAAC Cosmic Explorer: observations of the 21-cm power spectrum in the EDGES absorption trough}",
      journal = {\mnras},
         year = 2020,
        month = dec,
       volume = {499},
       number = {3},
        pages = {4158-4173},
          doi = {10.1093/mnras/staa3093},
archivePrefix = {arXiv},
       eprint = {2010.02269},
 primaryClass = {astro-ph.CO},
       adsurl = {https://ui.adsabs.harvard.edu/abs/2020MNRAS.499.4158G}
}

@ARTICLE{fialkov2017,
       author = {{Fialkov}, Anastasia and {Cohen}, Aviad and {Barkana}, Rennan and {Silk}, Joseph},
        title = "{Constraining the redshifted 21-cm signal with the unresolved soft X-ray background}",
      journal = {\mnras},
         year = 2017,
        month = jan,
       volume = {464},
       number = {3},
        pages = {3498-3508},
          doi = {10.1093/mnras/stw2540},
archivePrefix = {arXiv},
       eprint = {1602.07322},
 primaryClass = {astro-ph.CO},
       adsurl = {https://ui.adsabs.harvard.edu/abs/2017MNRAS.464.3498F}
}

@Inbook{Gilfanov2022,
author="Gilfanov, Marat
and Fabbiano, Giuseppina
and Lehmer, Bret
and Zezas, Andreas",
title="X-Ray Binaries in External Galaxies",
bookTitle="Handbook of X-ray and Gamma-ray Astrophysics",
year="2022",
publisher="Springer Nature Singapore",
address="Singapore",
pages="1--38",
isbn="978-981-16-4544-0",
doi="10.1007/978-981-16-4544-0_108-1",
url="https://doi.org/10.1007/978-981-16-4544-0_108-1"
}

@ARTICLE{Brorby2016,
       author = {{Brorby}, M. and {Kaaret}, P. and {Prestwich}, A. and {Mirabel}, I.~F.},
        title = "{Enhanced X-ray emission from Lyman break analogues and a possible L$_{X}$-SFR-metallicity plane}",
      journal = {\mnras},
         year = 2016,
        month = apr,
       volume = {457},
       number = {4},
        pages = {4081-4088},
          doi = {10.1093/mnras/stw284},
archivePrefix = {arXiv},
       eprint = {1602.01091},
 primaryClass = {astro-ph.HE},
       adsurl = {https://ui.adsabs.harvard.edu/abs/2016MNRAS.457.4081B}
}

@ARTICLE{Grimm2003,
       author = {{Grimm}, H. -J. and {Gilfanov}, M. and {Sunyaev}, R.},
        title = "{High-mass X-ray binaries as a star formation rate indicator in distant galaxies}",
      journal = {\mnras},
         year = 2003,
        month = mar,
       volume = {339},
       number = {3},
        pages = {793-809},
          doi = {10.1046/j.1365-8711.2003.06224.x},
archivePrefix = {arXiv},
       eprint = {astro-ph/0205371},
 primaryClass = {astro-ph},
       adsurl = {https://ui.adsabs.harvard.edu/abs/2003MNRAS.339..793G}
}

@ARTICLE{Antoniou2016,
       author = {{Antoniou}, V. and {Zezas}, A.},
        title = "{Star formation history and X-ray binary populations: the case of the Large Magellanic Cloud}",
      journal = {\mnras},
         year = 2016,
        month = jun,
       volume = {459},
       number = {1},
        pages = {528-553},
          doi = {10.1093/mnras/stw167},
archivePrefix = {arXiv},
       eprint = {1603.08011},
 primaryClass = {astro-ph.HE},
       adsurl = {https://ui.adsabs.harvard.edu/abs/2016MNRAS.459..528A}
}

@ARTICLE{Dhandha2025b,
       author = {{Dhandha}, Jiten and {Fialkov}, Anastasia and {Gessey-Jones}, Thomas and {Bevins}, Harry T.~J. and {Tacchella}, Sandro and {Pochinda}, Simon and {de Lera Acedo}, Eloy and {Singh}, Saurabh and {Barkana}, Rennan},
        title = "{Narrowing the discovery space of the cosmological 21-cm signal using multi-wavelength constraints}",
      journal = {arXiv e-prints},
         year = 2025,
        month = aug,
          eid = {arXiv:2508.13761},
        pages = {arXiv:2508.13761},
          doi = {10.48550/arXiv.2508.13761},
archivePrefix = {arXiv},
       eprint = {2508.13761},
 primaryClass = {astro-ph.CO},
       adsurl = {https://ui.adsabs.harvard.edu/abs/2025arXiv250813761D}
}

@ARTICLE{Philip2019,
       author = {{Philip}, L. and {Abdurashidova}, Z. and {Chiang}, H.~C. and {Ghazi}, N. and {Gumba}, A. and {Heilgendorff}, H.~M. and {J{\'a}uregui-Garc{\'\i}a}, J.~M. and {Malepe}, K. and {Nunhokee}, C.~D. and {Peterson}, J. and {Sievers}, J.~L. and {Simes}, V. and {Spann}, R.},
        title = "{Probing Radio Intensity at High-Z from Marion: 2017 Instrument}",
      journal = {Journal of Astronomical Instrumentation},
         year = 2019,
        month = jan,
       volume = {8},
       number = {2},
          eid = {1950004},
        pages = {1950004},
          doi = {10.1142/S2251171719500041},
archivePrefix = {arXiv},
       eprint = {1806.09531},
 primaryClass = {astro-ph.IM},
       adsurl = {https://ui.adsabs.harvard.edu/abs/2019JAI.....850004P}
}

@ARTICLE{planck14,
       author = {{Planck Collaboration} and {Ade}, P.~A.~R. and {Aghanim}, N. and {Armitage-Caplan}, C. and {Arnaud}, M. and {Ashdown}, M. and {Atrio-Barandela}, F. and {Aumont}, J. and {Baccigalupi}, C. and {Banday}, A.~J. and {Barreiro}, R.~B. and {Bartlett}, J.~G. and {Battaner}, E. and {Benabed}, K. and {Beno{\^\i}t}, A. and {Benoit-L{\'e}vy}, A. and {Bernard}, J. -P. and {Bersanelli}, M. and {Bielewicz}, P. and {Bobin}, J. and {Bock}, J.~J. and {Bonaldi}, A. and {Bond}, J.~R. and {Borrill}, J. and {Bouchet}, F.~R. and {Bridges}, M. and {Bucher}, M. and {Burigana}, C. and {Butler}, R.~C. and {Calabrese}, E. and {Cappellini}, B. and {Cardoso}, J. -F. and {Catalano}, A. and {Challinor}, A. and {Chamballu}, A. and {Chary}, R. -R. and {Chen}, X. and {Chiang}, H.~C. and {Chiang}, L. -Y. and {Christensen}, P.~R. and {Church}, S. and {Clements}, D.~L. and {Colombi}, S. and {Colombo}, L.~P.~L. and {Couchot}, F. and {Coulais}, A. and {Crill}, B.~P. and {Curto}, A. and {Cuttaia}, F. and {Danese}, L. and {Davies}, R.~D. and {Davis}, R.~J. and {de Bernardis}, P. and {de Rosa}, A. and {de Zotti}, G. and {Delabrouille}, J. and {Delouis}, J. -M. and {D{\'e}sert}, F. -X. and {Dickinson}, C. and {Diego}, J.~M. and {Dolag}, K. and {Dole}, H. and {Donzelli}, S. and {Dor{\'e}}, O. and {Douspis}, M. and {Dunkley}, J. and {Dupac}, X. and {Efstathiou}, G. and {Elsner}, F. and {En{\ss}lin}, T.~A. and {Eriksen}, H.~K. and {Finelli}, F. and {Forni}, O. and {Frailis}, M. and {Fraisse}, A.~A. and {Franceschi}, E. and {Gaier}, T.~C. and {Galeotta}, S. and {Galli}, S. and {Ganga}, K. and {Giard}, M. and {Giardino}, G. and {Giraud-H{\'e}raud}, Y. and {Gjerl{\o}w}, E. and {Gonz{\'a}lez-Nuevo}, J. and {G{\'o}rski}, K.~M. and {Gratton}, S. and {Gregorio}, A. and {Gruppuso}, A. and {Gudmundsson}, J.~E. and {Haissinski}, J. and {Hamann}, J. and {Hansen}, F.~K. and {Hanson}, D. and {Harrison}, D. and {Henrot-Versill{\'e}}, S. and {Hern{\'a}ndez-Monteagudo}, C. and {Herranz}, D. and {Hildebrandt}, S.~R. and {Hivon}, E. and {Hobson}, M. and {Holmes}, W.~A. and {Hornstrup}, A. and {Hou}, Z. and {Hovest}, W. and {Huffenberger}, K.~M. and {Jaffe}, A.~H. and {Jaffe}, T.~R. and {Jewell}, J. and {Jones}, W.~C. and {Juvela}, M. and {Keih{\"a}nen}, E. and {Keskitalo}, R. and {Kisner}, T.~S. and {Kneissl}, R. and {Knoche}, J. and {Knox}, L. and {Kunz}, M. and {Kurki-Suonio}, H. and {Lagache}, G. and {L{\"a}hteenm{\"a}ki}, A. and {Lamarre}, J. -M. and {Lasenby}, A. and {Lattanzi}, M. and {Laureijs}, R.~J. and {Lawrence}, C.~R. and {Leach}, S. and {Leahy}, J.~P. and {Leonardi}, R. and {Le{\'o}n-Tavares}, J. and {Lesgourgues}, J. and {Lewis}, A. and {Liguori}, M. and {Lilje}, P.~B. and {Linden-V{\o}rnle}, M. and {L{\'o}pez-Caniego}, M. and {Lubin}, P.~M. and {Mac{\'\i}as-P{\'e}rez}, J.~F. and {Maffei}, B. and {Maino}, D. and {Mandolesi}, N. and {Maris}, M. and {Marshall}, D.~J. and {Martin}, P.~G. and {Mart{\'\i}nez-Gonz{\'a}lez}, E. and {Masi}, S. and {Massardi}, M. and {Matarrese}, S. and {Matthai}, F. and {Mazzotta}, P. and {Meinhold}, P.~R. and {Melchiorri}, A. and {Melin}, J. -B. and {Mendes}, L. and {Menegoni}, E. and {Mennella}, A. and {Migliaccio}, M. and {Millea}, M. and {Mitra}, S. and {Miville-Desch{\^e}nes}, M. -A. and {Moneti}, A. and {Montier}, L. and {Morgante}, G. and {Mortlock}, D. and {Moss}, A. and {Munshi}, D. and {Murphy}, J.~A. and {Naselsky}, P. and {Nati}, F. and {Natoli}, P. and {Netterfield}, C.~B. and {N{\o}rgaard-Nielsen}, H.~U. and {Noviello}, F. and {Novikov}, D. and {Novikov}, I. and {O'Dwyer}, I.~J. and {Osborne}, S. and {Oxborrow}, C.~A. and {Paci}, F. and {Pagano}, L. and {Pajot}, F. and {Paladini}, R. and {Paoletti}, D. and {Partridge}, B. and {Pasian}, F. and {Patanchon}, G. and {Pearson}, D. and {Pearson}, T.~J. and {Peiris}, H.~V. and {Perdereau}, O. and {Perotto}, L. and {Perrotta}, F. and {Pettorino}, V. and {Piacentini}, F. and {Piat}, M. and {Pierpaoli}, E. and {Pietrobon}, D. and {Plaszczynski}, S. and {Platania}, P. and {Pointecouteau}, E.},
        title = "{Planck 2013 results. XVI. Cosmological parameters}",
      journal = {\aap},
         year = 2014,
        month = nov,
       volume = {571},
          eid = {A16},
        pages = {A16},
          doi = {10.1051/0004-6361/201321591},
archivePrefix = {arXiv},
       eprint = {1303.5076},
 primaryClass = {astro-ph.CO},
       adsurl = {https://ui.adsabs.harvard.edu/abs/2014A&A...571A..16P}
}

@ARTICLE{planck18,
       author = {{Planck Collaboration} and {Aghanim}, N. and {Akrami}, Y. and {Ashdown}, M. and {Aumont}, J. and {Baccigalupi}, C. and {Ballardini}, M. and {Banday}, A.~J. and {Barreiro}, R.~B. and {Bartolo}, N. and {Basak}, S. and {Battye}, R. and {Benabed}, K. and {Bernard}, J. -P. and {Bersanelli}, M. and {Bielewicz}, P. and {Bock}, J.~J. and {Bond}, J.~R. and {Borrill}, J. and {Bouchet}, F.~R. and {Boulanger}, F. and {Bucher}, M. and {Burigana}, C. and {Butler}, R.~C. and {Calabrese}, E. and {Cardoso}, J. -F. and {Carron}, J. and {Challinor}, A. and {Chiang}, H.~C. and {Chluba}, J. and {Colombo}, L.~P.~L. and {Combet}, C. and {Contreras}, D. and {Crill}, B.~P. and {Cuttaia}, F. and {de Bernardis}, P. and {de Zotti}, G. and {Delabrouille}, J. and {Delouis}, J. -M. and {Di Valentino}, E. and {Diego}, J.~M. and {Dor{\'e}}, O. and {Douspis}, M. and {Ducout}, A. and {Dupac}, X. and {Dusini}, S. and {Efstathiou}, G. and {Elsner}, F. and {En{\ss}lin}, T.~A. and {Eriksen}, H.~K. and {Fantaye}, Y. and {Farhang}, M. and {Fergusson}, J. and {Fernandez-Cobos}, R. and {Finelli}, F. and {Forastieri}, F. and {Frailis}, M. and {Fraisse}, A.~A. and {Franceschi}, E. and {Frolov}, A. and {Galeotta}, S. and {Galli}, S. and {Ganga}, K. and {G{\'e}nova-Santos}, R.~T. and {Gerbino}, M. and {Ghosh}, T. and {Gonz{\'a}lez-Nuevo}, J. and {G{\'o}rski}, K.~M. and {Gratton}, S. and {Gruppuso}, A. and {Gudmundsson}, J.~E. and {Hamann}, J. and {Handley}, W. and {Hansen}, F.~K. and {Herranz}, D. and {Hildebrandt}, S.~R. and {Hivon}, E. and {Huang}, Z. and {Jaffe}, A.~H. and {Jones}, W.~C. and {Karakci}, A. and {Keih{\"a}nen}, E. and {Keskitalo}, R. and {Kiiveri}, K. and {Kim}, J. and {Kisner}, T.~S. and {Knox}, L. and {Krachmalnicoff}, N. and {Kunz}, M. and {Kurki-Suonio}, H. and {Lagache}, G. and {Lamarre}, J. -M. and {Lasenby}, A. and {Lattanzi}, M. and {Lawrence}, C.~R. and {Le Jeune}, M. and {Lemos}, P. and {Lesgourgues}, J. and {Levrier}, F. and {Lewis}, A. and {Liguori}, M. and {Lilje}, P.~B. and {Lilley}, M. and {Lindholm}, V. and {L{\'o}pez-Caniego}, M. and {Lubin}, P.~M. and {Ma}, Y. -Z. and {Mac{\'\i}as-P{\'e}rez}, J.~F. and {Maggio}, G. and {Maino}, D. and {Mandolesi}, N. and {Mangilli}, A. and {Marcos-Caballero}, A. and {Maris}, M. and {Martin}, P.~G. and {Martinelli}, M. and {Mart{\'\i}nez-Gonz{\'a}lez}, E. and {Matarrese}, S. and {Mauri}, N. and {McEwen}, J.~D. and {Meinhold}, P.~R. and {Melchiorri}, A. and {Mennella}, A. and {Migliaccio}, M. and {Millea}, M. and {Mitra}, S. and {Miville-Desch{\^e}nes}, M. -A. and {Molinari}, D. and {Montier}, L. and {Morgante}, G. and {Moss}, A. and {Natoli}, P. and {N{\o}rgaard-Nielsen}, H.~U. and {Pagano}, L. and {Paoletti}, D. and {Partridge}, B. and {Patanchon}, G. and {Peiris}, H.~V. and {Perrotta}, F. and {Pettorino}, V. and {Piacentini}, F. and {Polastri}, L. and {Polenta}, G. and {Puget}, J. -L. and {Rachen}, J.~P. and {Reinecke}, M. and {Remazeilles}, M. and {Renzi}, A. and {Rocha}, G. and {Rosset}, C. and {Roudier}, G. and {Rubi{\~n}o-Mart{\'\i}n}, J.~A. and {Ruiz-Granados}, B. and {Salvati}, L. and {Sandri}, M. and {Savelainen}, M. and {Scott}, D. and {Shellard}, E.~P.~S. and {Sirignano}, C. and {Sirri}, G. and {Spencer}, L.~D. and {Sunyaev}, R. and {Suur-Uski}, A. -S. and {Tauber}, J.~A. and {Tavagnacco}, D. and {Tenti}, M. and {Toffolatti}, L. and {Tomasi}, M. and {Trombetti}, T. and {Valenziano}, L. and {Valiviita}, J. and {Van Tent}, B. and {Vibert}, L. and {Vielva}, P. and {Villa}, F. and {Vittorio}, N. and {Wandelt}, B.~D. and {Wehus}, I.~K. and {White}, M. and {White}, S.~D.~M. and {Zacchei}, A. and {Zonca}, A.},
        title = "{Planck 2018 results. VI. Cosmological parameters}",
      journal = {\aap},
         year = 2020,
        month = sep,
       volume = {641},
          eid = {A6},
        pages = {A6},
          doi = {10.1051/0004-6361/201833910},
archivePrefix = {arXiv},
       eprint = {1807.06209},
 primaryClass = {astro-ph.CO},
       adsurl = {https://ui.adsabs.harvard.edu/abs/2020A&A...641A...6P}
}

@ARTICLE{fialkov18b,
       author = {{Fialkov}, Anastasia and {Barkana}, Rennan and {Cohen}, Aviad},
        title = "{Constraining Baryon-Dark-Matter Scattering with the Cosmic Dawn 21-cm Signal}",
      journal = {\prl},
         year = 2018,
        month = jul,
       volume = {121},
       number = {1},
          eid = {011101},
        pages = {011101},
          doi = {10.1103/PhysRevLett.121.011101},
archivePrefix = {arXiv},
       eprint = {1802.10577},
 primaryClass = {astro-ph.CO},
       adsurl = {https://ui.adsabs.harvard.edu/abs/2018PhRvL.121a1101F}
}

@ARTICLE{bull24b,
       author = {{Bull}, Philip and {Guandalin}, Caroline and {Addis}, Chris},
        title = "{Modes of the Dark Ages 21 cm field accessible to a lunar radio interferometer}",
      journal = {Philosophical Transactions of the Royal Society of London Series A},
         year = 2024,
        month = may,
       volume = {382},
       number = {2271},
          eid = {20230072},
        pages = {20230072},
          doi = {10.1098/rsta.2023.0072},
archivePrefix = {arXiv},
       eprint = {2403.16955},
 primaryClass = {astro-ph.CO},
       adsurl = {https://ui.adsabs.harvard.edu/abs/2024RSPTA.38230072B}
}

@ARTICLE{greig20,
   author = {{Greig}, Bradley and {Mesinger}, Andrei and {Koopmans}, Leon V.~E.},
    title = "{Reionization and cosmic dawn astrophysics from the Square Kilometre Array: impact of observing strategies}",
  journal = {MNRAS},
     year = 2020,
   volume = {491},
   number = {1},
    pages = {1398-1407},
      doi = {10.1093/mnras/stz3138}
}

@INPROCEEDINGS{burns20,
       author = {{Burns}, J.~O. and {Hallinan}, G.},
        title = "{FARSIDE: Farside Array for Radio Science Investigations of the Dark ages and Exoplanets}",
    booktitle = {American Astronomical Society Meeting Abstracts \#235},
         year = 2020,
       series = {American Astronomical Society Meeting Abstracts},
       volume = {235},
        month = jan,
          eid = {130.01},
        pages = {130.01},
       adsurl = {https://ui.adsabs.harvard.edu/abs/2020AAS...23513001B}
}

@ARTICLE{rao23,
       author = {{Sathyanarayana Rao}, Mayuri and {Singh}, Saurabh and {K.~S.}, Srivani and {B.~S.}, Girish and {Sathish}, Keerthipriya and {Somashekar}, R. and {Agaram}, Raghunathan and {Kavitha}, K. and {Vishwapriya}, Gautam and {Anand}, Ashish and {Udaya Shankar}, N. and {Seetha}, S.},
        title = "{PRATUSH experiment concept and design overview}",
      journal = {Experimental Astronomy},
         year = 2023,
        month = sep,
       volume = {56},
       number = {2-3},
        pages = {741-778},
          doi = {10.1007/s10686-023-09909-5},
       adsurl = {https://ui.adsabs.harvard.edu/abs/2023ExA....56..741S}
}

@ARTICLE{hibbard25,
       author = {{Hibbard}, Joshua J. and {Burns}, Jack O. and {MacDowall}, Robert and {Gopalswamy}, Natchimuthuk and {Boardsen}, Scott A. and {Farrell}, William and {Bradley}, Damon and {Schulszas}, Thomas M. and {Dorigo Jones}, Johnny and {Rapetti}, David and {Turner}, Jake D.},
        title = "{Results from NASA's First Radio Telescope on the Moon: Terrestrial Technosignatures and the Low-Frequency Galactic Background Observed by ROLSES-1 Onboard the Odysseus Lander}",
      journal = {arXiv e-prints},
         year = 2025,
        month = mar,
          eid = {arXiv:2503.09842},
        pages = {arXiv:2503.09842},
          doi = {10.48550/arXiv.2503.09842},
archivePrefix = {arXiv},
       eprint = {2503.09842},
 primaryClass = {astro-ph.IM},
       adsurl = {https://ui.adsabs.harvard.edu/abs/2025arXiv250309842H}
}

@ARTICLE{polidan24,
       author = {{Polidan}, Ronald S. and {Burns}, Jack O. and {Ignatiev}, Alex and {Hegedus}, Alex and {Pober}, Jonathan and {Mahesh}, Nivedita and {Chang}, Tzu-Ching and {Hallinan}, Gregg and {Ning}, Yuhong and {Bowman}, Judd},
        title = "{FarView: An in-situ manufactured lunar far side radio array concept for 21-cm Dark Ages cosmology}",
      journal = {Advances in Space Research},
         year = 2024,
        month = jul,
       volume = {74},
       number = {1},
        pages = {528-546},
          doi = {10.1016/j.asr.2024.04.008},
archivePrefix = {arXiv},
       eprint = {2404.03840},
 primaryClass = {astro-ph.IM},
       adsurl = {https://ui.adsabs.harvard.edu/abs/2024AdSpR..74..528P}
}

@ARTICLE{brinkerink25,
       author = {{Brinkerink}, C.~D. and {Arts}, M.~J. and {Bentum}, M.~J. and {Boonstra}, A.~J. and {Cecconi}, B. and {Fialkov}, A. and {Garcia Guti{\'e}rrez}, J. and {Ghosh}, S. and {Grenouilleau}, J. and {Gurvits}, L.~I. and {Klein-Wolt}, M. and {Koopmans}, L.~V.~E. and {Lazendic-Galloway}, J. and {Paragi}, Z. and {Prinsloo}, D. and {Rajan}, R.~T. and {Rouill{\'e}}, E. and {Ruiter}, M. and {Tauber}, J.~A. and {Vedantham}, H.~K. and {Vecchio}, A. and {Vertegaal}, C.~J.~C. and {Zandboer}, J.~C.~F. and {Zucca}, P.},
        title = "{The Dark Ages Explorer (DEX): a filled-aperture ultra-long wavelength radio interferometer on the lunar far side}",
      journal = {arXiv e-prints},
         year = 2025,
        month = apr,
          eid = {arXiv:2504.03418},
        pages = {arXiv:2504.03418},
          doi = {10.48550/arXiv.2504.03418},
archivePrefix = {arXiv},
       eprint = {2504.03418},
 primaryClass = {astro-ph.IM},
       adsurl = {https://ui.adsabs.harvard.edu/abs/2025arXiv250403418B}
}

@ARTICLE{fialkov24,
       author = {{Fialkov}, A. and {Gessey-Jones}, T. and {Dhandha}, J.},
        title = "{Cosmic mysteries and the hydrogen 21-cm line: bridging the gap with lunar observations}",
      journal = {Philosophical Transactions of the Royal Society of London Series A},
         year = 2024,
        month = may,
       volume = {382},
       number = {2271},
          eid = {20230068},
        pages = {20230068},
          doi = {10.1098/rsta.2023.0068},
archivePrefix = {arXiv},
       eprint = {2311.05366},
 primaryClass = {astro-ph.CO},
       adsurl = {https://ui.adsabs.harvard.edu/abs/2024RSPTA.38230068F}
}

@ARTICLE{bale23,
       author = {{Bale}, Stuart D. and {Bassett}, Neil and {Burns}, Jack O. and {Dorigo Jones}, Johnny and {Goetz}, Keith and {Hellum-Bye}, Christian and {Hermann}, Sven and {Hibbard}, Joshua and {Maksimovic}, Milan and {McLean}, Ryan and {Monsalve}, Raul and {O'Connor}, Paul and {Parsons}, Aaron and {Pulupa}, Marc and {Pund}, Rugved and {Rapetti}, David and {Rotermund}, Kaja M. and {Saliwanchik}, Ben and {Slosar}, Anze and {Sundkvist}, David and {Suzuki}, Aritoki},
        title = "{LuSEE 'Night': The Lunar Surface Electromagnetics Experiment}",
      journal = {arXiv e-prints},
         year = 2023,
        month = jan,
          eid = {arXiv:2301.10345},
        pages = {arXiv:2301.10345},
          doi = {10.48550/arXiv.2301.10345},
archivePrefix = {arXiv},
       eprint = {2301.10345},
 primaryClass = {astro-ph.IM},
       adsurl = {https://ui.adsabs.harvard.edu/abs/2023arXiv230110345B}
}

@ARTICLE{burns21,
       author = {{Burns}, Jack O. and {MacDowall}, Robert and {Bale}, Stuart and {Hallinan}, Gregg and {Bassett}, Neil and {Hegedus}, Alex},
        title = "{Low Radio Frequency Observations from the Moon Enabled by NASA Landed Payload Missions}",
      journal = {PSJ},
         year = 2021,
        month = apr,
       volume = {2},
       number = {2},
          eid = {44},
        pages = {44},
          doi = {10.3847/PSJ/abdfc3},
archivePrefix = {arXiv},
       eprint = {2102.02331},
 primaryClass = {astro-ph.IM},
       adsurl = {https://ui.adsabs.harvard.edu/abs/2021PSJ.....2...44B}
}

@ARTICLE{chen21,
       author = {{Chen}, Xuelei and {Yan}, Jingye and {Deng}, Li and {Wu}, Fengquan and {Wu}, Lin and {Xu}, Yidong and {Zhou}, Li},
        title = "{Discovering the sky at the longest wavelengths with a lunar orbit array}",
      journal = {Philosophical Transactions of the Royal Society of London Series A},
         year = 2021,
        month = jan,
       volume = {379},
       number = {2188},
          eid = {20190566},
        pages = {20190566},
          doi = {10.1098/rsta.2019.0566},
archivePrefix = {arXiv},
       eprint = {2007.15794},
 primaryClass = {astro-ph.IM},
       adsurl = {https://ui.adsabs.harvard.edu/abs/2021RSPTA.37990566C}
}

@article{barkana05,
  title={The first stars in the Universe and cosmic reionization},
  author={Barkana, Rennan and Loeb, Abraham},
  journal={Monthly Notices of the Royal Astronomical Society},
  volume={363},
  number={2},
  pages={L36--L40},
  year={2005},
  publisher={Oxford University Press}
}

@article{ross17,
    author = {Ross, Hannah E. and Dixon, Keri L. and Iliev, Ilian T. and Mellema, Garrelt},
    title = "{Simulating the impact of X-ray heating during the cosmic dawn}",
    journal = {Monthly Notices of the Royal Astronomical Society},
    volume = {468},
    number = {4},
    pages = {3785-3797},
    year = {2017},
    month = {03},
    issn = {0035-8711},
    doi = {10.1093/mnras/stx649},
    url = {https://doi.org/10.1093/mnras/stx649},
    eprint = {https://academic.oup.com/mnras/article-pdf/468/4/3785/13944088/stx649.pdf},
}

@ARTICLE{barkana18,
       author = {{Barkana}, Rennan},
        title = "{Possible interaction between baryons and dark-matter particles revealed by the first stars}",
      journal = {\nat},
         year = 2018,
        month = mar,
       volume = {555},
       number = {7694},
        pages = {71-74},
          doi = {10.1038/nature25791},
archivePrefix = {arXiv},
       eprint = {1803.06698},
 primaryClass = {astro-ph.CO},
       adsurl = {https://ui.adsabs.harvard.edu/abs/2018Natur.555...71B}
}

@INPROCEEDINGS{koopmans15,
       author = {{Koopmans}, L. and {Pritchard}, J. and {Mellema}, G. and {Aguirre}, J. and {Ahn}, K. and {Barkana}, R. and {van Bemmel}, I. and {Bernardi}, G. and {Bonaldi}, A. and {Briggs}, F. and {de Bruyn}, A.~G. and {Chang}, T.~C. and {Chapman}, E. and {Chen}, X. and {Ciardi}, B. and {Dayal}, P. and {Ferrara}, A. and {Fialkov}, A. and {Fiore}, F. and {Ichiki}, K. and {Illiev}, I.~T. and {Inoue}, S. and {Jelic}, V. and {Jones}, M. and {Lazio}, J. and {Maio}, U. and {Majumdar}, S. and {Mack}, K.~J. and {Mesinger}, A. and {Morales}, M.~F. and {Parsons}, A. and {Pen}, U.~L. and {Santos}, M. and {Schneider}, R. and {Semelin}, B. and {de Souza}, R.~S. and {Subrahmanyan}, R. and {Takeuchi}, T. and {Vedantham}, H. and {Wagg}, J. and {Webster}, R. and {Wyithe}, S. and {Datta}, K.~K. and {Trott}, C.},
        title = "{The Cosmic Dawn and Epoch of Reionisation with SKA}",
    booktitle = {Advancing Astrophysics with the Square Kilometre Array (AASKA14)},
         year = 2015,
        month = apr,
          eid = {1},
        pages = {1},
          doi = {10.22323/1.215.0001},
archivePrefix = {arXiv},
       eprint = {1505.07568},
 primaryClass = {astro-ph.CO},
       adsurl = {https://ui.adsabs.harvard.edu/abs/2015aska.confE...1K}
}

@ARTICLE{wouthuysen52,
       author = {{Wouthuysen}, S.~A.},
        title = "{On the excitation mechanism of the 21-cm (radio-frequency) interstellar hydrogen emission line.}",
      journal = {\aj},
         year = 1952,
        month = jan,
       volume = {57},
        pages = {31-32},
          doi = {10.1086/106661},
       adsurl = {https://ui.adsabs.harvard.edu/abs/1952AJ.....57R..31W}
}

@ARTICLE{field58,
       author = {{Field}, George B.},
        title = "{Excitation of the Hydrogen 21-CM Line}",
      journal = {Proceedings of the IRE},
         year = 1958,
        month = jan,
       volume = {46},
        pages = {240-250},
          doi = {10.1109/JRPROC.1958.286741},
       adsurl = {https://ui.adsabs.harvard.edu/abs/1958PIRE...46..240F}
}

@ARTICLE{madau97,
       author = {{Madau}, Piero and {Meiksin}, Avery and {Rees}, Martin J.},
        title = "{21 Centimeter Tomography of the Intergalactic Medium at High Redshift}",
      journal = {\apj},
         year = 1997,
        month = feb,
       volume = {475},
       number = {2},
        pages = {429-444},
          doi = {10.1086/303549},
archivePrefix = {arXiv},
       eprint = {astro-ph/9608010},
 primaryClass = {astro-ph},
       adsurl = {https://ui.adsabs.harvard.edu/abs/1997ApJ...475..429M}
}

@ARTICLE{furlanetto04,
       author = {{Furlanetto}, Steven R. and {Zaldarriaga}, Matias and {Hernquist}, Lars},
        title = "{The Growth of H II Regions During Reionization}",
      journal = {\apj},
         year = 2004,
        month = sep,
       volume = {613},
       number = {1},
        pages = {1-15},
          doi = {10.1086/423025},
archivePrefix = {arXiv},
       eprint = {astro-ph/0403697},
 primaryClass = {astro-ph},
       adsurl = {https://ui.adsabs.harvard.edu/abs/2004ApJ...613....1F}
}

@ARTICLE{barkanaloeb05b,
       author = {{Barkana}, Rennan and {Loeb}, Abraham},
        title = "{A Method for Separating the Physics from the Astrophysics of High-Redshift 21 Centimeter Fluctuations}",
      journal = {\apjl},
         year = 2005,
        month = may,
       volume = {624},
       number = {2},
        pages = {L65-L68},
          doi = {10.1086/430599},
archivePrefix = {arXiv},
       eprint = {astro-ph/0409572},
 primaryClass = {astro-ph},
       adsurl = {https://ui.adsabs.harvard.edu/abs/2005ApJ...624L..65B}
}

@ARTICLE{furlanetto06,
       author = {{Furlanetto}, Steven R. and {Oh}, S. Peng and {Briggs}, Frank H.},
        title = "{Cosmology at low frequencies: The 21 cm transition and the high-redshift Universe}",
      journal = {Physics Reports},
         year = 2006,
        month = oct,
       volume = {433},
       number = {4-6},
        pages = {181-301},
          doi = {10.1016/j.physrep.2006.08.002},
       adsurl = {https://ui.adsabs.harvard.edu/abs/2006PhR...433..181F}
}

@ARTICLE{barkana04,
       author = {{Barkana}, Rennan and {Loeb}, Abraham},
        title = "{Unusually Large Fluctuations in the Statistics of Galaxy Formation at High Redshift}",
      journal = {\apj},
         year = 2004,
        month = jul,
       volume = {609},
       number = {2},
        pages = {474-481},
          doi = {10.1086/421079},
archivePrefix = {arXiv},
       eprint = {astro-ph/0310338},
 primaryClass = {astro-ph},
       adsurl = {https://ui.adsabs.harvard.edu/abs/2004ApJ...609..474B}
}

@ARTICLE{wyithe04,
       author = {{Wyithe}, J. Stuart B. and {Loeb}, Abraham},
        title = "{A characteristic size of \raisebox{-0.5ex}\textasciitilde10Mpc for the ionized bubbles at the end of cosmic reionization}",
      journal = {\nat},
         year = 2004,
        month = nov,
       volume = {432},
       number = {7014},
        pages = {194-196},
          doi = {10.1038/nature03033},
archivePrefix = {arXiv},
       eprint = {astro-ph/0409412},
 primaryClass = {astro-ph},
       adsurl = {https://ui.adsabs.harvard.edu/abs/2004Natur.432..194W}
}

@ARTICLE{mesinger11,
       author = {{Mesinger}, Andrei and {Furlanetto}, Steven and {Cen}, Renyue},
        title = "{21CMFAST: a fast, seminumerical simulation of the high-redshift 21-cm signal}",
      journal = {\mnras},
         year = 2011,
        month = feb,
       volume = {411},
       number = {2},
        pages = {955-972},
          doi = {10.1111/j.1365-2966.2010.17731.x},
archivePrefix = {arXiv},
       eprint = {1003.3878},
 primaryClass = {astro-ph.CO},
       adsurl = {https://ui.adsabs.harvard.edu/abs/2011MNRAS.411..955M}
}

@ARTICLE{tseliakhovich10,
       author = {{Tseliakhovich}, Dmitriy and {Hirata}, Christopher},
        title = "{Relative velocity of dark matter and baryonic fluids and the formation of the first structures}",
      journal = {\prd},
         year = 2010,
        month = oct,
       volume = {82},
       number = {8},
          eid = {083520},
        pages = {083520},
          doi = {10.1103/PhysRevD.82.083520},
archivePrefix = {arXiv},
       eprint = {1005.2416},
 primaryClass = {astro-ph.CO},
       adsurl = {https://ui.adsabs.harvard.edu/abs/2010PhRvD..82h3520T}
}

@misc{lewis11,
       author = {{Lewis}, Antony and {Challinor}, Anthony},
        title = "{CAMB: Code for Anisotropies in the Microwave Background}",
 howpublished = {Astrophysics Source Code Library, record ascl:1102.026},
         year = 2011,
        month = feb,
          eid = {ascl:1102.026},
archivePrefix = {ascl},
       eprint = {1102.026},
       adsurl = {https://ui.adsabs.harvard.edu/abs/2011ascl.soft02026L}
}

@ARTICLE{fialkov12,
       author = {{Fialkov}, Anastasia and {Barkana}, Rennan and {Tseliakhovich}, Dmitriy and {Hirata}, Christopher M.},
        title = "{Impact of the relative motion between the dark matter and baryons on the first stars: semi-analytical modelling}",
      journal = {\mnras},
         year = 2012,
        month = aug,
       volume = {424},
       number = {2},
        pages = {1335-1345},
          doi = {10.1111/j.1365-2966.2012.21318.x},
archivePrefix = {arXiv},
       eprint = {1110.2111},
 primaryClass = {astro-ph.CO},
       adsurl = {https://ui.adsabs.harvard.edu/abs/2012MNRAS.424.1335F}
}

@ARTICLE{fialkov13,
       author = {{Fialkov}, Anastasia and {Barkana}, Rennan and {Visbal}, Eli and {Tseliakhovich}, Dmitriy and {Hirata}, Christopher M.},
        title = "{The 21-cm signature of the first stars during the Lyman-Werner feedback era}",
      journal = {\mnras},
         year = 2013,
        month = jul,
       volume = {432},
       number = {4},
        pages = {2909-2916},
          doi = {10.1093/mnras/stt650},
archivePrefix = {arXiv},
       eprint = {1212.0513},
 primaryClass = {astro-ph.CO},
       adsurl = {https://ui.adsabs.harvard.edu/abs/2013MNRAS.432.2909F}
}

@ARTICLE{sobacchi13,
       author = {{Sobacchi}, Emanuele and {Mesinger}, Andrei},
        title = "{How does radiative feedback from an ultraviolet background impact reionization?}",
      journal = {\mnras},
         year = 2013,
        month = jul,
       volume = {432},
       number = {4},
        pages = {3340-3348},
          doi = {10.1093/mnras/stt693},
archivePrefix = {arXiv},
       eprint = {1301.6781},
 primaryClass = {astro-ph.CO},
       adsurl = {https://ui.adsabs.harvard.edu/abs/2013MNRAS.432.3340S}
}

@ARTICLE{tseliakhovich11,
       author = {{Tseliakhovich}, Dmitriy and {Barkana}, Rennan and {Hirata}, Christopher M.},
        title = "{Suppression and spatial variation of early galaxies and minihaloes}",
      journal = {\mnras},
         year = 2011,
        month = dec,
       volume = {418},
       number = {2},
        pages = {906-915},
          doi = {10.1111/j.1365-2966.2011.19541.x},
archivePrefix = {arXiv},
       eprint = {1012.2574},
 primaryClass = {astro-ph.CO},
       adsurl = {https://ui.adsabs.harvard.edu/abs/2011MNRAS.418..906T}
}

@ARTICLE{visbal12,
       author = {{Visbal}, Eli and {Barkana}, Rennan and {Fialkov}, Anastasia and {Tseliakhovich}, Dmitriy and {Hirata}, Christopher M.},
        title = "{The signature of the first stars in atomic hydrogen at redshift 20}",
      journal = {\nat},
         year = 2012,
        month = jul,
       volume = {487},
       number = {7405},
        pages = {70-73},
          doi = {10.1038/nature11177},
archivePrefix = {arXiv},
       eprint = {1201.1005},
 primaryClass = {astro-ph.CO},
       adsurl = {https://ui.adsabs.harvard.edu/abs/2012Natur.487...70V}
}

@ARTICLE{reis22,
       author = {{Reis}, Itamar and {Barkana}, Rennan and {Fialkov}, Anastasia},
        title = "{Shot noise and scatter in the star formation efficiency as a source of 21-cm fluctuations}",
      journal = {\mnras},
         year = 2022,
        month = apr,
       volume = {511},
       number = {4},
        pages = {5265-5273},
          doi = {10.1093/mnras/stac411},
archivePrefix = {arXiv},
       eprint = {2106.13111},
 primaryClass = {astro-ph.CO},
       adsurl = {https://ui.adsabs.harvard.edu/abs/2022MNRAS.511.5265R}
}

@ARTICLE{munoz22,
       author = {{Mu{\~n}oz}, Julian B. and {Qin}, Yuxiang and {Mesinger}, Andrei and {Murray}, Steven G. and {Greig}, Bradley and {Mason}, Charlotte},
        title = "{The impact of the first galaxies on cosmic dawn and reionization}",
      journal = {\mnras},
         year = 2022,
        month = apr,
       volume = {511},
       number = {3},
        pages = {3657-3681},
          doi = {10.1093/mnras/stac185},
archivePrefix = {arXiv},
       eprint = {2110.13919},
 primaryClass = {astro-ph.CO},
       adsurl = {https://ui.adsabs.harvard.edu/abs/2022MNRAS.511.3657M}
}

@ARTICLE{cohen16,
       author = {{Cohen}, Aviad and {Fialkov}, Anastasia and {Barkana}, Rennan},
        title = "{The 21-cm BAO signature of enriched low-mass galaxies during cosmic reionization}",
      journal = {\mnras},
         year = 2016,
        month = jun,
       volume = {459},
       number = {1},
        pages = {L90-L94},
          doi = {10.1093/mnrasl/slw047},
archivePrefix = {arXiv},
       eprint = {1508.04138},
 primaryClass = {astro-ph.CO},
       adsurl = {https://ui.adsabs.harvard.edu/abs/2016MNRAS.459L..90C}
}

@ARTICLE{gesseyjones22,
       author = {{Gessey-Jones}, T. and {Sartorio}, N.~S. and {Fialkov}, A. and {Mirouh}, G.~M. and {Magg}, M. and {Izzard}, R.~G. and {de Lera Acedo}, E. and {Handley}, W.~J. and {Barkana}, R.},
        title = "{Impact of the primordial stellar initial mass function on the 21-cm signal}",
      journal = {\mnras},
         year = 2022,
        month = oct,
       volume = {516},
       number = {1},
        pages = {841-860},
          doi = {10.1093/mnras/stac2049},
archivePrefix = {arXiv},
       eprint = {2202.02099},
 primaryClass = {astro-ph.CO},
       adsurl = {https://ui.adsabs.harvard.edu/abs/2022MNRAS.516..841G}
}

@ARTICLE{gesseyjones23,
       author = {{Gessey-Jones}, T. and {Fialkov}, A. and {de Lera Acedo}, E. and {Handley}, W.~J. and {Barkana}, R.},
        title = "{Signatures of cosmic ray heating in 21-cm observables}",
      journal = {\mnras},
         year = 2023,
        month = dec,
       volume = {526},
       number = {3},
        pages = {4262-4284},
          doi = {10.1093/mnras/stad3014},
archivePrefix = {arXiv},
       eprint = {2304.07201},
 primaryClass = {astro-ph.CO},
       adsurl = {https://ui.adsabs.harvard.edu/abs/2023MNRAS.526.4262G}
}

@ARTICLE{nikolic24,
       author = {{Nikoli{\'c}}, Ivan and {Mesinger}, Andrei and {Davies}, James E. and {Prelogovi{\'c}}, David},
        title = "{The importance of stochasticity in determining galaxy emissivities and UV LFs during cosmic dawn and reionization}",
      journal = {\aap},
         year = 2024,
        month = dec,
       volume = {692},
          eid = {A142},
        pages = {A142},
          doi = {10.1051/0004-6361/202451213},
archivePrefix = {arXiv},
       eprint = {2406.15237},
 primaryClass = {astro-ph.CO},
       adsurl = {https://ui.adsabs.harvard.edu/abs/2024A&A...692A.142N}
}

@ARTICLE{murmu24,
       author = {{Murmu}, Chandra Shekhar and {Datta}, Kanan K. and {Majumdar}, Suman and {Greve}, Thomas R.},
        title = "{Impact of astrophysical scatter on the epoch of reionization [H I]$_{21}$ bispectrum}",
      journal = {\jcap},
         year = 2024,
        month = aug,
       volume = {2024},
       number = {8},
          eid = {032},
        pages = {032},
          doi = {10.1088/1475-7516/2024/08/032},
archivePrefix = {arXiv},
       eprint = {2311.17062},
 primaryClass = {astro-ph.CO},
       adsurl = {https://ui.adsabs.harvard.edu/abs/2024JCAP...08..032M}
}

@ARTICLE{pochinda24,
       author = {{Pochinda}, S. and {Gessey-Jones}, T. and {Bevins}, H.~T.~J. and {Fialkov}, A. and {Heimersheim}, S. and {Abril-Cabezas}, I. and {de Lera Acedo}, E. and {Singh}, S. and {Sikder}, S. and {Barkana}, R.},
        title = "{Constraining the properties of Population III galaxies with multiwavelength observations}",
      journal = {\mnras},
         year = 2024,
        month = jun,
       volume = {531},
       number = {1},
        pages = {1113-1132},
          doi = {10.1093/mnras/stae1185},
archivePrefix = {arXiv},
       eprint = {2312.08095},
 primaryClass = {astro-ph.CO},
       adsurl = {https://ui.adsabs.harvard.edu/abs/2024MNRAS.531.1113P}
}

@ARTICLE{gesseyjones25,
       author = {{Gessey-Jones}, T. and {Sartorio}, N.~S. and {Bevins}, H.~T.~J. and {Fialkov}, A. and {Handley}, W.~J. and {de Lera Acedo}, E. and {Mirouh}, G.~M. and {Izzard}, R.~G. and {Barkana}, R.},
        title = "{Determination of the mass distribution of the first stars from the 21-cm signal}",
      journal = {Nature Astronomy},
         year = 2025,
        month = aug,
       volume = {9},
        pages = {1268-1279},
          doi = {10.1038/s41550-025-02575-x},
archivePrefix = {arXiv},
       eprint = {2502.18098},
 primaryClass = {astro-ph.CO},
       adsurl = {https://ui.adsabs.harvard.edu/abs/2025NatAs...9.1268G}
}

@ARTICLE{leitherer99,
       author = {{Leitherer}, Claus and {Schaerer}, Daniel and {Goldader}, Jeffrey D. and {Delgado}, Rosa M. Gonz{\'a}lez and {Robert}, Carmelle and {Kune}, Denis Foo and {de Mello}, Du{\'\i}lia F. and {Devost}, Daniel and {Heckman}, Timothy M.},
        title = "{Starburst99: Synthesis Models for Galaxies with Active Star Formation}",
      journal = {\apjs},
         year = 1999,
        month = jul,
       volume = {123},
       number = {1},
        pages = {3-40},
          doi = {10.1086/313233},
archivePrefix = {arXiv},
       eprint = {astro-ph/9902334},
 primaryClass = {astro-ph},
       adsurl = {https://ui.adsabs.harvard.edu/abs/1999ApJS..123....3L}
}

@ARTICLE{lehmer21,
       author = {{Lehmer}, Bret D. and {Eufrasio}, Rafael T. and {Basu-Zych}, Antara and {Doore}, Keith and {Fragos}, Tassos and {Garofali}, Kristen and {Kovlakas}, Konstantinos and {Williams}, Benjamin F. and {Zezas}, Andreas and {Santana-Silva}, Luidhy},
        title = "{The Metallicity Dependence of the High-mass X-Ray Binary Luminosity Function}",
      journal = {\apj},
         year = 2021,
        month = jan,
       volume = {907},
       number = {1},
          eid = {17},
        pages = {17},
          doi = {10.3847/1538-4357/abcec1},
archivePrefix = {arXiv},
       eprint = {2011.09476},
 primaryClass = {astro-ph.GA},
       adsurl = {https://ui.adsabs.harvard.edu/abs/2021ApJ...907...17L}
}

@ARTICLE{pritchard07,
       author = {{Pritchard}, Jonathan R. and {Furlanetto}, Steven R.},
        title = "{21-cm fluctuations from inhomogeneous X-ray heating before reionization}",
      journal = {\mnras},
         year = 2007,
        month = apr,
       volume = {376},
       number = {4},
        pages = {1680-1694},
          doi = {10.1111/j.1365-2966.2007.11519.x},
archivePrefix = {arXiv},
       eprint = {astro-ph/0607234},
 primaryClass = {astro-ph},
       adsurl = {https://ui.adsabs.harvard.edu/abs/2007MNRAS.376.1680P}
}

@ARTICLE{venumadhav18,
       author = {{Venumadhav}, Tejaswi and {Dai}, Liang and {Kaurov}, Alexander and {Zaldarriaga}, Matias},
        title = "{Heating of the intergalactic medium by the cosmic microwave background during cosmic dawn}",
      journal = {\prd},
         year = 2018,
        month = nov,
       volume = {98},
       number = {10},
          eid = {103513},
        pages = {103513},
          doi = {10.1103/PhysRevD.98.103513},
archivePrefix = {arXiv},
       eprint = {1804.02406},
 primaryClass = {astro-ph.CO},
       adsurl = {https://ui.adsabs.harvard.edu/abs/2018PhRvD..98j3513V}
}

@ARTICLE{dhandha25,
       author = {{Dhandha}, Jiten and {Fialkov}, Anastasia and {Gessey-Jones}, Thomas and {Bevins}, Harry T.~J. and {Tacchella}, Sandro and {Pochinda}, Simon and {de Lera Acedo}, Eloy and {Singh}, Saurabh and {Barkana}, Rennan},
        title = "{Exploiting synergies between JWST and cosmic 21-cm observations to uncover star formation in the early Universe}",
      journal = {\mnras},
         year = 2025,
        month = sep,
       volume = {542},
       number = {3},
        pages = {2292-2322},
          doi = {10.1093/mnras/staf1359},
archivePrefix = {arXiv},
       eprint = {2503.21687},
 primaryClass = {astro-ph.GA},
       adsurl = {https://ui.adsabs.harvard.edu/abs/2025MNRAS.542.2292D}
}

@ARTICLE{fragos13a,
       author = {{Fragos}, T. and {Lehmer}, B. and {Tremmel}, M. and {Tzanavaris}, P. and {Basu-Zych}, A. and {Belczynski}, K. and {Hornschemeier}, A. and {Jenkins}, L. and {Kalogera}, V. and {Ptak}, A. and {Zezas}, A.},
        title = "{X-Ray Binary Evolution Across Cosmic Time}",
      journal = {\apj},
         year = 2013,
        month = feb,
       volume = {764},
       number = {1},
          eid = {41},
        pages = {41},
          doi = {10.1088/0004-637X/764/1/41},
archivePrefix = {arXiv},
       eprint = {1206.2395},
 primaryClass = {astro-ph.HE},
       adsurl = {https://ui.adsabs.harvard.edu/abs/2013ApJ...764...41F}
}

@ARTICLE{fragos13b,
       author = {{Fragos}, T. and {Lehmer}, B.~D. and {Naoz}, S. and {Zezas}, A. and {Basu-Zych}, A.},
        title = "{Energy Feedback from X-Ray Binaries in the Early Universe}",
      journal = {\apjl},
         year = 2013,
        month = oct,
       volume = {776},
       number = {2},
          eid = {L31},
        pages = {L31},
          doi = {10.1088/2041-8205/776/2/L31},
archivePrefix = {arXiv},
       eprint = {1306.1405},
 primaryClass = {astro-ph.CO},
       adsurl = {https://ui.adsabs.harvard.edu/abs/2013ApJ...776L..31F}
}

@ARTICLE{fialkov14a,
       author = {{Fialkov}, Anastasia and {Barkana}, Rennan and {Visbal}, Eli},
        title = "{The observable signature of late heating of the Universe during cosmic reionization}",
      journal = {\nat},
         year = 2014,
        month = feb,
       volume = {506},
       number = {7487},
        pages = {197-199},
          doi = {10.1038/nature12999},
archivePrefix = {arXiv},
       eprint = {1402.0940},
 primaryClass = {astro-ph.CO},
       adsurl = {https://ui.adsabs.harvard.edu/abs/2014Natur.506..197F}
}

@ARTICLE{eide18,
       author = {{Eide}, Marius B. and {Graziani}, Luca and {Ciardi}, Benedetta and {Feng}, Yu and {Kakiichi}, Koki and {Di Matteo}, Tiziana},
        title = "{The epoch of cosmic heating by early sources of X-rays}",
      journal = {\mnras},
         year = 2018,
        month = may,
       volume = {476},
       number = {1},
        pages = {1174-1190},
          doi = {10.1093/mnras/sty272},
archivePrefix = {arXiv},
       eprint = {1801.09719},
 primaryClass = {astro-ph.CO},
       adsurl = {https://ui.adsabs.harvard.edu/abs/2018MNRAS.476.1174E}
}

@ARTICLE{izzard04,
       author = {{Izzard}, Robert G. and {Tout}, Christopher A. and {Karakas}, Amanda I. and {Pols}, Onno R.},
        title = "{A new synthetic model for asymptotic giant branch stars}",
      journal = {\mnras},
         year = 2004,
        month = may,
       volume = {350},
       number = {2},
        pages = {407-426},
          doi = {10.1111/j.1365-2966.2004.07446.x},
archivePrefix = {arXiv},
       eprint = {astro-ph/0402403},
 primaryClass = {astro-ph},
       adsurl = {https://ui.adsabs.harvard.edu/abs/2004MNRAS.350..407I}
}

@ARTICLE{izzard06,
       author = {{Izzard}, R.~G. and {Dray}, L.~M. and {Karakas}, A.~I. and {Lugaro}, M. and {Tout}, C.~A.},
        title = "{Population nucleosynthesis in single and binary stars. I. Model}",
      journal = {\aap},
         year = 2006,
        month = dec,
       volume = {460},
       number = {2},
        pages = {565-572},
          doi = {10.1051/0004-6361:20066129},
       adsurl = {https://ui.adsabs.harvard.edu/abs/2006A&A...460..565I}
}

@ARTICLE{liu23,
       author = {{Liu}, Boyuan and {Sartorio}, Nina S. and {Izzard}, Robert G. and {Fialkov}, Anastasia},
        title = "{Population synthesis of Be X-ray binaries: metallicity dependence of total X-ray outputs}",
      journal = {\mnras},
         year = 2024,
        month = jan,
       volume = {527},
       number = {3},
        pages = {5023-5048},
          doi = {10.1093/mnras/stad3475},
archivePrefix = {arXiv},
       eprint = {2308.06154},
 primaryClass = {astro-ph.HE},
       adsurl = {https://ui.adsabs.harvard.edu/abs/2024MNRAS.527.5023L}
}

@ARTICLE{klessen23,
       author = {{Klessen}, Ralf S. and {Glover}, Simon C.~O.},
        title = "{The First Stars: Formation, Properties, and Impact}",
      journal = {\araa},
         year = 2023,
        month = aug,
       volume = {61},
        pages = {65-130},
          doi = {10.1146/annurev-astro-071221-053453},
archivePrefix = {arXiv},
       eprint = {2303.12500},
 primaryClass = {astro-ph.CO},
       adsurl = {https://ui.adsabs.harvard.edu/abs/2023ARA&A..61...65K}
}

@ARTICLE{kaur22,
       author = {{Kaur}, Harman Deep and {Qin}, Yuxiang and {Mesinger}, Andrei and {Pallottini}, Andrea and {Fragos}, Tassos and {Basu-Zych}, Antara},
        title = "{The 21-cm signal from the cosmic dawn: metallicity dependence of high-mass X-ray binaries}",
      journal = {\mnras},
         year = 2022,
        month = jul,
       volume = {513},
       number = {4},
        pages = {5097-5108},
          doi = {10.1093/mnras/stac1226},
archivePrefix = {arXiv},
       eprint = {2203.10851},
 primaryClass = {astro-ph.GA},
       adsurl = {https://ui.adsabs.harvard.edu/abs/2022MNRAS.513.5097K}
}

@article{sartorio23,
    author = {Sartorio, N. S. and others},
    title = {Population III X-ray binaries and their impact on the early universe},
    journal = {Monthly Notices of the Royal Astronomical Society},
    volume = {521},
    number = {3},
    pages = {4039-4056},
    year = {2023},
    doi = {10.1093/mnras/stad819},
    url = {https://academic.oup.com/mnras/article/521/3/4039/7075887}
}

@ARTICLE{grimm03,
       author = {{Grimm}, H. -J. and {Gilfanov}, M. and {Sunyaev}, R.},
        title = "{High-mass X-ray binaries as a star formation rate indicator in distant galaxies}",
      journal = {\mnras},
         year = 2003,
        month = mar,
       volume = {339},
       number = {3},
        pages = {793-809},
          doi = {10.1046/j.1365-8711.2003.06224.x},
archivePrefix = {arXiv},
       eprint = {astro-ph/0205371},
 primaryClass = {astro-ph},
       adsurl = {https://ui.adsabs.harvard.edu/abs/2003MNRAS.339..793G}
}

@ARTICLE{mineo12,
       author = {{Mineo}, S. and {Gilfanov}, M. and {Sunyaev}, R.},
        title = "{X-ray emission from star-forming galaxies - I. High-mass X-ray binaries}",
      journal = {\mnras},
         year = 2012,
        month = jan,
       volume = {419},
       number = {3},
        pages = {2095-2115},
          doi = {10.1111/j.1365-2966.2011.19862.x},
archivePrefix = {arXiv},
       eprint = {1105.4610},
 primaryClass = {astro-ph.HE},
       adsurl = {https://ui.adsabs.harvard.edu/abs/2012MNRAS.419.2095M}
}

@ARTICLE{pacucci14,
       author = {{Pacucci}, Fabio and {Mesinger}, Andrei and {Mineo}, Stefano and {Ferrara}, Andrea},
        title = "{The X-ray spectra of the first galaxies: 21 cm signatures}",
      journal = {\mnras},
         year = 2014,
        month = sep,
       volume = {443},
       number = {1},
        pages = {678-686},
          doi = {10.1093/mnras/stu1240},
archivePrefix = {arXiv},
       eprint = {1403.6125},
 primaryClass = {astro-ph.CO},
       adsurl = {https://ui.adsabs.harvard.edu/abs/2014MNRAS.443..678P}
}

@ARTICLE{madau18,
       author = {{Madau}, Piero},
        title = "{Constraints on early star formation from the 21-cm global signal}",
      journal = {\mnras},
         year = 2018,
        month = oct,
       volume = {480},
       number = {1},
        pages = {L43-L47},
          doi = {10.1093/mnrasl/sly125},
archivePrefix = {arXiv},
       eprint = {1807.01316},
 primaryClass = {astro-ph.CO},
       adsurl = {https://ui.adsabs.harvard.edu/abs/2018MNRAS.480L..43M}
}

@misc{izzard23,
  author       = {Izzard, Robert},
  title        = {binary\_c},
  month        = nov,
  year         = 2023,
  publisher    = {Zenodo},
  doi          = {10.5281/zenodo.10030398},
  url          = {https://doi.org/10.5281/zenodo.10030398}
}

@ARTICLE{monslave24,
       author = {{Monsalve}, R.~A. and {Altamirano}, C. and {Bidula}, V. and {Bustos}, R. and {Bye}, C.~H. and {Chiang}, H.~C. and {D{\'\i}az}, M. and {Fern{\'a}ndez}, B. and {Guo}, X. and {Hendricksen}, I. and {Hornecker}, E. and {Lucero}, F. and {Mani}, H. and {McGee}, F. and {Mena}, F.~P. and {Pess{\^o}a}, M. and {Prabhakar}, G. and {Restrepo}, O. and {Sievers}, J.~L. and {Thyagarajan}, N.},
        title = "{Mapper of the IGM spin temperature: instrument overview}",
      journal = {\mnras},
         year = 2024,
        month = jun,
       volume = {530},
       number = {4},
        pages = {4125-4147},
          doi = {10.1093/mnras/stae1138},
archivePrefix = {arXiv},
       eprint = {2309.02996},
 primaryClass = {astro-ph.IM},
       adsurl = {https://ui.adsabs.harvard.edu/abs/2024MNRAS.530.4125M}
}

@ARTICLE{bull24,
       author = {{Bull}, Philip and {El-Makadema}, Ahmed and {Garsden}, Hugh and {Edgley}, John and {Roddis}, Neil and {Chluba}, Jens and {Conselice}, Christopher J. and {Dutta}, Sohini and {Glasscock}, Katrine A. and {Nasirudin}, Ainulnabilah and {Norris}, Jordan and {Wilensky}, Michael J. and {Ye}, Isabelle and {Zhang}, Zheng},
        title = "{RHINO: A large horn antenna for detecting the 21cm global signal}",
      journal = {arXiv e-prints},
         year = 2024,
        month = sep,
          eid = {arXiv:2410.00076},
        pages = {arXiv:2410.00076},
          doi = {10.48550/arXiv.2410.00076},
archivePrefix = {arXiv},
       eprint = {2410.00076},
 primaryClass = {astro-ph.IM},
       adsurl = {https://ui.adsabs.harvard.edu/abs/2024arXiv241000076B}
}

@ARTICLE{singh22,
       author = {{Singh}, Saurabh and {Jishnu}, Nambissan T. and {Subrahmanyan}, Ravi and {Udaya Shankar}, N. and {Girish}, B.~S. and {Raghunathan}, A. and {Somashekar}, R. and {Srivani}, K.~S. and {Sathyanarayana Rao}, Mayuri},
        title = "{On the detection of a cosmic dawn signal in the radio background}",
      journal = {Nature Astronomy},
         year = 2022,
        month = feb,
       volume = {6},
        pages = {607-617},
          doi = {10.1038/s41550-022-01610-5},
archivePrefix = {arXiv},
       eprint = {2112.06778},
 primaryClass = {astro-ph.CO},
       adsurl = {https://ui.adsabs.harvard.edu/abs/2022NatAs...6..607S}
}

@ARTICLE{bowman18,
       author = {{Bowman}, Judd D. and {Rogers}, Alan E.~E. and {Monsalve}, Raul A. and {Mozdzen}, Thomas J. and {Mahesh}, Nivedita},
        title = "{An absorption profile centred at 78 megahertz in the sky-averaged spectrum}",
      journal = {\nat},
         year = 2018,
        month = mar,
       volume = {555},
       number = {7694},
        pages = {67-70},
          doi = {10.1038/nature25792},
archivePrefix = {arXiv},
       eprint = {1810.05912},
 primaryClass = {astro-ph.CO},
       adsurl = {https://ui.adsabs.harvard.edu/abs/2018Natur.555...67B}
}

@ARTICLE{fialkov19,
       author = {{Fialkov}, Anastasia and {Barkana}, Rennan},
        title = "{Signature of excess radio background in the 21-cm global signal and power spectrum}",
      journal = {\mnras},
         year = 2019,
        month = jun,
       volume = {486},
       number = {2},
        pages = {1763-1773},
          doi = {10.1093/mnras/stz873},
archivePrefix = {arXiv},
       eprint = {1902.02438},
 primaryClass = {astro-ph.CO},
       adsurl = {https://ui.adsabs.harvard.edu/abs/2019MNRAS.486.1763F}
}

@ARTICLE{hills18b,
       author = {{Hills}, Richard and {Kulkarni}, Girish and {Meerburg}, P. Daniel and {Puchwein}, Ewald},
        title = "{Concerns about modelling of the EDGES data}",
      journal = {\nat},
         year = 2018,
        month = dec,
       volume = {564},
       number = {7736},
        pages = {E32-E34},
          doi = {10.1038/s41586-018-0796-5},
archivePrefix = {arXiv},
       eprint = {1805.01421},
 primaryClass = {astro-ph.CO},
       adsurl = {https://ui.adsabs.harvard.edu/abs/2018Natur.564E..32H}
}

@ARTICLE{singh19,
       author = {{Singh}, Saurabh and {Subrahmanyan}, Ravi},
        title = "{The Redshifted 21 cm Signal in the EDGES Low-band Spectrum}",
      journal = {\apj},
         year = 2019,
        month = jul,
       volume = {880},
       number = {1},
          eid = {26},
        pages = {26},
          doi = {10.3847/1538-4357/ab2879},
archivePrefix = {arXiv},
       eprint = {1903.04540},
 primaryClass = {astro-ph.CO},
       adsurl = {https://ui.adsabs.harvard.edu/abs/2019ApJ...880...26S}
}

@ARTICLE{bradley19,
       author = {{Bradley}, Richard F. and {Tauscher}, Keith and {Rapetti}, David and {Burns}, Jack O.},
        title = "{A Ground Plane Artifact that Induces an Absorption Profile in Averaged Spectra from Global 21 cm Measurements, with Possible Application to EDGES}",
      journal = {\apj},
         year = 2019,
        month = apr,
       volume = {874},
       number = {2},
          eid = {153},
        pages = {153},
          doi = {10.3847/1538-4357/ab0d8b},
archivePrefix = {arXiv},
       eprint = {1810.09015},
 primaryClass = {astro-ph.IM},
       adsurl = {https://ui.adsabs.harvard.edu/abs/2019ApJ...874..153B}
}

@ARTICLE{munoz18,
       author = {{Mu{\~n}oz}, Julian B. and {Loeb}, Abraham},
        title = "{A small amount of mini-charged dark matter could cool the baryons in the early Universe}",
      journal = {\nat},
         year = 2018,
        month = may,
       volume = {557},
       number = {7707},
        pages = {684-686},
          doi = {10.1038/s41586-018-0151-x},
archivePrefix = {arXiv},
       eprint = {1802.10094},
 primaryClass = {astro-ph.CO},
       adsurl = {https://ui.adsabs.harvard.edu/abs/2018Natur.557..684M}
}

@ARTICLE{berlin18,
       author = {{Berlin}, Asher and {Hooper}, Dan and {Krnjaic}, Gordan and {McDermott}, Samuel D.},
        title = "{Severely Constraining Dark-Matter Interpretations of the 21-cm Anomaly}",
      journal = {\prl},
         year = 2018,
        month = jul,
       volume = {121},
       number = {1},
          eid = {011102},
        pages = {011102},
          doi = {10.1103/PhysRevLett.121.011102},
archivePrefix = {arXiv},
       eprint = {1803.02804},
 primaryClass = {hep-ph},
       adsurl = {https://ui.adsabs.harvard.edu/abs/2018PhRvL.121a1102B}
}

@ARTICLE{feng18,
       author = {{Feng}, Chang and {Holder}, Gilbert},
        title = "{Enhanced Global Signal of Neutral Hydrogen Due to Excess Radiation at Cosmic Dawn}",
      journal = {\apjl},
         year = 2018,
        month = may,
       volume = {858},
       number = {2},
          eid = {L17},
        pages = {L17},
          doi = {10.3847/2041-8213/aac0fe},
archivePrefix = {arXiv},
       eprint = {1802.07432},
 primaryClass = {astro-ph.CO},
       adsurl = {https://ui.adsabs.harvard.edu/abs/2018ApJ...858L..17F}
}

@ARTICLE{brandenberger19,
       author = {{Brandenberger}, Robert and {Cyr}, Bryce and {Shi}, Rui},
        title = "{Constraints on superconducting cosmic strings from the global 21-cm signal before reionization}",
      journal = {\jcap},
         year = 2019,
        month = sep,
       volume = {2019},
       number = {9},
          eid = {009},
        pages = {009},
          doi = {10.1088/1475-7516/2019/09/009},
archivePrefix = {arXiv},
       eprint = {1902.08282},
 primaryClass = {astro-ph.CO},
       adsurl = {https://ui.adsabs.harvard.edu/abs/2019JCAP...09..009B}
}

@ARTICLE{reis20,
       author = {{Reis}, Itamar and {Fialkov}, Anastasia and {Barkana}, Rennan},
        title = "{High-redshift radio galaxies: a potential new source of 21-cm fluctuations}",
      journal = {\mnras},
         year = 2020,
        month = dec,
       volume = {499},
       number = {4},
        pages = {5993-6008},
          doi = {10.1093/mnras/staa3091},
archivePrefix = {arXiv},
       eprint = {2008.04315},
 primaryClass = {astro-ph.CO},
       adsurl = {https://ui.adsabs.harvard.edu/abs/2020MNRAS.499.5993R}
}

@ARTICLE{mittal22,
       author = {{Mittal}, Shikhar and {Ray}, Anupam and {Kulkarni}, Girish and {Dasgupta}, Basudeb},
        title = "{Constraining primordial black holes as dark matter using the global 21-cm signal with X-ray heating and excess radio background}",
      journal = {\jcap},
         year = 2022,
        month = mar,
       volume = {2022},
       number = {3},
          eid = {030},
        pages = {030},
          doi = {10.1088/1475-7516/2022/03/030},
archivePrefix = {arXiv},
       eprint = {2107.02190},
 primaryClass = {astro-ph.CO},
       adsurl = {https://ui.adsabs.harvard.edu/abs/2022JCAP...03..030M}
}

@ARTICLE{eloy22,
       author = {{de Lera Acedo}, E. and {de Villiers}, D.~I.~L. and {Razavi-Ghods}, N. and {Handley}, W. and {Fialkov}, A. and {Magro}, A. and {Anstey}, D. and {Bevins}, H.~T.~J. and {Chiello}, R. and {Cumner}, J. and {Josaitis}, A.~T. and {Roque}, I.~L.~V. and {Sims}, P.~H. and {Scheutwinkel}, K.~H. and {Alexander}, P. and {Bernardi}, G. and {Carey}, S. and {Cavillot}, J. and {Croukamp}, W. and {Ely}, J.~A. and {Gessey-Jones}, T. and {Gueuning}, Q. and {Hills}, R. and {Kulkarni}, G. and {Maiolino}, R. and {Meerburg}, P.~D. and {Mittal}, S. and {Pritchard}, J.~R. and {Puchwein}, E. and {Saxena}, A. and {Shen}, E. and {Smirnov}, O. and {Spinelli}, M. and {Zarb-Adami}, K.},
        title = "{The REACH radiometer for detecting the 21-cm hydrogen signal from redshift z {\ensuremath{\approx}} 7.5-28}",
      journal = {Nature Astronomy},
         year = 2022,
        month = jul,
       volume = {6},
        pages = {984-998},
          doi = {10.1038/s41550-022-01709-9},
archivePrefix = {arXiv},
       eprint = {2210.07409},
 primaryClass = {astro-ph.CO},
       adsurl = {https://ui.adsabs.harvard.edu/abs/2022NatAs...6..984D}
}

@INPROCEEDINGS{deboer15,
       author = {{DeBoer}, David R. and {HERA}},
        title = "{Hydrogen Epoch of Reionization Array (HERA)}",
    booktitle = {American Astronomical Society Meeting Abstracts \#225},
         year = 2015,
       series = {American Astronomical Society Meeting Abstracts},
       volume = {225},
        month = jan,
          eid = {328.03},
        pages = {328.03},
       adsurl = {https://ui.adsabs.harvard.edu/abs/2015AAS...22532803D}
}

@ARTICLE{heraupper22,
       author = {{Abdurashidova}, Zara and {Aguirre}, James E. and {Alexander}, Paul and {Ali}, Zaki S. and {Balfour}, Yanga and {Barkana}, Rennan and {Beardsley}, Adam P. and {Bernardi}, Gianni and {Billings}, Tashalee S. and {Bowman}, Judd D. and {Bradley}, Richard F. and {Bull}, Philip and {Burba}, Jacob and {Carey}, Steve and {Carilli}, Chris L. and {Cheng}, Carina and {DeBoer}, David R. and {Dexter}, Matt and {de Lera Acedo}, Eloy and {Dillon}, Joshua S. and {Ely}, John and {Ewall-Wice}, Aaron and {Fagnoni}, Nicolas and {Fialkov}, Anastasia and {Fritz}, Randall and {Furlanetto}, Steven R. and {Gale-Sides}, Kingsley and {Glendenning}, Brian and {Gorthi}, Deepthi and {Greig}, Bradley and {Grobbelaar}, Jasper and {Halday}, Ziyaad and {Hazelton}, Bryna J. and {Heimersheim}, Stefan and {Hewitt}, Jacqueline N. and {Hickish}, Jack and {Jacobs}, Daniel C. and {Julius}, Austin and {Kern}, Nicholas S. and {Kerrigan}, Joshua and {Kittiwisit}, Piyanat and {Kohn}, Saul A. and {Kolopanis}, Matthew and {Lanman}, Adam and {La Plante}, Paul and {Lekalake}, Telalo and {Lewis}, David and {Liu}, Adrian and {Ma}, Yin-Zhe and {MacMahon}, David and {Malan}, Lourence and {Malgas}, Cresshim and {Maree}, Matthys and {Martinot}, Zachary E. and {Matsetela}, Eunice and {Mesinger}, Andrei and {Mirocha}, Jordan and {Molewa}, Mathakane and {Morales}, Miguel F. and {Mosiane}, Tshegofalang and {Mu{\~n}oz}, Julian B. and {Murray}, Steven G. and {Neben}, Abraham R. and {Nikolic}, Bojan and {Nunhokee}, Chuneeta D. and {Parsons}, Aaron R. and {Patra}, Nipanjana and {Pieterse}, Samantha and {Pober}, Jonathan C. and {Qin}, Yuxiang and {Razavi-Ghods}, Nima and {Reis}, Itamar and {Ringuette}, Jon and {Robnett}, James and {Rosie}, Kathryn and {Santos}, Mario G. and {Sikder}, Sudipta and {Sims}, Peter and {Smith}, Craig and {Syce}, Angelo and {Thyagarajan}, Nithyanandan and {Williams}, Peter K.~G. and {Zheng}, Haoxuan},
        title = "{HERA Phase I Limits on the Cosmic 21 cm Signal: Constraints on Astrophysics and Cosmology during the Epoch of Reionization}",
      journal = {\apj},
         year = 2022,
        month = jan,
       volume = {924},
       number = {2},
          eid = {51},
        pages = {51},
          doi = {10.3847/1538-4357/ac2ffc},
archivePrefix = {arXiv},
       eprint = {2108.07282},
 primaryClass = {astro-ph.CO},
       adsurl = {https://ui.adsabs.harvard.edu/abs/2022ApJ...924...51A}
}

@ARTICLE{van15,
       author = {{van Haarlem}, M.~P. and {Wise}, M.~W. and {Gunst}, A.~W. and {Heald}, G. and {McKean}, J.~P. and {Hessels}, J.~W.~T. and {de Bruyn}, A.~G. and {Nijboer}, R. and {Swinbank}, J. and {Fallows}, R. and {Brentjens}, M. and {Nelles}, A. and {Beck}, R. and {Falcke}, H. and {Fender}, R. and {H{\"o}randel}, J. and {Koopmans}, L.~V.~E. and {Mann}, G. and {Miley}, G. and {R{\"o}ttgering}, H. and {Stappers}, B.~W. and {Wijers}, R.~A.~M.~J. and {Zaroubi}, S. and {van den Akker}, M. and {Alexov}, A. and {Anderson}, J. and {Anderson}, K. and {van Ardenne}, A. and {Arts}, M. and {Asgekar}, A. and {Avruch}, I.~M. and {Batejat}, F. and {B{\"a}hren}, L. and {Bell}, M.~E. and {Bell}, M.~R. and {van Bemmel}, I. and {Bennema}, P. and {Bentum}, M.~J. and {Bernardi}, G. and {Best}, P. and {B{\^\i}rzan}, L. and {Bonafede}, A. and {Boonstra}, A. -J. and {Braun}, R. and {Bregman}, J. and {Breitling}, F. and {van de Brink}, R.~H. and {Broderick}, J. and {Broekema}, P.~C. and {Brouw}, W.~N. and {Br{\"u}ggen}, M. and {Butcher}, H.~R. and {van Cappellen}, W. and {Ciardi}, B. and {Coenen}, T. and {Conway}, J. and {Coolen}, A. and {Corstanje}, A. and {Damstra}, S. and {Davies}, O. and {Deller}, A.~T. and {Dettmar}, R. -J. and {van Diepen}, G. and {Dijkstra}, K. and {Donker}, P. and {Doorduin}, A. and {Dromer}, J. and {Drost}, M. and {van Duin}, A. and {Eisl{\"o}ffel}, J. and {van Enst}, J. and {Ferrari}, C. and {Frieswijk}, W. and {Gankema}, H. and {Garrett}, M.~A. and {de Gasperin}, F. and {Gerbers}, M. and {de Geus}, E. and {Grie{\ss}meier}, J. -M. and {Grit}, T. and {Gruppen}, P. and {Hamaker}, J.~P. and {Hassall}, T. and {Hoeft}, M. and {Holties}, H.~A. and {Horneffer}, A. and {van der Horst}, A. and {van Houwelingen}, A. and {Huijgen}, A. and {Iacobelli}, M. and {Intema}, H. and {Jackson}, N. and {Jelic}, V. and {de Jong}, A. and {Juette}, E. and {Kant}, D. and {Karastergiou}, A. and {Koers}, A. and {Kollen}, H. and {Kondratiev}, V.~I. and {Kooistra}, E. and {Koopman}, Y. and {Koster}, A. and {Kuniyoshi}, M. and {Kramer}, M. and {Kuper}, G. and {Lambropoulos}, P. and {Law}, C. and {van Leeuwen}, J. and {Lemaitre}, J. and {Loose}, M. and {Maat}, P. and {Macario}, G. and {Markoff}, S. and {Masters}, J. and {McFadden}, R.~A. and {McKay-Bukowski}, D. and {Meijering}, H. and {Meulman}, H. and {Mevius}, M. and {Middelberg}, E. and {Millenaar}, R. and {Miller-Jones}, J.~C.~A. and {Mohan}, R.~N. and {Mol}, J.~D. and {Morawietz}, J. and {Morganti}, R. and {Mulcahy}, D.~D. and {Mulder}, E. and {Munk}, H. and {Nieuwenhuis}, L. and {van Nieuwpoort}, R. and {Noordam}, J.~E. and {Norden}, M. and {Noutsos}, A. and {Offringa}, A.~R. and {Olofsson}, H. and {Omar}, A. and {Orr{\'u}}, E. and {Overeem}, R. and {Paas}, H. and {Pandey-Pommier}, M. and {Pandey}, V.~N. and {Pizzo}, R. and {Polatidis}, A. and {Rafferty}, D. and {Rawlings}, S. and {Reich}, W. and {de Reijer}, J. -P. and {Reitsma}, J. and {Renting}, G.~A. and {Riemers}, P. and {Rol}, E. and {Romein}, J.~W. and {Roosjen}, J. and {Ruiter}, M. and {Scaife}, A. and {van der Schaaf}, K. and {Scheers}, B. and {Schellart}, P. and {Schoenmakers}, A. and {Schoonderbeek}, G. and {Serylak}, M. and {Shulevski}, A. and {Sluman}, J. and {Smirnov}, O. and {Sobey}, C. and {Spreeuw}, H. and {Steinmetz}, M. and {Sterks}, C.~G.~M. and {Stiepel}, H. -J. and {Stuurwold}, K. and {Tagger}, M. and {Tang}, Y. and {Tasse}, C. and {Thomas}, I. and {Thoudam}, S. and {Toribio}, M.~C. and {van der Tol}, B. and {Usov}, O. and {van Veelen}, M. and {van der Veen}, A. -J. and {ter Veen}, S. and {Verbiest}, J.~P.~W. and {Vermeulen}, R. and {Vermaas}, N. and {Vocks}, C. and {Vogt}, C. and {de Vos}, M. and {van der Wal}, E. and {van Weeren}, R. and {Weggemans}, H. and {Weltevrede}, P. and {White}, S. and {Wijnholds}, S.~J. and {Wilhelmsson}, T. and {Wucknitz}, O. and {Yatawatta}, S. and {Zarka}, P. and {Zensus}, A. and {van Zwieten}, J.},
        title = "{LOFAR: The LOw-Frequency ARray}",
      journal = {\aap},
         year = 2013,
        month = aug,
       volume = {556},
          eid = {A2},
        pages = {A2},
          doi = {10.1051/0004-6361/201220873},
archivePrefix = {arXiv},
       eprint = {1305.3550},
 primaryClass = {astro-ph.IM},
       adsurl = {https://ui.adsabs.harvard.edu/abs/2013A&A...556A...2V}
}

@INPROCEEDINGS{mertens21,
       author = {{Mertens}, F.~G. and {Semelin}, B. and {Koopmans}, L.~V.~E.},
        title = "{Exploring the Cosmic Dawn with NenuFAR}",
    booktitle = {SF2A-2021: Proceedings of the Annual meeting of the French Society of Astronomy and Astrophysics},
         year = 2021,
       editor = {{Siebert}, A. and {Bailli{\'e}}, K. and {Lagadec}, E. and {Lagarde}, N. and {Malzac}, J. and {Marquette}, J. -B. and {N'Diaye}, M. and {Richard}, J. and {Venot}, O.},
        month = dec,
        pages = {211-214},
          doi = {10.48550/arXiv.2109.10055},
archivePrefix = {arXiv},
       eprint = {2109.10055},
 primaryClass = {astro-ph.CO},
       adsurl = {https://ui.adsabs.harvard.edu/abs/2021sf2a.conf..211M}
}

@ARTICLE{munshi25,
       author = {{Munshi}, S. and {Mertens}, F.~G. and {Chege}, J.~K. and {Koopmans}, L.~V.~E. and {Offringa}, A.~R. and {Semelin}, B. and {Barkana}, R. and {Dhandha}, J. and {Fialkov}, A. and {M{\'e}riot}, R. and {Sikder}, S. and {Bracco}, A. and {Brackenhoff}, S.~A. and {Ceccotti}, E. and {Ghara}, R. and {Ghosh}, S. and {Hothi}, I. and {Mevius}, M. and {Ocvirk}, P. and {Shaw}, A.~K. and {Yatawatta}, S. and {Zarka}, P.},
        title = "{Improved upper limits on the 21-cm signal power spectrum at z = 17.0 and z = 20.3 from an optimal field observed with NenuFAR}",
      journal = {\mnras},
         year = 2025,
        month = oct,
       volume = {542},
       number = {4},
        pages = {2785-2807},
          doi = {10.1093/mnras/staf1386},
archivePrefix = {arXiv},
       eprint = {2507.10533},
 primaryClass = {astro-ph.CO},
       adsurl = {https://ui.adsabs.harvard.edu/abs/2025MNRAS.542.2785M}
}

@ARTICLE{tingay13,
       author = {{Tingay}, S.~J. and {Goeke}, R. and {Bowman}, J.~D. and {Emrich}, D. and {Ord}, S.~M. and {Mitchell}, D.~A. and {Morales}, M.~F. and {Booler}, T. and {Crosse}, B. and {Wayth}, R.~B. and {Lonsdale}, C.~J. and {Tremblay}, S. and {Pallot}, D. and {Colegate}, T. and {Wicenec}, A. and {Kudryavtseva}, N. and {Arcus}, W. and {Barnes}, D. and {Bernardi}, G. and {Briggs}, F. and {Burns}, S. and {Bunton}, J.~D. and {Cappallo}, R.~J. and {Corey}, B.~E. and {Deshpande}, A. and {Desouza}, L. and {Gaensler}, B.~M. and {Greenhill}, L.~J. and {Hall}, P.~J. and {Hazelton}, B.~J. and {Herne}, D. and {Hewitt}, J.~N. and {Johnston-Hollitt}, M. and {Kaplan}, D.~L. and {Kasper}, J.~C. and {Kincaid}, B.~B. and {Koenig}, R. and {Kratzenberg}, E. and {Lynch}, M.~J. and {Mckinley}, B. and {Mcwhirter}, S.~R. and {Morgan}, E. and {Oberoi}, D. and {Pathikulangara}, J. and {Prabu}, T. and {Remillard}, R.~A. and {Rogers}, A.~E.~E. and {Roshi}, A. and {Salah}, J.~E. and {Sault}, R.~J. and {Udaya-Shankar}, N. and {Schlagenhaufer}, F. and {Srivani}, K.~S. and {Stevens}, J. and {Subrahmanyan}, R. and {Waterson}, M. and {Webster}, R.~L. and {Whitney}, A.~R. and {Williams}, A. and {Williams}, C.~L. and {Wyithe}, J.~S.~B.},
        title = "{The Murchison Widefield Array: The Square Kilometre Array Precursor at Low Radio Frequencies}",
      journal = {\pasa},
         year = 2013,
        month = jan,
       volume = {30},
          eid = {e007},
        pages = {e007},
          doi = {10.1017/pasa.2012.007},
archivePrefix = {arXiv},
       eprint = {1206.6945},
 primaryClass = {astro-ph.IM},
       adsurl = {https://ui.adsabs.harvard.edu/abs/2013PASA...30....7T}
}

@ARTICLE{shaver99,
       author = {{Shaver}, P.~A. and {Windhorst}, R.~A. and {Madau}, P. and {de Bruyn}, A.~G.},
        title = "{Can the reionization epoch be detected as a global signature in the cosmic background?}",
      journal = {\aap},
         year = 1999,
        month = may,
       volume = {345},
        pages = {380-390},
          doi = {10.48550/arXiv.astro-ph/9901320},
archivePrefix = {arXiv},
       eprint = {astro-ph/9901320},
 primaryClass = {astro-ph},
       adsurl = {https://ui.adsabs.harvard.edu/abs/1999A&A...345..380S}
}

@ARTICLE{Bharadwaj05,
       author = {{Bharadwaj}, Somnath and {Ali}, Sk. Saiyad},
        title = "{On using visibility correlations to probe the HI distribution from the dark ages to the present epoch - I. Formalism and the expected signal}",
      journal = {\mnras},
         year = 2005,
        month = feb,
       volume = {356},
       number = {4},
        pages = {1519-1528},
          doi = {10.1111/j.1365-2966.2004.08604.x},
archivePrefix = {arXiv},
       eprint = {astro-ph/0406676},
 primaryClass = {astro-ph},
       adsurl = {https://ui.adsabs.harvard.edu/abs/2005MNRAS.356.1519B}
}

@ARTICLE{jelic08,
       author = {{Jeli{\'c}}, V. and {Zaroubi}, S. and {Labropoulos}, P. and {Thomas}, R.~M. and {Bernardi}, G. and {Brentjens}, M.~A. and {de Bruyn}, A.~G. and {Ciardi}, B. and {Harker}, G. and {Koopmans}, L.~V.~E. and {Pandey}, V.~N. and {Schaye}, J. and {Yatawatta}, S.},
        title = "{Foreground simulations for the LOFAR-epoch of reionization experiment}",
      journal = {\mnras},
         year = 2008,
        month = sep,
       volume = {389},
       number = {3},
        pages = {1319-1335},
          doi = {10.1111/j.1365-2966.2008.13634.x},
archivePrefix = {arXiv},
       eprint = {0804.1130},
 primaryClass = {astro-ph},
       adsurl = {https://ui.adsabs.harvard.edu/abs/2008MNRAS.389.1319J}
}

@ARTICLE{jelic10,
       author = {{Jeli{\'c}}, Vibor and {Zaroubi}, Saleem and {Labropoulos}, Panagiotis and {Bernardi}, Gianni and {de Bruyn}, A.~G. and {Koopmans}, L{\'e}on V.~E.},
        title = "{Realistic simulations of the Galactic polarized foreground: consequences for 21-cm reionization detection experiments}",
      journal = {\mnras},
         year = 2010,
        month = dec,
       volume = {409},
       number = {4},
        pages = {1647-1659},
          doi = {10.1111/j.1365-2966.2010.17407.x},
archivePrefix = {arXiv},
       eprint = {1007.4135},
 primaryClass = {astro-ph.GA},
       adsurl = {https://ui.adsabs.harvard.edu/abs/2010MNRAS.409.1647J}
}

@ARTICLE{datta10,
       author = {{Datta}, A. and {Bowman}, J.~D. and {Carilli}, C.~L.},
        title = "{Bright Source Subtraction Requirements for Redshifted 21 cm Measurements}",
      journal = {\apj},
         year = 2010,
        month = nov,
       volume = {724},
       number = {1},
        pages = {526-538},
          doi = {10.1088/0004-637X/724/1/526},
archivePrefix = {arXiv},
       eprint = {1005.4071},
 primaryClass = {astro-ph.CO},
       adsurl = {https://ui.adsabs.harvard.edu/abs/2010ApJ...724..526D}
}

@ARTICLE{datta16,
       author = {{Datta}, Abhirup and {Bradley}, Richard and {Burns}, Jack O. and {Harker}, Geraint and {Komjathy}, Attila and {Lazio}, T. Joseph W.},
        title = "{The Effects of the Ionosphere on Ground-based Detection of the Global 21 cm Signal from the Cosmic Dawn and the Dark Ages}",
      journal = {\apj},
         year = 2016,
        month = nov,
       volume = {831},
       number = {1},
          eid = {6},
        pages = {6},
          doi = {10.3847/0004-637X/831/1/6},
       adsurl = {https://ui.adsabs.harvard.edu/abs/2016ApJ...831....6D}
}

@ARTICLE{shen21,
       author = {{Shen}, Emma and {Anstey}, Dominic and {de Lera Acedo}, Eloy and {Fialkov}, Anastasia and {Handley}, Will},
        title = "{Quantifying ionospheric effects on global 21-cm observations}",
      journal = {\mnras},
         year = 2021,
        month = may,
       volume = {503},
       number = {1},
        pages = {344-353},
          doi = {10.1093/mnras/stab429},
archivePrefix = {arXiv},
       eprint = {2011.10517},
 primaryClass = {astro-ph.IM},
       adsurl = {https://ui.adsabs.harvard.edu/abs/2021MNRAS.503..344S}
}

@ARTICLE{nasirudin20,
       author = {{Nasirudin}, A. and {Murray}, S.~G. and {Trott}, C.~M. and {Greig}, B. and {Joseph}, R.~C. and {Power}, C.},
        title = "{The Impact of Realistic Foreground and Instrument Models on 21 cm Epoch of Reionization Experiments}",
      journal = {\apj},
         year = 2020,
        month = apr,
       volume = {893},
       number = {2},
          eid = {118},
        pages = {118},
          doi = {10.3847/1538-4357/ab8003},
archivePrefix = {arXiv},
       eprint = {2003.08552},
 primaryClass = {astro-ph.CO},
       adsurl = {https://ui.adsabs.harvard.edu/abs/2020ApJ...893..118N}
}

@ARTICLE{pattinson24,
       author = {{Pattison}, Joe H.~N. and {Anstey}, Dominic J. and {de Lera Acedo}, Eloy},
        title = "{Modelling a hot horizon in global 21-cm experimental foregrounds}",
      journal = {\mnras},
         year = 2024,
        month = jan,
       volume = {527},
       number = {2},
        pages = {2413-2425},
          doi = {10.1093/mnras/stad3378},
archivePrefix = {arXiv},
       eprint = {2307.02908},
 primaryClass = {astro-ph.CO},
       adsurl = {https://ui.adsabs.harvard.edu/abs/2024MNRAS.527.2413P}
}

@ARTICLE{kim22,
       author = {{Kim}, Honggeun and {Nhan}, Bang D. and {Hewitt}, Jacqueline N. and {Kern}, Nicholas S. and {Dillon}, Joshua S. and {de Lera Acedo}, Eloy and {Dynes}, Scott B.~C. and {Mahesh}, Nivedita and {Fagnoni}, Nicolas and {DeBoer}, David R.},
        title = "{The Impact of Beam Variations on Power Spectrum Estimation for 21 cm Cosmology. I. Simulations of Foreground Contamination for HERA}",
      journal = {\apj},
         year = 2022,
        month = dec,
       volume = {941},
       number = {2},
          eid = {207},
        pages = {207},
          doi = {10.3847/1538-4357/ac9eaf},
archivePrefix = {arXiv},
       eprint = {2210.16421},
 primaryClass = {astro-ph.IM},
       adsurl = {https://ui.adsabs.harvard.edu/abs/2022ApJ...941..207K}
}

@ARTICLE{sims20,
       author = {{Sims}, Peter H. and {Pober}, Jonathan C.},
        title = "{Testing for calibration systematics in the EDGES low-band data using Bayesian model selection}",
      journal = {\mnras},
         year = 2020,
        month = feb,
       volume = {492},
       number = {1},
        pages = {22-38},
          doi = {10.1093/mnras/stz3388},
archivePrefix = {arXiv},
       eprint = {1910.03165},
 primaryClass = {astro-ph.CO},
       adsurl = {https://ui.adsabs.harvard.edu/abs/2020MNRAS.492...22S}
}

@ARTICLE{dasgupta23,
       author = {{Dasgupta}, Saswata and {Pal}, Samit Kumar and {Bag}, Satadru and {Dutta}, Sohini and {Majumdar}, Suman and {Datta}, Abhirup and {Pathak}, Aadarsh and {Kamran}, Mohd and {Mondal}, Rajesh and {Sarkar}, Prakash},
        title = "{Interpreting the HI 21-cm cosmology maps through Largest Cluster Statistics. Part I. Impact of the synthetic SKA1-Low observations}",
      journal = {\jcap},
         year = 2023,
        month = may,
       volume = {2023},
       number = {5},
          eid = {014},
        pages = {014},
          doi = {10.1088/1475-7516/2023/05/014},
archivePrefix = {arXiv},
       eprint = {2302.02727},
 primaryClass = {astro-ph.CO},
       adsurl = {https://ui.adsabs.harvard.edu/abs/2023JCAP...05..014D}
}

@ARTICLE{hills18,
       author = {{Hill}, J. Colin and {Baxter}, Eric J.},
        title = "{Can early dark energy explain EDGES?}",
      journal = {\jcap},
         year = 2018,
        month = aug,
       volume = {2018},
       number = {8},
          eid = {037},
        pages = {037},
          doi = {10.1088/1475-7516/2018/08/037},
archivePrefix = {arXiv},
       eprint = {1803.07555},
 primaryClass = {astro-ph.CO},
       adsurl = {https://ui.adsabs.harvard.edu/abs/2018JCAP...08..037H}
}

@ARTICLE{pal25,
       author = {{Pal}, Samit Kumar and {Dasgupta}, Saswata and {Datta}, Abhirup and {Majumdar}, Suman and {Bag}, Satadru and {Sarkar}, Prakash},
        title = "{Interpreting the HI 21-cm cosmology maps through Largest Cluster Statistics. Part II. Impact of the realistic foreground and instrumental noise on synthetic SKA1-Low observations}",
      journal = {arXiv e-prints},
         year = 2025,
        month = mar,
          eid = {arXiv:2503.00919},
        pages = {arXiv:2503.00919},
          doi = {10.48550/arXiv.2503.00919},
archivePrefix = {arXiv},
       eprint = {2503.00919},
 primaryClass = {astro-ph.CO},
       adsurl = {https://ui.adsabs.harvard.edu/abs/2025arXiv250300919P}
}

@ARTICLE{ohara24,
       author = {{O'Hara}, Oscar S.~D. and {Dulwich}, Fred and {de Lera Acedo}, Eloy and {Dhandha}, Jiten and {Gessey-Jones}, Thomas and {Anstey}, Dominic and {Fialkov}, Anastasia},
        title = "{Understanding spectral artefacts in SKA-Low 21-cm cosmology experiments: the impact of cable reflections}",
      journal = {\mnras},
         year = 2024,
        month = sep,
       volume = {533},
       number = {3},
        pages = {2876-2892},
          doi = {10.1093/mnras/stae1952},
archivePrefix = {arXiv},
       eprint = {2402.04008},
 primaryClass = {astro-ph.CO},
       adsurl = {https://ui.adsabs.harvard.edu/abs/2024MNRAS.533.2876O}
}

@ARTICLE{cunnington21,
       author = {{Cunnington}, Steven and {Irfan}, Melis O. and {Carucci}, Isabella P. and {Pourtsidou}, Alkistis and {Bobin}, J{\'e}r{\^o}me},
        title = "{21-cm foregrounds and polarization leakage: cleaning and mitigation strategies}",
      journal = {\mnras},
         year = 2021,
        month = jun,
       volume = {504},
       number = {1},
        pages = {208-227},
          doi = {10.1093/mnras/stab856},
archivePrefix = {arXiv},
       eprint = {2010.02907},
 primaryClass = {astro-ph.CO},
       adsurl = {https://ui.adsabs.harvard.edu/abs/2021MNRAS.504..208C}
}

@ARTICLE{carucci20,
       author = {{Carucci}, Isabella P. and {Irfan}, Melis O. and {Bobin}, J{\'e}r{\^o}me},
        title = "{Recovery of 21-cm intensity maps with sparse component separation}",
      journal = {\mnras},
         year = 2020,
        month = nov,
       volume = {499},
       number = {1},
        pages = {304-319},
          doi = {10.1093/mnras/staa2854},
archivePrefix = {arXiv},
       eprint = {2006.05996},
 primaryClass = {astro-ph.CO},
       adsurl = {https://ui.adsabs.harvard.edu/abs/2020MNRAS.499..304C}
}

@ARTICLE{rath25,
       author = {{Rath}, E. and {Pascua}, R. and {Josaitis}, A.~T. and {Ewall-Wice}, A. and {Fagnoni}, N. and {de Lera Acedo}, E. and {Martinot}, Z.~E. and {Abdurashidova}, Z. and {Adams}, T. and {Aguirre}, J.~E. and {Baartman}, R. and {Beardsley}, A.~P. and {Berkhout}, L.~M. and {Bernardi}, G. and {Billings}, T.~S. and {Bowman}, J.~D. and {Bull}, P. and {Burba}, J. and {Byrne}, R. and {Carey}, S. and {Chen}, K. -F. and {Choudhuri}, S. and {Cox}, T. and {DeBoer}, D.~R. and {Dexter}, M. and {Dillon}, J.~S. and {Dynes}, S. and {Eksteen}, N. and {Ely}, J. and {Fritz}, R. and {Furlanetto}, S.~R. and {Gale-Sides}, K. and {Garsden}, H. and {Gehlot}, B.~K. and {Ghosh}, A. and {Gorce}, A. and {Gorthi}, D. and {Halday}, Z. and {Hazelton}, B.~J. and {Hewitt}, J.~N. and {Hickish}, J. and {Huang}, T. and {Jacobs}, D.~C. and {Kern}, N.~S. and {Kerrigan}, J. and {Kittiwisit}, P. and {Kolopanis}, M. and {Lanman}, A. and {Liu}, A. and {Ma}, Y. -Z. and {MacMahon}, D.~H.~E. and {Malan}, L. and {Malgas}, C. and {Malgas}, K. and {Marero}, B. and {McBride}, L. and {Mesinger}, A. and {Mohamed-Hinds}, N. and {Molewa}, M. and {Morales}, M.~F. and {Murray}, S.~G. and {Nikolic}, B. and {Nuwegeld}, H. and {Parsons}, A.~R. and {Patra}, N. and {Plante}, P. La and {Qin}, Y. and {Razavi-Ghods}, N. and {Riley}, D. and {Robnett}, J. and {Rosie}, K. and {Santos}, M.~G. and {Sims}, P. and {Singh}, S. and {Storer}, D. and {Swarts}, H. and {Tan}, J. and {Wilensky}, M.~J. and {Williams}, P.~K.~G. and {van Wyngaarden}, P. and {Zheng}, H.},
        title = "{Investigating mutual coupling in the hydrogen epoch of reionization array and mitigating its effects on the 21-cm power spectrum}",
      journal = {\mnras},
         year = 2025,
        month = aug,
       volume = {541},
       number = {2},
        pages = {1125-1144},
          doi = {10.1093/mnras/staf1012},
archivePrefix = {arXiv},
       eprint = {2406.08549},
 primaryClass = {astro-ph.CO},
       adsurl = {https://ui.adsabs.harvard.edu/abs/2025MNRAS.541.1125R}
}

@ARTICLE{ohara25,
       author = {{O'Hara}, Oscar S.~D. and {Gueuning}, Quentin and {de Lera Acedo}, Eloy and {Dulwich}, Fred and {Cumner}, John and {Anstey}, Dominic and {Brown}, Anthony and {Fialkov}, Anastasia and {Dhandha}, Jiten and {Faulkner}, Andrew and {Liu}, Yuchen},
        title = "{Uncovering the effects of array mutual coupling in 21-cm experiments with the SKA-Low radio telescope}",
      journal = {\mnras},
         year = 2025,
        month = mar,
       volume = {538},
       number = {1},
        pages = {31-48},
          doi = {10.1093/mnras/staf264},
archivePrefix = {arXiv},
       eprint = {2412.01699},
 primaryClass = {astro-ph.CO},
       adsurl = {https://ui.adsabs.harvard.edu/abs/2025MNRAS.538...31O}
}

@ARTICLE{Schauer2019,
       author = {{Schauer}, Anna T.~P. and {Liu}, Boyuan and {Bromm}, Volker},
        title = "{Constraining First Star Formation with 21 cm Cosmology}",
      journal = {\apjl},
         year = 2019,
        month = may,
       volume = {877},
       number = {1},
          eid = {L5},
        pages = {L5},
          doi = {10.3847/2041-8213/ab1e51},
archivePrefix = {arXiv},
       eprint = {1901.03344},
 primaryClass = {astro-ph.GA},
       adsurl = {https://ui.adsabs.harvard.edu/abs/2019ApJ...877L...5S}
}

@ARTICLE{Liu2019,
       author = {{Liu}, Boyuan and {Schauer}, Anna T.~P. and {Bromm}, Volker},
        title = "{Constraining the non-gravitational scattering of baryons and dark matter with early cosmic structure formation}",
      journal = {\mnras},
         year = 2019,
        month = aug,
       volume = {487},
       number = {4},
        pages = {4711-4720},
          doi = {10.1093/mnras/stz1583},
archivePrefix = {arXiv},
       eprint = {1902.09650},
 primaryClass = {astro-ph.GA},
       adsurl = {https://ui.adsabs.harvard.edu/abs/2019MNRAS.487.4711L}
}

@ARTICLE{Grudic2023,
       author = {{Grudic}, Michael Y. and {Offner}, Stella S.~R. and {Guszejnov}, D{\'a}vid and {Faucher-Gigu{\`e}re}, Claude-Andr{\'e} and {Hopkins}, Philip F.},
        title = "{Does God play dice with star clusters?}",
      journal = {The Open Journal of Astrophysics},
         year = 2023,
        month = dec,
       volume = {6},
          eid = {48},
        pages = {48},
          doi = {10.21105/astro.2307.00052},
archivePrefix = {arXiv},
       eprint = {2307.00052},
 primaryClass = {astro-ph.GA},
       adsurl = {https://ui.adsabs.harvard.edu/abs/2023OJAp....6E..48G}
}

@ARTICLE{Yan2023,
       author = {{Yan}, Zhiqiang and {Jerabkova}, Tereza and {Kroupa}, Pavel},
        title = "{The most massive stars in very young star clusters with a limited mass: Evidence favours significant self-regulation in the star formation processes}",
      journal = {\aap},
         year = 2023,
        month = feb,
       volume = {670},
          eid = {A151},
        pages = {A151},
          doi = {10.1051/0004-6361/202244919},
archivePrefix = {arXiv},
       eprint = {2211.13244},
 primaryClass = {astro-ph.GA},
       adsurl = {https://ui.adsabs.harvard.edu/abs/2023A&A...670A.151Y}
}

@ARTICLE{Smith2023,
       author = {{Smith}, Jamie D. and {Jaffa}, Sarah E. and {Krause}, Martin G.~H.},
        title = "{Can massive stars form in low mass clouds?}",
      journal = {\mnras},
         year = 2023,
        month = nov,
       volume = {525},
       number = {4},
        pages = {6182-6190},
          doi = {10.1093/mnras/stad2689},
archivePrefix = {arXiv},
       eprint = {2309.05635},
 primaryClass = {astro-ph.GA},
       adsurl = {https://ui.adsabs.harvard.edu/abs/2023MNRAS.525.6182S}
}

@ARTICLE{shull85,
       author = {{Shull}, J.~M. and {van Steenberg}, M.~E.},
        title = "{X-ray secondary heating and ionization in quasar emission-line clouds}",
      journal = {\apj},
         year = 1985,
        month = nov,
       volume = {298},
        pages = {268-274},
          doi = {10.1086/163605},
       adsurl = {https://ui.adsabs.harvard.edu/abs/1985ApJ...298..268S}
}

@ARTICLE{mirocha17,
       author = {{Mirocha}, Jordan and {Furlanetto}, Steven R. and {Sun}, Guochao},
        title = "{The global 21-cm signal in the context of the high- z galaxy luminosity function}",
      journal = {\mnras},
         year = 2017,
        month = jan,
       volume = {464},
       number = {2},
        pages = {1365-1379},
          doi = {10.1093/mnras/stw2412},
archivePrefix = {arXiv},
       eprint = {1607.00386},
 primaryClass = {astro-ph.GA},
       adsurl = {https://ui.adsabs.harvard.edu/abs/2017MNRAS.464.1365M}
}

@ARTICLE{gurian24,
       author = {{Gurian}, James and {Jeong}, Donghui and {Liu}, Boyuan},
        title = "{Zero Metallicity with Zero CPU Hours: Masses of the First Stars on the Laptop}",
      journal = {\apj},
         year = 2024,
        month = mar,
       volume = {963},
       number = {1},
          eid = {33},
        pages = {33},
          doi = {10.3847/1538-4357/ad1e5b},
archivePrefix = {arXiv},
       eprint = {2309.05758},
 primaryClass = {astro-ph.CO},
       adsurl = {https://ui.adsabs.harvard.edu/abs/2024ApJ...963...33G}
}

@ARTICLE{chiaki18,
       author = {{Chiaki}, Gen and {Susa}, Hajime and {Hirano}, Shingo},
        title = "{Metal-poor star formation triggered by the feedback effects from Pop III stars}",
      journal = {\mnras},
         year = 2018,
        month = apr,
       volume = {475},
       number = {4},
        pages = {4378-4395},
          doi = {10.1093/mnras/sty040},
archivePrefix = {arXiv},
       eprint = {1801.01583},
 primaryClass = {astro-ph.GA},
       adsurl = {https://ui.adsabs.harvard.edu/abs/2018MNRAS.475.4378C}
}

@ARTICLE{bromm02,
       author = {{Bromm}, Volker and {Coppi}, Paolo S. and {Larson}, Richard B.},
        title = "{The Formation of the First Stars. I. The Primordial Star-forming Cloud}",
      journal = {\apj},
         year = 2002,
        month = jan,
       volume = {564},
       number = {1},
        pages = {23-51},
          doi = {10.1086/323947},
archivePrefix = {arXiv},
       eprint = {astro-ph/0102503},
 primaryClass = {astro-ph},
       adsurl = {https://ui.adsabs.harvard.edu/abs/2002ApJ...564...23B}
}

@ARTICLE{hirano14,
       author = {{Hirano}, Shingo and {Hosokawa}, Takashi and {Yoshida}, Naoki and {Umeda}, Hideyuki and {Omukai}, Kazuyuki and {Chiaki}, Gen and {Yorke}, Harold W.},
        title = "{One Hundred First Stars: Protostellar Evolution and the Final Masses}",
      journal = {\apj},
         year = 2014,
        month = feb,
       volume = {781},
       number = {2},
          eid = {60},
        pages = {60},
          doi = {10.1088/0004-637X/781/2/60},
archivePrefix = {arXiv},
       eprint = {1308.4456},
 primaryClass = {astro-ph.CO},
       adsurl = {https://ui.adsabs.harvard.edu/abs/2014ApJ...781...60H}
}

@ARTICLE{schauer21,
       author = {{Schauer}, Anna T.~P. and {Glover}, Simon C.~O. and {Klessen}, Ralf S. and {Clark}, Paul},
        title = "{The influence of streaming velocities and Lyman-Werner radiation on the formation of the first stars}",
      journal = {\mnras},
         year = 2021,
        month = oct,
       volume = {507},
       number = {2},
        pages = {1775-1787},
          doi = {10.1093/mnras/stab1953},
archivePrefix = {arXiv},
       eprint = {2008.05663},
 primaryClass = {astro-ph.GA},
       adsurl = {https://ui.adsabs.harvard.edu/abs/2021MNRAS.507.1775S}
}

@ARTICLE{barkana16,
       author = {{Barkana}, Rennan},
        title = "{The rise of the first stars: Supersonic streaming, radiative feedback, and 21-cm cosmology}",
      journal = {\physrep},
         year = 2016,
        month = jul,
       volume = {645},
        pages = {1-59},
          doi = {10.1016/j.physrep.2016.06.006},
archivePrefix = {arXiv},
       eprint = {1605.04357},
 primaryClass = {astro-ph.CO},
       adsurl = {https://ui.adsabs.harvard.edu/abs/2016PhR...645....1B}
}

@ARTICLE{amara08,
       author = {{Amara}, Adam and {R{\'e}fr{\'e}gier}, Alexandre},
        title = "{Systematic bias in cosmic shear: extending the Fisher matrix}",
      journal = {Monthly Notices of the Royal Astronomical Society},
         year = 2008,
        month = nov,
       volume = {391},
       number = {1},
        pages = {228-236},
          doi = {10.1111/j.1365-2966.2008.13880.x},
archivePrefix = {arXiv},
       eprint = {0710.5171},
 primaryClass = {astro-ph},
       adsurl = {https://ui.adsabs.harvard.edu/abs/2008MNRAS.391..228A},
}

@ARTICLE{taylor07,
       author = {{Taylor}, A.~N. and {Kitching}, T.~D. and {Bacon}, D.~J. and {Heavens}, A.~F.},
        title = "{Probing dark energy with the shear-ratio geometric test}",
      journal = {Monthly Notices of the Royal Astronomical Society},
         year = 2007,
        month = feb,
       volume = {374},
       number = {4},
        pages = {1377-1403},
          doi = {10.1111/j.1365-2966.2006.11257.x},
       adsurl = {https://ui.adsabs.harvard.edu/abs/2007MNRAS.374.1377T},
}

%%%%%%%%%%%%%%%%%%%%%%%%%%%%%%%%%%%%%%%%%%%%%%%%%%

%%%%%%%%%%%%%%%%% APPENDICES %%%%%%%%%%%%%%%%%%%%%

\appendix

\section{Supplementary Results}
\label{apdx:t_X}
In this appendix, we present additional plots and analyses that complement the main results of the paper and reinforce our conclusions.

\subsection{Impact of XRB lifetime on stochastic X-ray heating}
\label{app:tx_fixedparams}

To explore the dependence of the results on the typical XRB lifetime, we vary $t_X$ at fixed values of other parameters, assuming   $\alpha=0.2$ and $\fxiii=1$, showing the resulting heating rate plots in Figure \ref{fig:tx_eps_maps_fixed}. 

As indicated by Eq.~\eqref{eq:Nhat_tx_new}, 
$t_X$ only impacts the sampling variance, leaving the mean luminosity unchanged: shorter $t_X$ increases $\Delta t/t_X$ and therefore $\hat N$, which suppresses fluctuations via averaging over multiple short-lived XRBs within a snapshot; longer $t_X$ drives the pre-factor in Eq.~\eqref{eq:Nhat_tx_new} to unity, minimizing $\hat N$ per step and enhancing the stochastic variance of the emissivity. Longer $t_X$ also means a longer correlation timescale in X-ray emission, which, through the propagation of X-rays, enhances the small-scale spatial correlations in the heating rate fields.  These two effects imply that the stochasticity features in the X-ray emissivity map have a complex dependence on $t_X$:
\begin{itemize}
\item \textbf{Short lifetimes:} In this regime, many short-lived XRBs contribute within a single timestep, so their contributions are averaged together. The sampling variance of $l=L_X/\hat{L}_X$ is therefore suppressed. 
This case is shown in the leftmost panel of Figure \ref{fig:tx_eps_maps_fixed} where $t_X = 0.7~\text{Myr}$.
\item \textbf{Long lifetimes:} In this case, the same sources persist over many simulation timesteps. The resulting strong correlation in the $N_{\rm XRB}$ field for different redshift (simulation time) steps smooths the (stochastic) emissivity history. 
This could be seen for the case of $t_X = 100~\mathrm{Myr}$ in the rightmost panel of Figure \ref{fig:tx_eps_maps_fixed}.
\item \textbf{Intermediate lifetimes:}
In this case, $t_X$ is comparable to the simulation timestep, which results in the most prominent visibility of the ring-like structures in the X-ray heating rate maps as shown in the middle two panels of Figure~\ref{fig:tx_eps_maps_fixed} for $t_X=3.4$ and $10~\rm Myr$. 

\end{itemize}

\begin{figure*}
    \centering
    % update filename as needed
    \includegraphics[width=\textwidth]{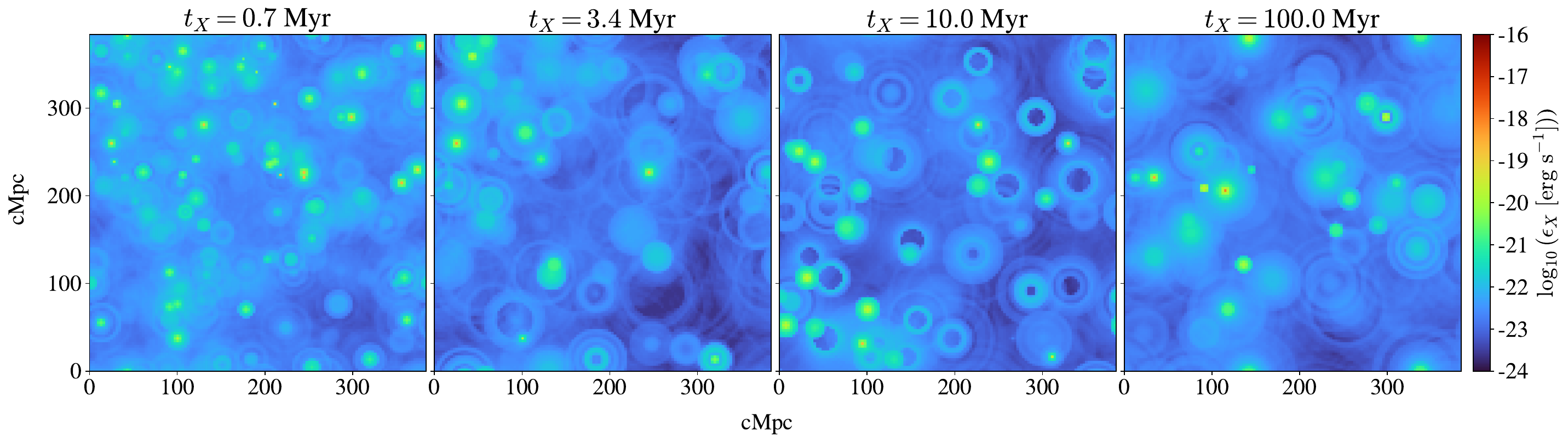}
    \caption{Heating rate per baryon at $z=32$ for scenarios with varying $t_X$ and fixed $\alpha=0.2$ and $\fxiii=1$. In each case, we show one slice of depth $3\,\rm cMpc$ of the heating rate fields. From left to right: $t_X=0.7$, $3.4$, $10$ and $100~\mathrm{Myr}$. These life-times correspond to redshift steps $\Delta z \sim 0.17$, 0.79, 1.1 and 13, respectively. In the first case, $t_X$ is much shorter than the simulation timestep (corresponding to $\Delta z = 1$), leading to averaging of the stochastic X-ray effects. In the two intermediate cases with $t_X = 3.4$ Myr and $t_X = 10$ Myr, the lifetime is very close to the simulation timestep, and the stochasticity is the most apparent. Finally, for $t_x = 100$ Myrs, XRBs are long-lived and keep contributing to the X-ray background for a long time, leading to bigger heating bubbles around individual sources.} 
    \label{fig:tx_eps_maps_fixed}
\end{figure*}

We next examine the impact of $t_X$ on the resulting 21-cm brightness temperature showing the snapshots in Figure  \ref{fig:tx_tb_maps_fixed} for $\alpha=0.2$ and $\fxiii=1$. Varying $t_X$ affects small-scale structure in $\delta T_b$ leaving its mean and large-scale morphology unchanged. The small-scale morphology of the 21-cm brightness temperature maps reflects the time-integrated thermal history of the IGM rather than the
instantaneous X-ray heating rate.
The differential brightness temperature which is governed by the spin temperature $T_S$ is driven towards the kinetic temperature $T_K$ once Ly-$\alpha$ coupling is efficient. In turn, the gas kinetic temperature depends on the cumulative X-ray energy deposited along the past light cone, \begin{equation}
T_K(\mathbf{x}, z) \propto \int^{t(z)} dt'\, \epsilon_X(\mathbf{x}, t') ,
\end{equation}
with $\epsilon_X$ being the heating rate per baryon. 
As a result, while intermediate X-ray lifetimes ($t_X \sim 10\,\mathrm{Myr}$) create clearer ringlike structures in the heating rate maps at high redshift, longer lifetimes ($t_X \sim 100\,\mathrm{Myr}$) produce stronger
spatial fluctuations in the brightness temperature field by lower redshift. At $z \sim 20$, small differences in the accumulated heating history can therefore translate into pronounced local fluctuations leading to speckles in the 21-cm maps.

\begin{figure*}
    \centering
    % update filename as needed
    \includegraphics[width=\textwidth]{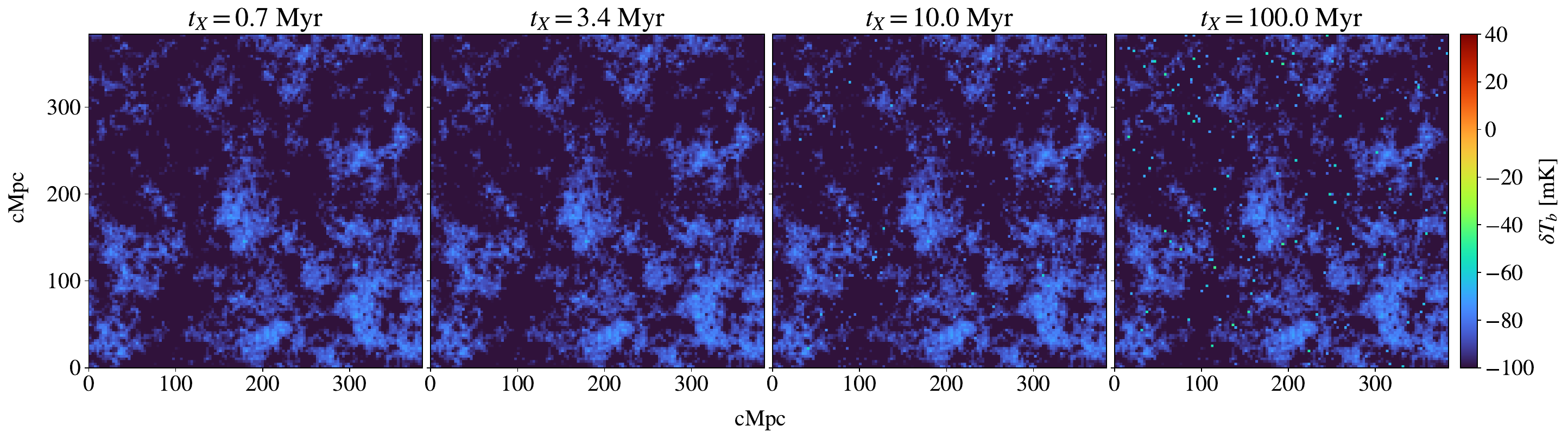}
    \caption{21-cm brightness temperature maps (with a slice depth of 3~cMpc) shown at $z=20$ for models with different $t_X$ and fixed values of $\alpha=0.2$ and $\fxiii=1$. From left to right: $t_X=0.7$, $3.4$, $10$ and $100~\mathrm{Myr}$. Here, the color scale is chosen to emphasize small-scale fluctuations driven by stochastic X-ray heating, which is different from that in Fig.~\ref{fig:temperature_maps}. In contrast to the heating rate field, the corresponding $\delta T_b$ maps show only modest sensitivity to $t_X$: the large-scale fluctuations remain nearly unchanged, while small-scale fluctuations increase with increasing average lifetime of XRBs. This is expected because $\delta T_b$ depends on the integrated heating history, which smooths short-timescale fluctuations. }
    \label{fig:tx_tb_maps_fixed}
\end{figure*}

In Figures \ref{fig:ps_tx_evolution} and \ref{fig:ps_tx_k} we quantify the impact of the XRB lifetime $t_X$ on the 21-cm power spectrum for a fixed XLF ($\alpha=0.2$) and our fiducial Pop~III parameter $\fxiii=1$. As noted earlier, $t_X$ controls both the effective number of independent XRB populations contributing within a simulation snapshot through the factor $\Delta t/t_X$ and the degree of temporal correlation across these snapshots. These two effects work together to produce the trend in Fig.~\ref{fig:ps_tx_evolution}: the longest lifetime ($t_X\sim10-100~\mathrm{Myr}$) produces the largest small-scale power, with $t_X=3.4~\mathrm{Myr}$ intermediate, and $t_X=0.7~\mathrm{Myr}$ having the lowest power.

\begin{figure}
    \centering
    % replace filename with your plotted figure
    \includegraphics[width=\columnwidth]{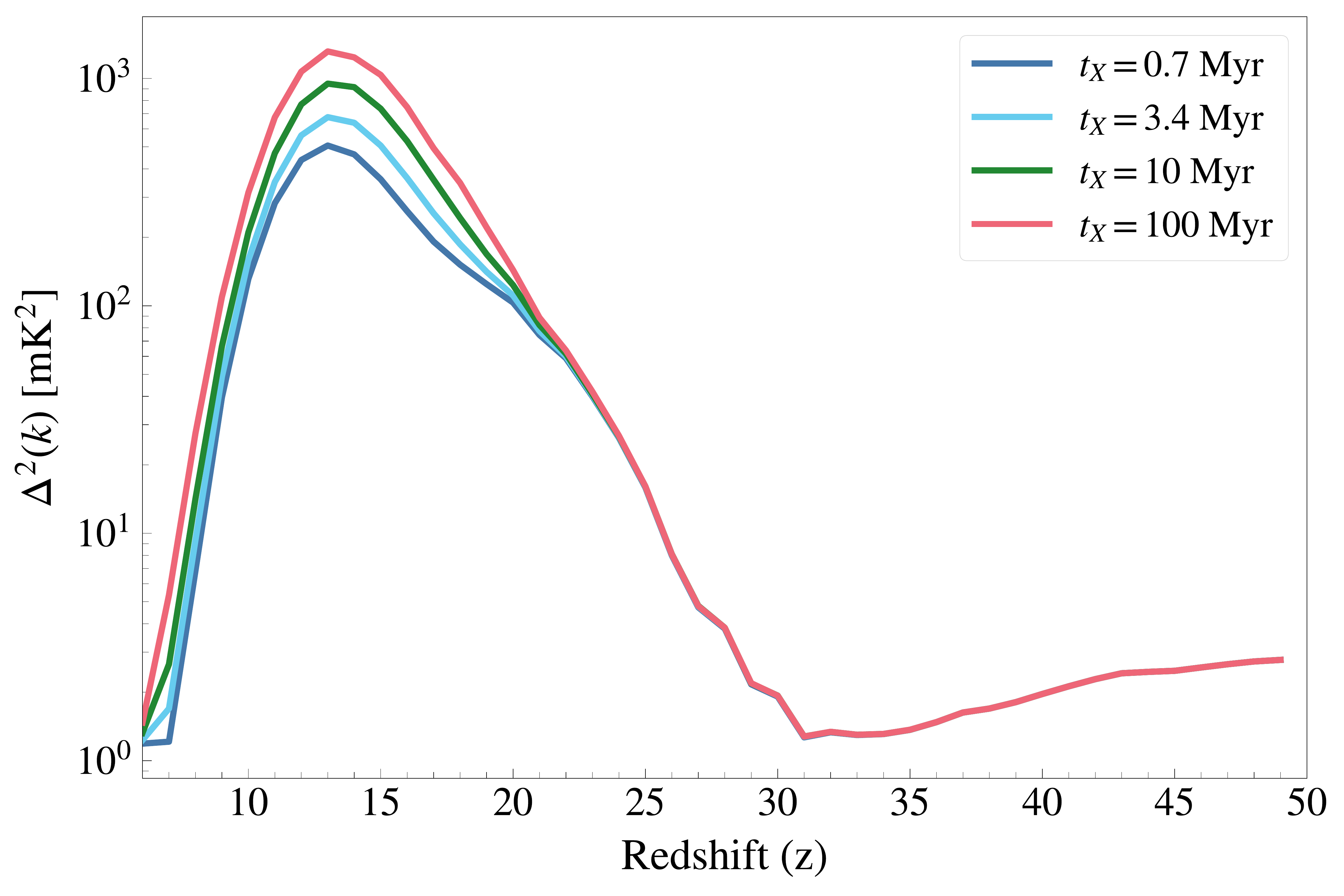}
    \caption{Redshift evolution of the 21-cm power at $k = 1~\mathrm{cMpc^{-1}}$ . We compare different stochastic runs using $t_X=0.7$, $3.4$, $10$ and $100~\mathrm{Myr}$ at fixed values of $\alpha=0.2$ and  $\fxiii=1$.
    Longer lifetimes systematically yield higher power, with the curves converging at the earliest times ($z\gtrsim 40$, corresponding to the star-less cosmic dark age before CD)%(very low $\hat N$ everywhere {\color{magenta} not clear what you mean by everywhere}) 
    and at late times when heating saturates.}
    \label{fig:ps_tx_evolution}
\end{figure}

\begin{figure}
    \centering
    % replace filename with your plotted figure
    \includegraphics[width=\columnwidth]{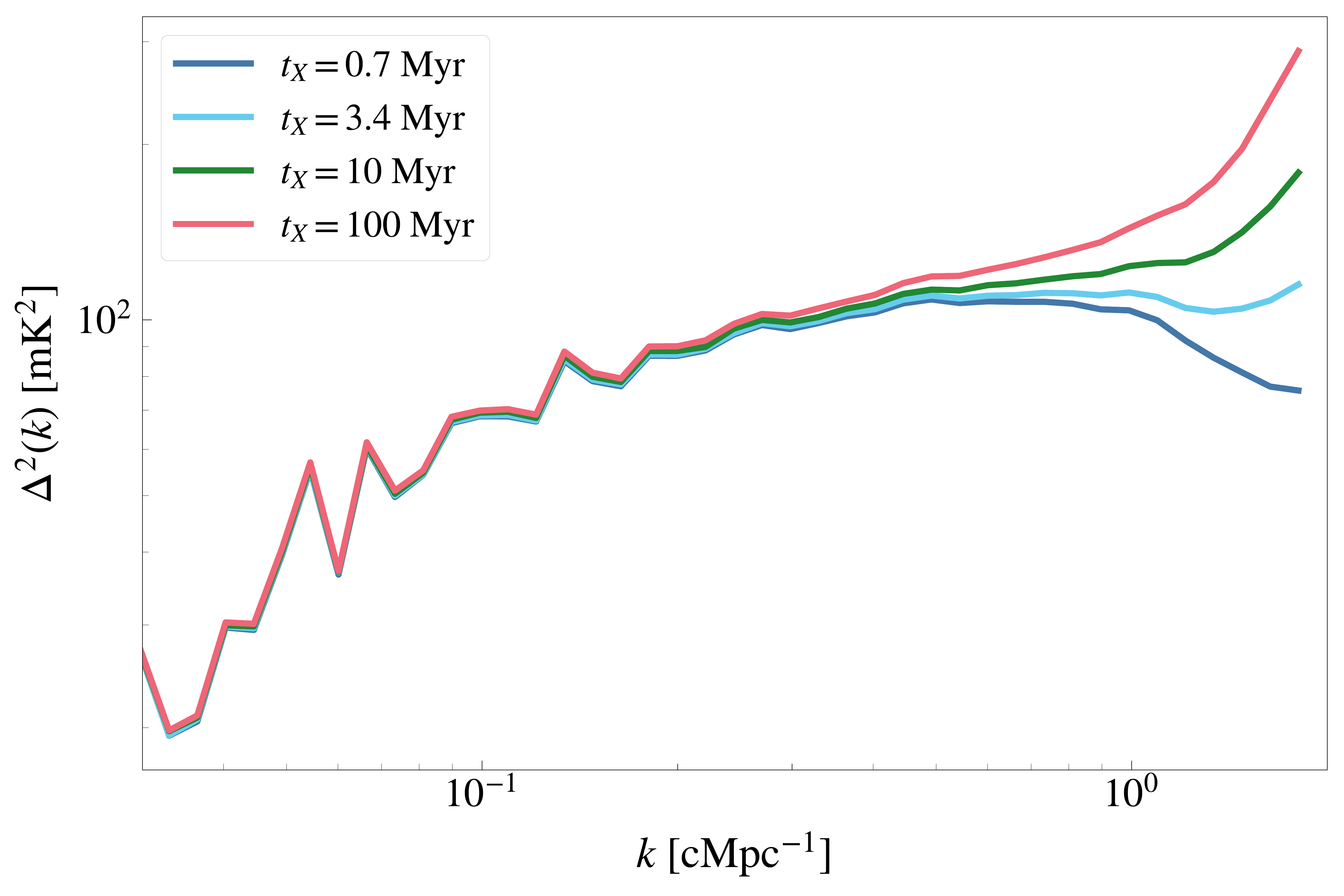}
    \caption{Power spectrum $\Delta^2(k)$ at $z=25$ as a function of wavenumber $k$ shown for models with $t_X=0.7$, $3.4$, $10$, and $100~\mathrm{Myr}$, given $\alpha=0.2$, $\fxiii=1$. Differences between the scenarios are negligible at large scales ($k\lesssim 0.3~\mathrm{cMpc}^{-1}$), but become apparent at small scales ($k\gtrsim 0.3~\mathrm{cMpc}^{-1}$), with the strongest fluctuations observed for $100~\mathrm{Myr}$.}
    \label{fig:ps_tx_k}
\end{figure}

In  Figure~\ref{fig:ps_tx_k},

we observe that at $k\gtrsim 0.3~\mathrm{cMpc}^{-1}$, longer $t_X$ boosts the power as fewer independent draws per step are required. This result highlights that it might be possible to constrain the mean XRB lifetime using the high-redshift small-scale 21-cm  power spectrum, although the effect could be degenerate with that of $\alpha$.  

\subsection{Potential Roles of Pop~II XRBs in Stochastic Heating}
\label{apdx:pop2}

\begin{figure}
    \centering
    \includegraphics[width=\columnwidth,height=\textheight,keepaspectratio]{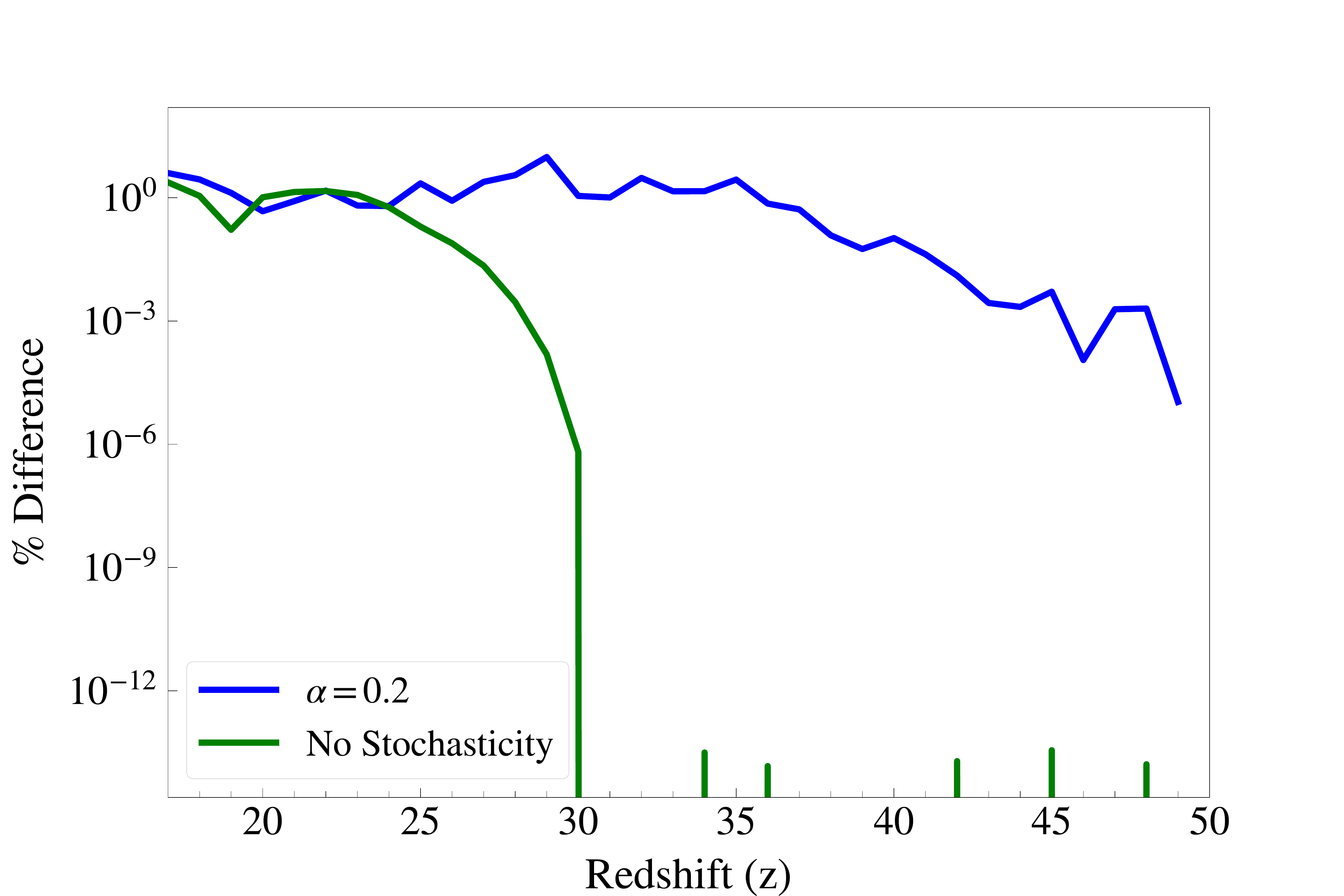}
    \caption{We show percentage difference in the 21-cm power spectra at $k=1~\mathrm{cMpc^{-1}}$ with ($f_{X,\mathrm{II}} = 1$) and without ($f_{X,\mathrm{II}} = 0$) the contribution of Pop~II XRBs. 
    Both cases are considered with and without stochastically modeled Pop~III XRBs keeping $\fxiii = 100$,  and $\alpha=0.2$. The difference remains small ($\lesssim 1\%$) in both scenarios, indicating that the contribution of Pop~II XRBs is unimportant in the regime where strong, stochastic X-ray heating regulates the 21-cm signal at $z\gtrsim 16$.}
    \label{fig:ps_fxii}
\end{figure}

In our analysis, we chose not to include stochastic modeling for Pop II X-ray binaries. This is justified by the typical small masses of Pop II stars formed in massive galaxies (relative to their Pop III counterparts), which results in an abundant population of Pop II XRBs. Furthermore, Pop II sources emerge when the IGM has already been preheated by the preceding Pop III XRBs; consequently, the relative thermal impact of individual Pop II sources is diminished. Finally, Pop III XRBs are expected to be much more efficient with $f_{X}=100$ for a log-flat Pop III IMF \citep{sartorio23}, while in the Pop II case the typical value is around $f_{X,\rm II}=1$ \citep{fragos13a}.  

To validate this decision, we present in Figure~\ref{fig:ps_fxii} the difference in the 21-cm power spectra at $k=1~\mathrm{cMpc^{-1}}$ between simulations with and without Pop II X-ray contributions, comparing the cases with $f_{X, \rm II}=1$ and  $f_{X,\rm II}=0$. We show the results for both stochastic and deterministic models of Pop~III XRBs, with $\fxiii = 100$, keeping the XLF slope $\alpha = 0.2$.
The results show that the impact of Pop~II XRBs on the power spectrum is marginal.  The percentage difference is at most a little more than $1 \%$ with either stochastic or deterministic Pop~III X-ray heating. This means that the inclusion of Pop~II XRBs does not have a dominant effect on the signal and does not 
enhance/reduce stochastic fluctuations in any observable way. This supports our assumption that Pop~II XRBs, though they significantly contribute to global heating at lower redshifts, hardly affect the stochasticity-driven variance at high redshifts ($z\gtrsim 16$) in the regime of early X-ray heating that we are concerned with here. In our case, this early heating is dominated by Pop~III XRBs given their enhanced efficiency $\fxiii = 100$ motivated by recent BPS simulations \citep{sartorio23}. If $\fxiii$ is smaller, as in a case of either the bottom-heavy Salpeter-like Pop III stellar IMF or an extremely top-heavy IMF \citep{gesseyjones25}, the epoch of heating is delayed, and/or instead driven by Pop~II XRBs. In this case, the stochastic effects are expected to be weaker and shifted to lower redshifts, as the global formation rate of Pop~II stars increases more rapidly over time than the Pop~III case, leading to faster late-time convergence to the deterministic case with rapidly increasing numbers of XRBs. A detailed investigation of the effects of Pop~II stochastic XRBs is deferred to future work.

\subsection{Inference Bias from Neglecting XRB Stochasticity}
\label{apdx:inference_bias}
In the main analysis we quantified the effect of Pop~III XRB stochasticity by comparing stochastic and non-stochastic simulations directly. A real observation, however, provides only one realization of the sky, which must be interpreted through an assumed model. If that model neglects stochasticity, the inferred astrophysical parameters can be biased. We therefore estimate this bias for the case identified in the main text as marginally detectable: the lunar Stage~III array, restricted to $k=0.5$-$1.0~\mathrm{cMpc^{-1}}$ and $z=18$-$22$. We take our most strongly stochastic run, with $\alpha=0.2$, $\fsiii=0.003$, and $\fxiii=100$, as the fiducial ``true'' signal.

The analysis is restricted to the two Pop~III parameters varied across our present simulation grid, $\fsiii$ and $\fxiii$. All other parameters in Table \ref{table:params} are held fixed at their fiducial values. A complete treatment of the full parameter vector would require additional derivative simulations for each parameter, and is left for future work. We use two independent methods, described below, and find consistent conclusions.

\subsubsection*{$\chi^2$ best-fit recovery}

We first fit the deterministic ($l\equiv1$) simulation grid directly to the fiducial stochastic power spectrum by minimizing
\begin{equation}
\chi^2(\fsiii,\fxiii) = \sum_{k,z}\frac{\left[\Delta^2_{\rm stoch}(k,z) - \Delta^2_{\rm det}(k,z;\fsiii,\fxiii)\right]^2}{\sigma_P^2(k,z)},
\end{equation}
where the sum is restricted to the lunar Stage~III region defined above. The uncertainty $\sigma_P(k,z)$ is evaluated from the fiducial stochastic signal using the lunar noise formalism of Section~\ref{sec:lunar} and the following subsection. We compute $\chi^2$ at each deterministic grid point, then fit a local quadratic surface to estimate the recovered ($\fsiii$, $\fxiii$) and the curvature around the minimum, in close analogy with a discretized Fisher matrix. The results are shown in Figure~\ref{fig:chi2_fit_lunar}.

The recovery of $\fxiii$ is stable. Restricting the fit to the well-bracketed neighbourhood $\fxiii\in\{10,100,1000\}$, where the injected value lies in the interior of the grid, the deterministic model recovers $\fxiii\approx64$ for an injected value of $100$. This corresponds to a bias of $-0.20$~dex, or $3.2$ times the marginal uncertainty from the fit. The result is essentially unchanged whether $\fsiii$ is held fixed or fitted jointly: in both cases the recovered value is $\fxiii\approx64$. We exclude $\fxiii=1$ from this local fit because $\chi^2$ is not monotonic between $\fxiii=1$ and $\fxiii=10$ at fixed $\fsiii=0.003$. This behaviour is physically plausible: as X-ray heating increases, the small-scale contrast can first grow and then saturate, in a manner analogous to the saturation with XRB lifetime $t_X$ discussed in Appendix~\ref{app:tx_fixedparams}. A single local parabola is therefore not a good description over the full range $\fxiii=1$-$1000$, but it is appropriate in the bracketed neighbourhood used here.

The recovery of $\fsiii$ is less well constrained. At fixed $\fxiii=100$, $\chi^2(\fsiii)$ has no interior minimum across the sampled interval, $0.003$-$0.01$; it remains lowest at the smallest simulated value, which is also the injected value, $\fsiii=0.003$. This is consistent with little or no bias in $\fsiii$, but the present grid does not allow us to place a precise bound on that bias. Simulations below $\fsiii=0.003$ would be needed to do so.

\begin{figure*}
    \centering
    \includegraphics[width=\textwidth]{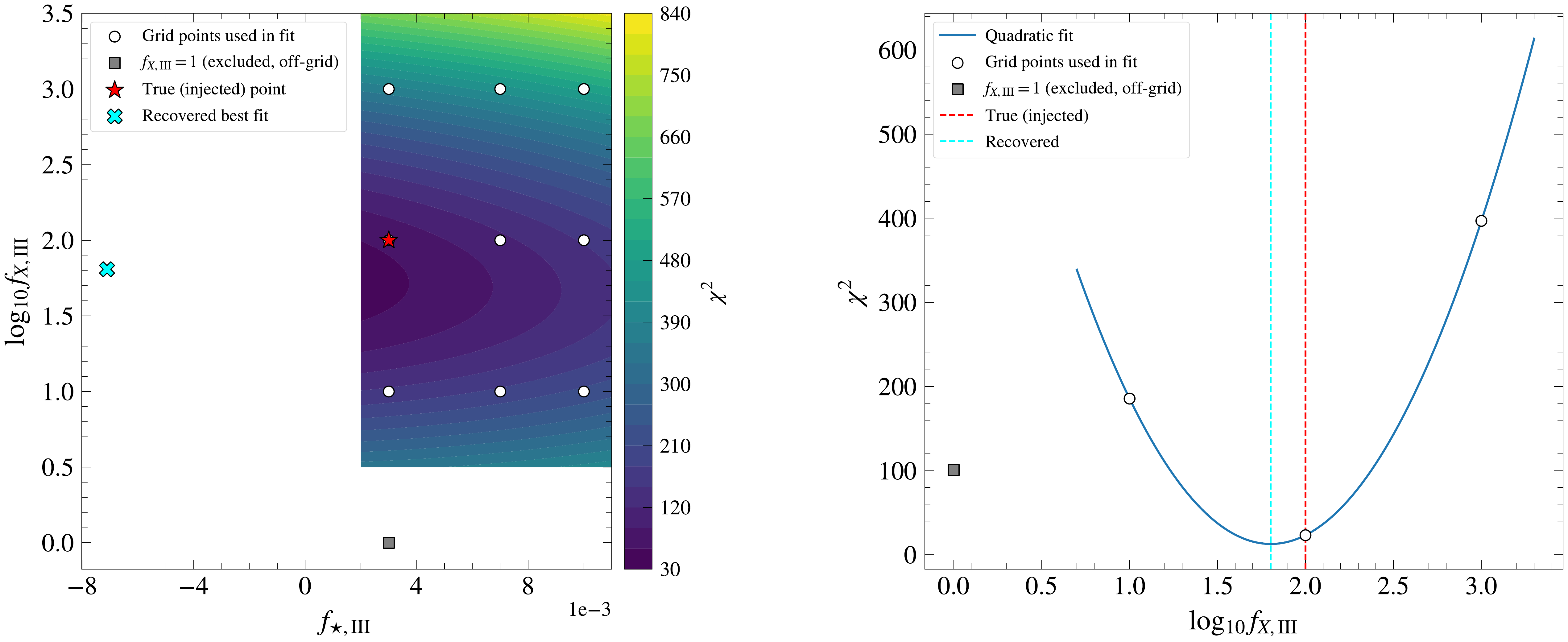}
    \caption{Recovery of ($\fsiii$, $\fxiii$) from a $\chi^2$ fit of deterministic-only simulations to our fiducial stochastic realization ($\alpha=0.2$, $\fsiii=0.003$, $\fxiii=100$), restricted to the lunar Stage~III marginally-detectable region ($k=0.5$-$1.0~\mathrm{cMpc^{-1}}$, $z=18$-$22$; $t_{\rm obs}=22{,}000$~hr total). \textit{Left:} $\chi^2$ surface (colour scale) from a quadratic fit to the nine deterministic grid points spanning $\fsiii\in\{0.003,0.007,0.010\}$ and $\fxiii\in\{10,100,1000\}$ (open circles), together with the injected parameter values (red star) and the recovered best fit (cyan cross). The grey square marks $\fxiii=1$, excluded from the fit (see text) but shown for reference; the recovered $\fsiii$ falls outside the sampled range, reflecting the absence of an interior minimum along that axis. \textit{Right:} 1-D parabola along $\fxiii$ at fixed $\fsiii=0.003$, using only the well-bracketed $\fxiii\in\{10,100,1000\}$ neighbourhood (open circles); the quadratic fit's minimum (recovered, cyan dashed line) is offset from the injected value (red dashed line) by $3.2\sigma$. The excluded $\fxiii=1$ point (grey square) lies above its neighbour at $\fxiii=10$, indicating a non-monotonic response of the power spectrum to $\fxiii$ over this range.}
    \label{fig:chi2_fit_lunar}
\end{figure*}

\subsubsection*{Fisher linear-bias estimate}

As an independent check, we also compute the linear inference bias following \citet{amara08} and \citep{taylor07}. If a deterministic model is fitted to data containing an unmodelled stochastic contribution $\Delta\Delta^2(k,z) \equiv \Delta^2_{\rm stoch}(k,z)-\Delta^2_{\rm det}(k,z)$, the parameter bias is $b_i = (F^{-1}B)_i$, with
\begin{align}
F_{ij} &= \sum_{k,z}\frac{1}{\sigma_P^2(k,z)}\frac{\partial \Delta^2}{\partial\theta_i}\frac{\partial \Delta^2}{\partial\theta_j}, \\
B_{i} &= \sum_{k,z}\frac{1}{\sigma_P^2(k,z)}\frac{\partial \Delta^2}{\partial\theta_i}\,\Delta\Delta^2(k,z),
\end{align}
for $\theta = (\fsiii,\log_{10}\fxiii)$\footnote{$\fxiii$ is treated logarithmically here since its simulation grid is evenly spaced in $\log_{10}\fxiii$ over two decades ($10,100,1000$), whereas the $\fsiii$ grid spans less than a factor of $4$ and is evenly spaced in linear $\fsiii$}. The calculation uses the same $(k,z)$ range, noise model, and fiducial point as the $\chi^2$ analysis. We estimate $\partial \Delta^2/\partial\fsiii$ from a quadratic fit through $\fsiii\in\{0.003,0.007,0.010\}$ at fixed $\fxiii=100$. The derivative $\partial \Delta^2/\partial\log_{10}\fxiii$ is estimated from a centred difference between $\fxiii=10$ and $\fxiii=1000$, matching the bracketed neighbourhood used above.

The two parameters are substantially correlated over this restricted region, with a correlation coefficient of $-0.78$ between $\fsiii$ and $\log_{10}\fxiii$. The Fisher matrix has a condition number of $\sim6\times10^{5}$ in this parameterization, although this value depends on the relative scaling of the parameters and is therefore not, by itself, a reliable measure of degeneracy. More importantly, with the present sparse simulation grid, the individual components of the joint bias $b=F^{-1}B$ are sensitive to the numerical derivative estimates. We therefore do not quote the joint displacement of either parameter as a robust bias estimate. Instead, we calculate the single-parameter bias for each parameter while holding the other fixed at its fiducial value, which does not require inversion of the two-parameter Fisher matrix. In this calculation, $\fxiii$ is biased low by $0.33$ in $\log_{10}\fxiii$, corresponding to a recovered value $\fxiii\approx47$ and a $2.0\sigma$ shift relative to the corresponding single-parameter Fisher uncertainty. By contrast, the bias in $\fsiii$ is only $0.46\sigma$ and is not formally significant. These results are consistent with the direct $\chi^2$ recovery: neglecting stochasticity preferentially shifts the inferred $\fxiii$ to lower values, while no significant bias in $\fsiii$ is resolved with the current simulation grid. Figures~\ref{fig:sigmap_lunar} and \ref{fig:ellipse_lunar} show the underlying significance map and the formal joint Fisher displacement relative to the Fisher uncertainties.

\begin{figure}
    \centering
    \includegraphics[width=\columnwidth]{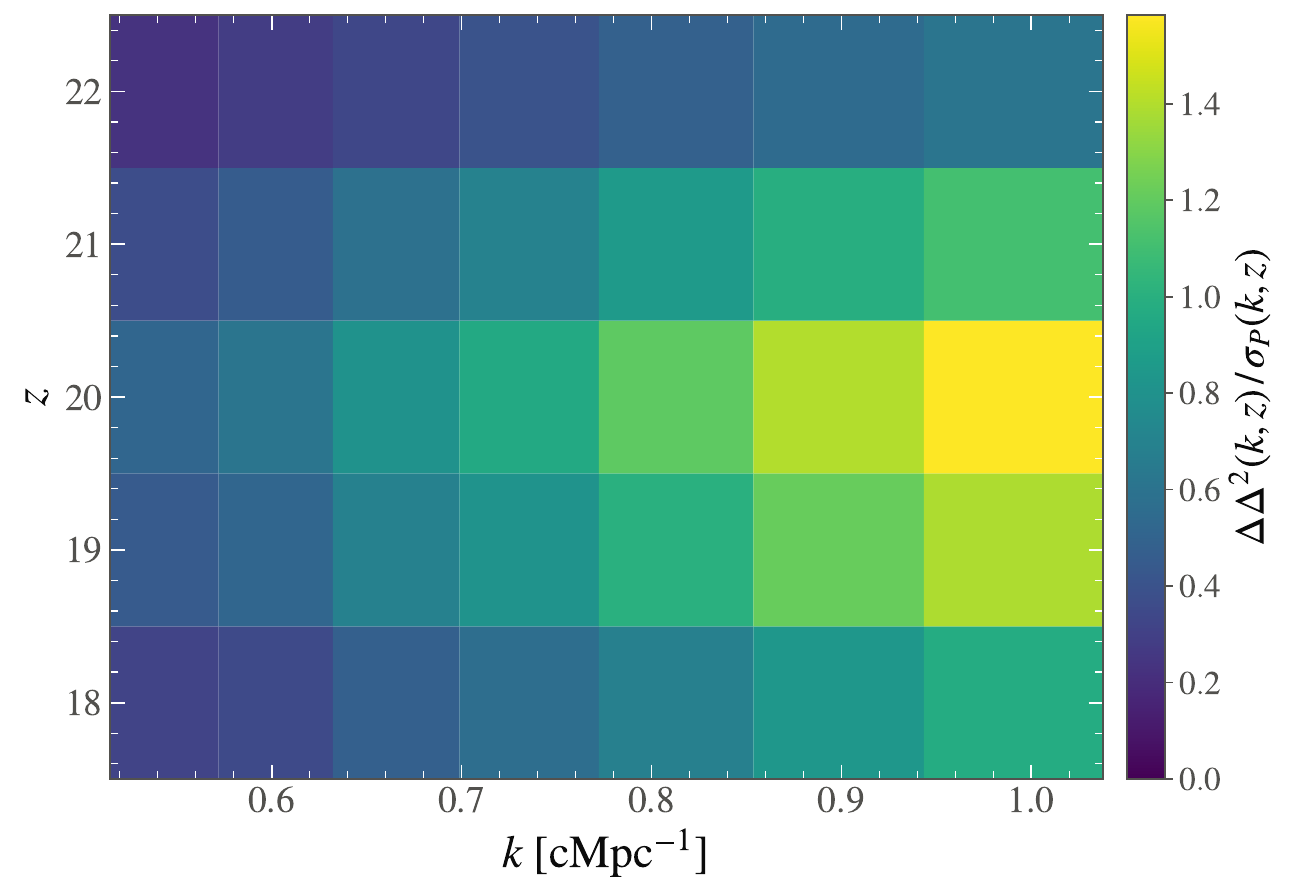}
    \caption{Difference between the stochastic and deterministic $21$-cm power spectra at our fiducial point ($\alpha=0.2$, $\fsiii=0.003$, $\fxiii=100$), $\Delta\Delta^2(k,z)$, in units of the lunar Stage~III $1$-$\sigma$ uncertainty $\sigma_P(k,z)$, restricted to the marginally-detectable region used throughout this appendix ($k=0.5$-$1.0~\mathrm{cMpc^{-1}}$, $z=18$-$22$; $t_{\rm obs}=22{,}000$~hr total). The difference is positive and peaks around $z\simeq20$ at the smallest scales shown.}
    \label{fig:sigmap_lunar}
\end{figure}

\begin{figure}
    \centering
    \includegraphics[width=\columnwidth]{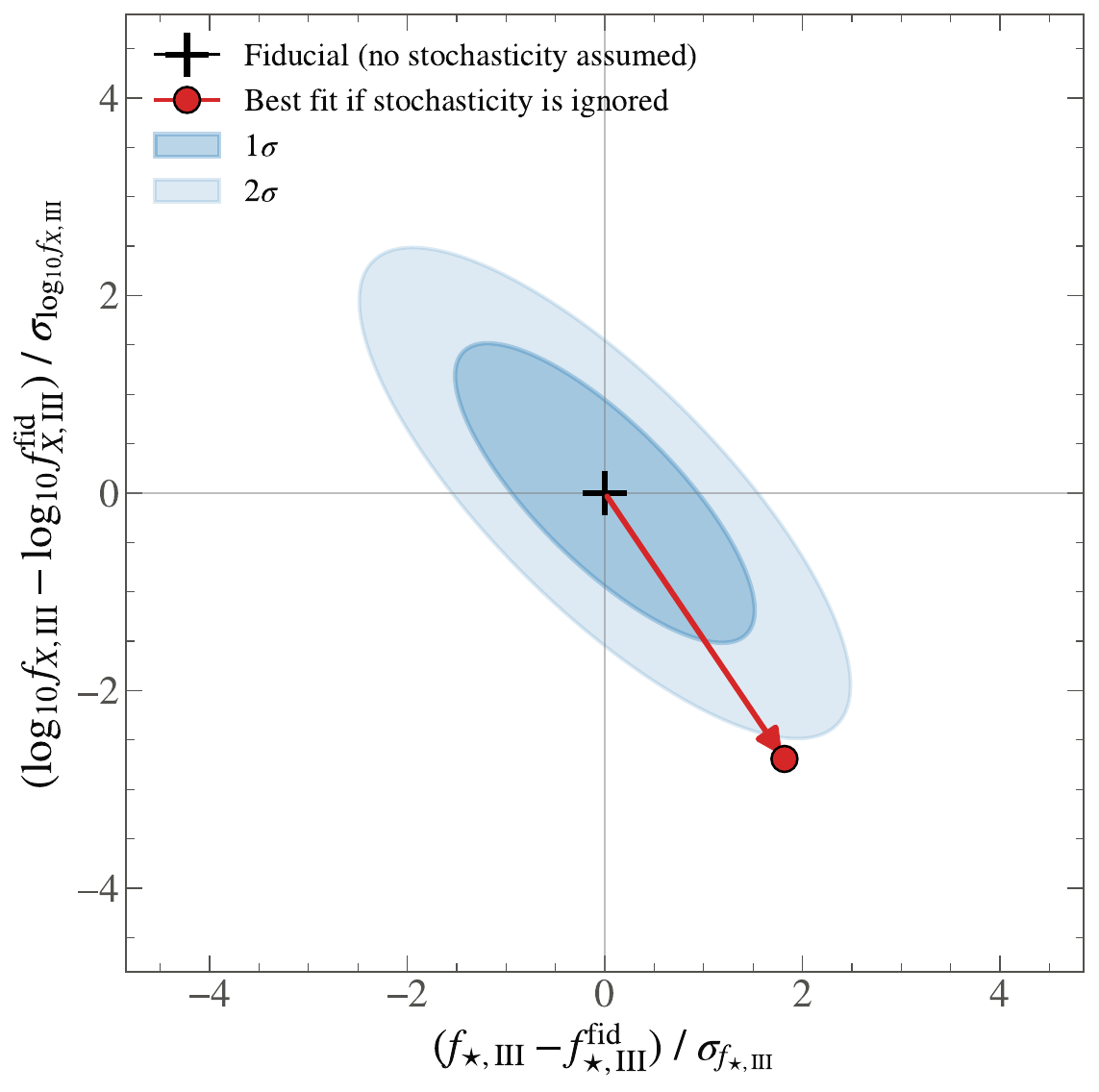}
    \caption{Fisher-matrix marginal $1$-$\sigma$ and $2$-$\sigma$ error contours (shaded) for ($\fsiii$, $\log_{10}\fxiii$) about the fiducial point (black cross), each axis normalized by its own marginal $1$-$\sigma$ uncertainty from the joint two-parameter fit. The red point, connected to the fiducial by an arrow, marks the parameter values that would be recovered by fitting a deterministic model to data containing the fiducial stochastic signal. The pronounced elongation of the contours reflects the strong correlation ($-0.78$) between the two parameters over this restricted region, which makes the joint bias estimate for either parameter individually unreliable (see text); the single-parameter estimates, which do not require inverting this near-degenerate matrix, are $0.46\sigma$ for $\fsiii$ and $2.0\sigma$ for $\log_{10}\fxiii$.}
    \label{fig:ellipse_lunar}
\end{figure}

Taken together, the two estimates indicate that neglecting XRB stochasticity in parameter inference can bias the recovered Pop~III X-ray efficiency $\fxiii$ at the few-$\sigma$ level under the observing conditions where the effect is marginally detectable in our model. We do not resolve a corresponding bias in the Pop~III star-formation efficiency $\fsiii$ with the present simulation grid. These estimates apply to the restricted two-parameter problem considered here, with all other astrophysical parameters held fixed. Allowing the full parameter set in Table~\ref{table:params} to vary would introduce additional degeneracies and could redistribute the model-mismatch bias among correlated parameters, while also broadening the inferred constraints. The magnitude and direction of the resulting parameter shifts therefore require a full multi-parameter analysis, which we leave for future work.

\subsection{Survival of the XLF-Stochasticity Signature under Shot Noise}
\label{apdx:poisson_survival}

Throughout this paper, XRB stochasticity is studied in isolation from other stochastic channels known to affect the 21-cm signal, such as Poisson shot noise and lognormal scatter in the star-formation efficiency \citep{reis22}. Isolating one channel at a time is standard practice when a new source of stochasticity is first introduced, but it leaves open the question of whether the XLF stochasticity signature is robust once a second stochastic channel is switched on, or whether it is instead washed out or dominated by it. We address this here.

At the fiducial parameters from Table~\ref{table:params} along with $\fsiii=0.003$ and $\fxiii=100$, we generated two additional simulation pairs with galaxy-Poisson shot noise switched on, one deterministic and one stochastic ($\alpha=0.2$) in the Pop~III XLF. The Poisson draw acts on the halo-count field used to compute star formation rates, and is applied globally in these runs. It perturbs the Pop~II and Pop~III SFR fields simultaneously, while the XRB stochasticity itself remains Pop~III-only, as elsewhere in the paper. Both members of each pair share the same Poisson random seed, so that the only difference between them is the presence or absence of XRB stochasticity. We compare the resulting power spectra to the paper's existing non-Poisson deterministic and stochastic baselines at the same fiducial parameters, over the region where the XRB stochasticity signature is strongest, $k=0.3$-$1.8~\mathrm{cMpc^{-1}}$ and $z=15$-$35$.

Figure~\ref{fig:poisson_survival} shows the redshift evolution of $\Delta^2(k,z)$ at $k=0.3$, $0.5$, and $1.0~\mathrm{cMpc^{-1}}$ for all four cases. Averaged over the tested region, the fractional difference between the stochastic and deterministic power spectra, $(\Delta^2_{\rm stoch}-\Delta^2_{\rm det})/\Delta^2_{\rm det}$, increases from $27\%$ without Poisson shot noise to $43\%$ with it switched on. At $k=0.3~\mathrm{cMpc^{-1}}$ the fractional signature grows from $2\%$ to $29\%$, while at $k=1.0~\mathrm{cMpc^{-1}}$, where the XLF-stochasticity signal is intrinsically strongest, it is reduced from $65\%$ to $51\%$ but remains large. The sign of the stochastic excess agrees between the Poisson-off and Poisson-on cases in $93\%$ of the tested $(k,z)$ cells, and the two difference fields are positively correlated.

This test isolates only the Poisson shot-noise component of the $f_\star$-stochasticity model of \citet{reis22}, and does not include the additional lognormal scatter in $f_\star$ used in that work, nor other possible stochastic channels such as IMF or SED variations. A fully joint treatment combining XLF stochasticity with these channels, accounting for their physical co-dependencies, remains a valuable future work. Even with this limitation, the present test shows that the XRB stochasticity signal, in several regimes becomes fractionally more prominent, once a second stochastic channel is also turned on.

\begin{figure*}
    \centering
    \includegraphics[width=\textwidth]{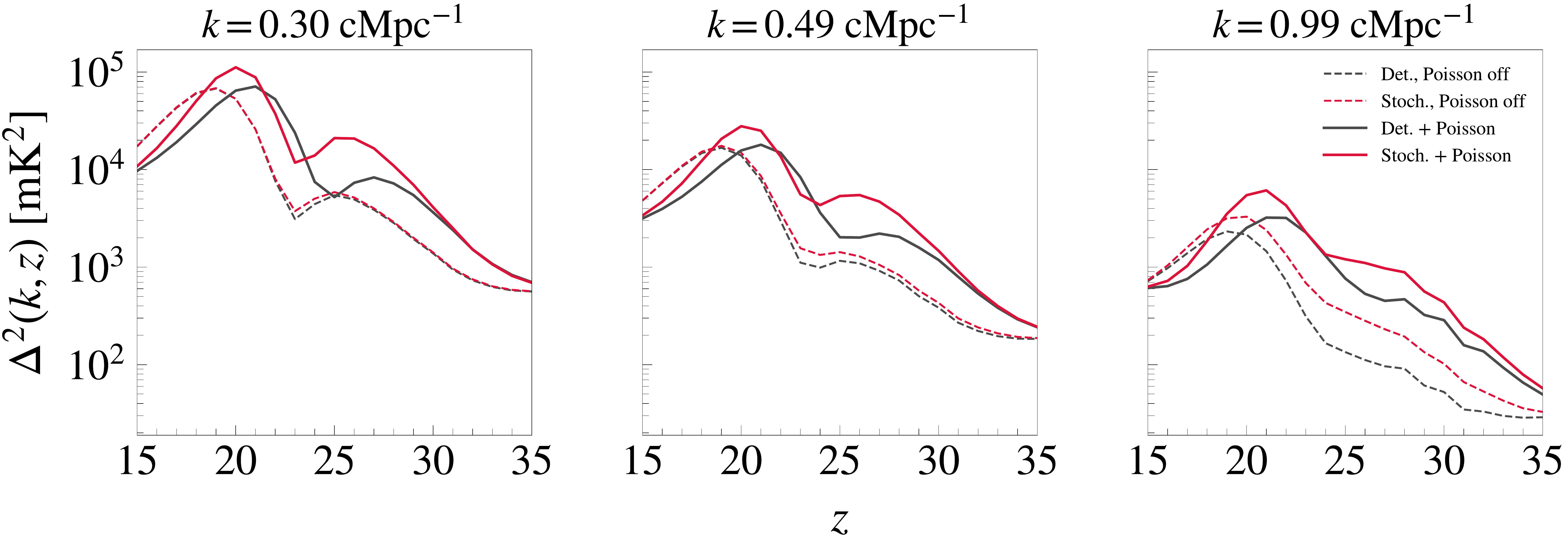}
    \caption{Redshift evolution of the 21-cm power spectrum, $\Delta^2(k,z)$, at $k=0.3$, $0.5$, and $1.0~\mathrm{cMpc^{-1}}$ (left to right), for the deterministic and stochastic ($\alpha=0.2$) models with Poisson shot noise switched off (dashed) and on (solid), at $\fsiii=0.003$, $\fxiii=100$. The stochastic excess over the deterministic case is visible in both the Poisson-off and Poisson-on pairs at every scale shown.}
    \label{fig:poisson_survival}
\end{figure*}
%%%%%%%%%%%%%%%%%%%%%%%%%%%%%%%%%%%%%%%%%%%%%%%%%%

% Don't change these lines
\bsp	% typesetting comment
\label{lastpage}
\end{document}